\documentclass[aps,pre,letterpaper,10pt,superscriptaddress,showpacs,nofootinbib,twocolumn]{revtex4-2}
\usepackage{amsfonts,amsmath,amssymb,bm,euscript,mathrsfs}
\usepackage[utf8]{inputenc}
\usepackage[T1]{fontenc}
\usepackage{txfonts}
\usepackage{textcomp}
\usepackage{lmodern}
\usepackage{graphicx}
\usepackage[tight,sf]{subfigure}
\usepackage{hyperref}
\usepackage{bookmark}
\usepackage{color,soul}
\usepackage{url}
\newcommand{\rma}{\mathrm{a}}
\newcommand{\rmT}{\mathrm{T}}
\newcommand{\rmL}{\mathrm{L}}
\newcommand{\rmk}{\mathrm{k}}
\newcommand{\mrmp}{\mathrm{p}}
\newcommand{\rmP}{\mathrm{P}}
\newcommand{\scrL}{\mathscr{L}}
\newcommand{\scrH}{\mathscr{H}}

\newcommand{\bfX}{\mathbf{X}}
\newcommand{\bfY}{\mathbf{Y}}
\newcommand{\bfx}{\mathbf{x}}
\newcommand{\bfe}{\mathbf{e}}
\newcommand{\bfn}{\mathbf{n}}
\newcommand{\bff}{\mathbf{f}}
\newcommand{\bfz}{\mathbf{z}}
\newcommand{\bfT}{\mathbf{T}}
\newcommand{\bfL}{\mathbf{L}}

\newcommand{\tell}{\tilde{\ell}}
\DeclareMathOperator{\arcsinh}{arcsinh}
\DeclareMathOperator{\arctanh}{arctanh}
\DeclareMathOperator{\arccot}{arccot}
\DeclareMathOperator{\sech}{sech}

\begin{document}

\title{Equatorial deformation of homogeneous spherical fluid vesicles by a rigid ring}

\author{Pablo Vázquez-Montejo}
\email[]{pablo.vazquez@secihti.mx}
\email[]{pablo.vazquez@correo.uady.mx}
\affiliation{SECIHTI - Facultad de Matemáticas, Universidad Autónoma de Yucatán, Periférico Norte, Tablaje 13615, 97110, Mérida, Yucatán, México}
\author{Bojan Bo\v{z}i\v{c}}
\email[]{bojan.bozic@mf.uni-lj.si}
\affiliation{Institute of Biophysics, Faculty of Medicine, University of Ljubljana, Vrazov trg 2, SI-1000 Ljubljana, Slovenia}
\author{Jemal Guven}
\email[]{jemal@nucleares.unam.mx}
\affiliation{Instituto de Ciencias Nucleares, Universidad Nacional Autónoma de México, Apdo. Postal
70-543, 04510, Coyoacán, Ciudad de México, México}

\begin{abstract}
We examine the deformation of homogeneous spherical fluid vesicles along their equator by a circular rigid ring. We consider deformations preserving the axial and equatorial mirror symmetries of the vesicles. The configurations of the vesicle are determined employing the spontaneous curvature model subject to the constraints imposed by the ring as well as of having constant area or volume. We determine two expressions of the force exerted by the ring, one involving a discontinuity in the derivative of the curvature of the membrane across the ring, and another one in terms of the global quantities of the vesicle. For small enough values of the spontaneous curvature there is only one sequence of configurations for either fixed area or volume. The behavior of constricted vesicles is similar for both constraints, they follow a transition from prolate to dumbbell shapes, which culminates in two quasispherical vesicles connected by a small catenoid-like neck. We analyze the geometry and the force of the small neck employing a perturbative analysis about the catenoid. A stretched vesicle initially adopts an oblate shape for either constraint. If the area is fixed the vesicle increasingly flattens until it attains a disclike shape, which we examine using an asymptotic analysis. If the volume is fixed the poles approach until they touch and the vesicle adopts a discocyte shape. When the spontaneous curvature of the vesicle is close to the mean curvature of the constricted quasispherical vesicles, the sequences of configurations of both constraints develop bifurcations, and some of the configurations corresponding to one of their branches have the lowest energy.
 \end{abstract}

\maketitle

\section{Introduction}

\begin{figure}[t]
\begin{center}
\begin{tabular}{ccc}
  $\vcenter{\hbox{\includegraphics[scale=0.35]{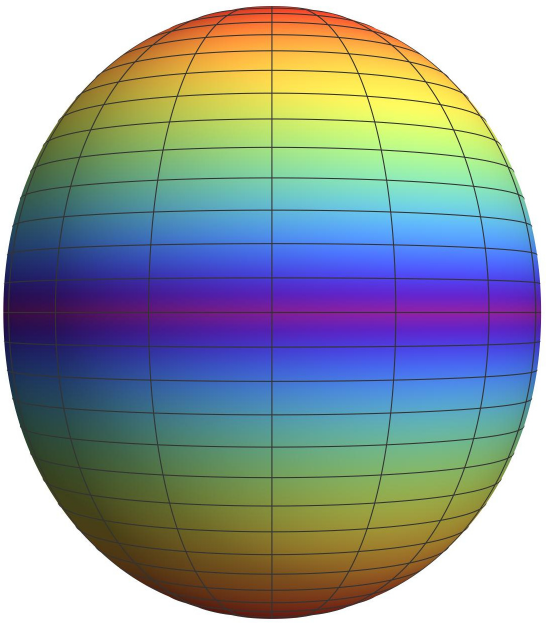}}}$ &
  $\vcenter{\hbox{\includegraphics[scale=0.35]{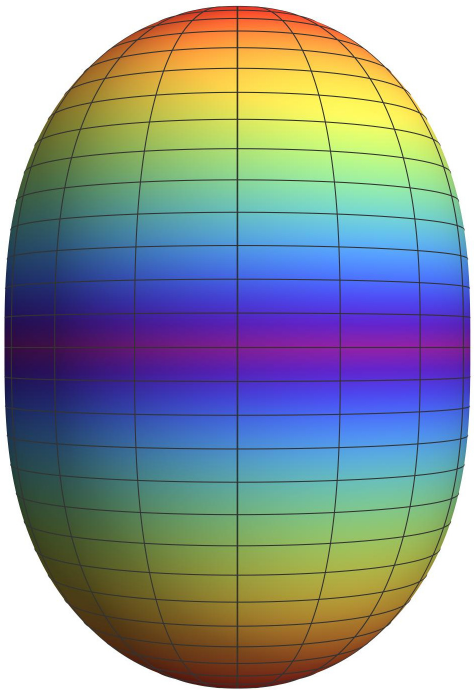}}}$ &
  $\vcenter{\hbox{\includegraphics[scale=0.35]{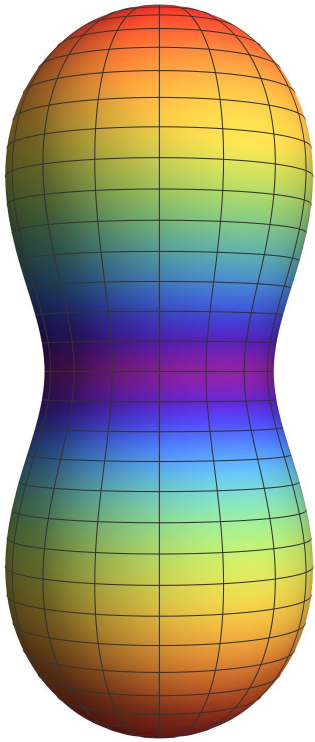}}}$ \\
  {\small (a) $r_0=0.95$} & {\small (b) $r_0=0.845$} & {\small (c) $r_0=0.5$}
  \end{tabular}
\end{center}
\begin{center}
\begin{tabular}{ccc}
   $\vcenter{\hbox{\includegraphics[scale=0.35]{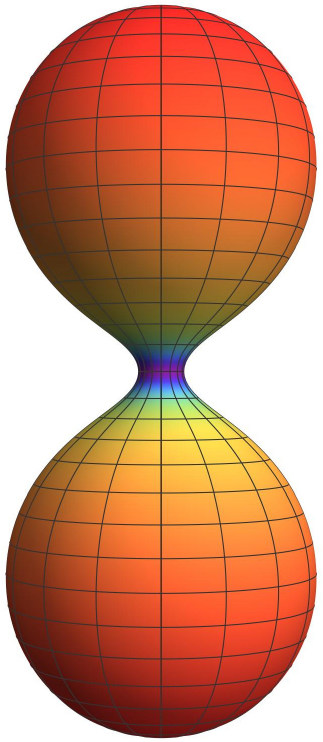}}}$ &
  $\vcenter{\hbox{\includegraphics[scale=0.35]{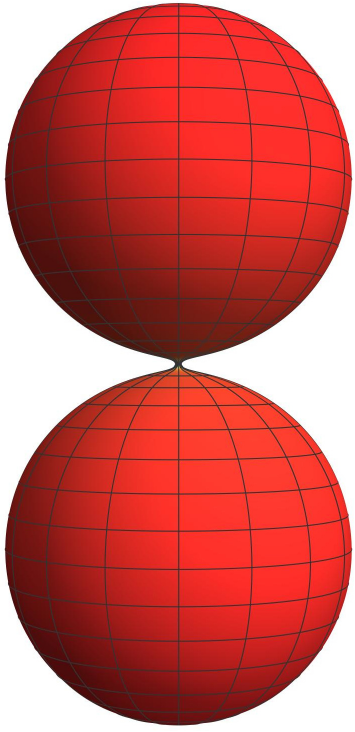}}}$ &
  $\vcenter{\hbox{\includegraphics[scale=0.4]{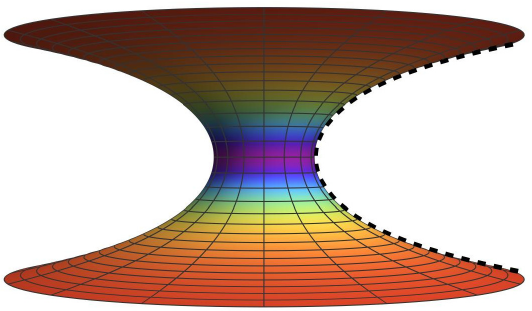}}}$ \quad \\
  {\small (d) $r_0=10^{-1}$} & {\small (e) $r_0=10^{-2}$} & {\small (f) $r_0=10^{-3}$}
    \end{tabular}
\end{center}
    \includegraphics[scale=0.6]{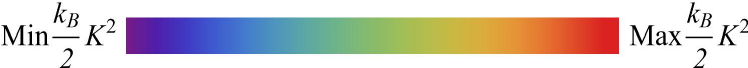}
\caption{(a)-(e) Sequence of constricted vesicles with fixed area and $C_s=0$. (e) At the maximum constriction, the vesicle consists of two quasispherical surfaces connected by a small catenoid-like neck. (f) Magnification of the small neck. The bending energy density is color coded.}
\label{fig:1}
\end{figure}

\begin{figure}[t]
\begin{center}
 \begin{tabular}{cc}
  $\vcenter{\hbox{\includegraphics[scale=0.425]{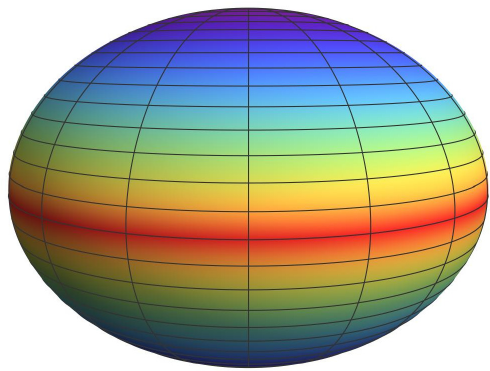}}}$ &
  $\vcenter{\hbox{\includegraphics[scale=0.425]{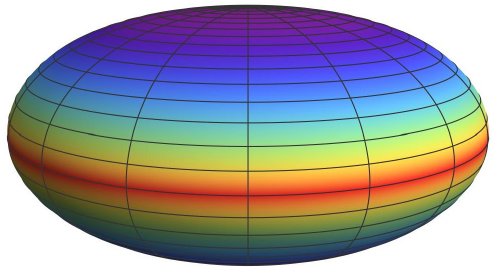}}}$ \\
 {\small (a) $r_0=1.1$} & {\small (b) $r_0=1.2$} \\
  $\vcenter{\hbox{\includegraphics[scale=0.425]{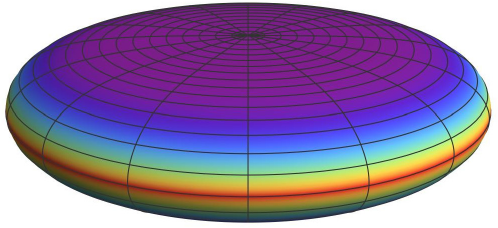}}}$ &
  $\vcenter{\hbox{\includegraphics[scale=0.35]{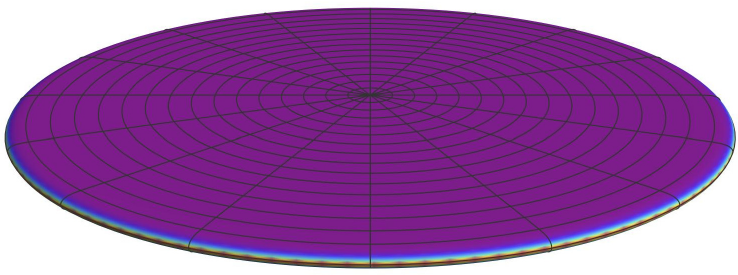}}}$ \\
  {\small (c) $r_0=1.3$} & {\small (d) $r_0=1.4$} 
  \end{tabular}
\end{center}
  \includegraphics[scale=0.6]{fig1g}
\caption{(a)-(d) Sequence of stretched vesicles with fixed area and $C_s=0$. (d) At the maximum stretching, the vesicle consists of two discs joined along their boundary. The bending energy density is color coded.}
\label{fig:2}
\end{figure}

\begin{figure}[htb]
\begin{center}
 \begin{tabular}{cc}
  $\vcenter{\hbox{\includegraphics[scale=0.425]{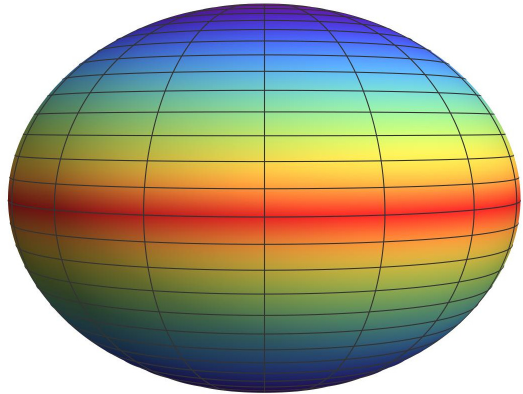}}}$ &
  $\vcenter{\hbox{\includegraphics[scale=0.425]{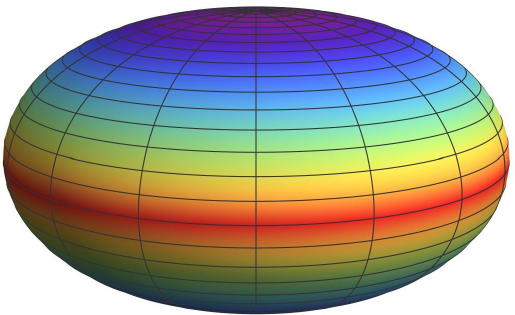}}}$ \\
 {\small (a) $r_0=1.1$} & {\small (b) $r_0=1.2$} \\
  $\vcenter{\hbox{\includegraphics[scale=0.425]{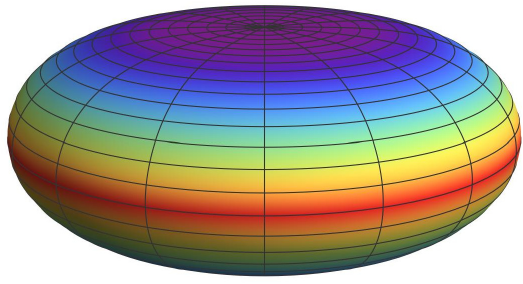}}}$ &
  $\vcenter{\hbox{\includegraphics[scale=0.425]{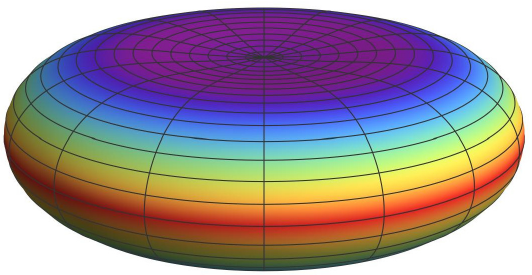}}}$ \\
  {\small (c) $r_0=1.3$} & {\small (d) $r_0=1.4$} \\
  $\vcenter{\hbox{\includegraphics[scale=0.425]{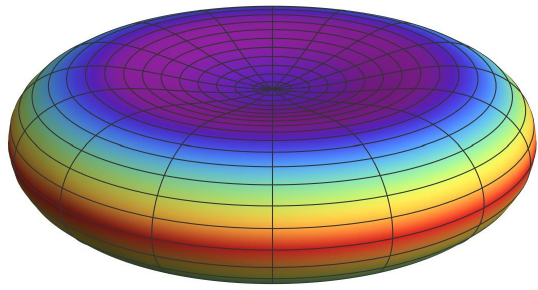}}}$ &
  $\vcenter{\hbox{\includegraphics[scale=0.425]{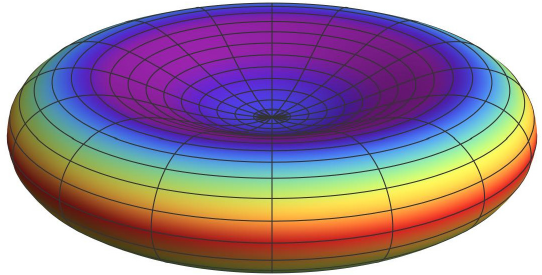}}}$ \\
{\small (e) $r_0=1.5$} & {\small (f) $r_0=1.63$} 
  \end{tabular}
  \end{center}
  \includegraphics[scale=0.6]{fig1g}
\caption{(a)-(f) Sequence of stretched vesicles with fixed volume and $C_s=0$. (f) At the maximum stretching, the vesicle adopts a discocyte shape. The bending energy density is color coded.}\label{fig:3}
\end{figure}

Many biophysical processes involve the interaction of biopolymers with membranes \cite{ThalmannBook}. Transmembrane proteins as well as proteins docking on the membrane surface will tend to deform the membrane geometry locally, which in turn will mediate interactions between proteins. Dramatic large but still localized deformations may also occur, as is strikingly illustrated by tethers formation \cite{Heinrich1999, Powers2002, Derenyi2002}. Proteins can also bind to the membrane and assemble into a filament or a protein complex, giving rise to stresses that induce large deformations, which may play a role in shaping the global morphology of the membrane, as well as change its topology. The prototypical example in the formation of vesicles is the endocytosis, driven by the helical protein dynamin, which is also involved in the process of cellular division in certain bacteria. The dynamin self-assembles into a helical elastic filament on the narrow neck of a membrane, and by applying an inward traction promotes its fission \cite{Kozlov2001, Morlot2013}. Another important example is the cytokinesis, in which the ESCRT-III protein polymerizes on the cell membrane forming a contractile ring, which  exerts a force that produces its division into two new cells \cite{Yang2017, Harker2021}. Also, by employing Monte Carlo simulations it has been shown that the mechanical constriction of vesicles could be driven by synthetic contractile rings composed of nanoparticles \cite{Bahrami2019}.
\\
In this work, as an approach to model the relevant physical aspects of this kind of cellular processes, we examine the response of an homogeneous spherical fluid membrane of radius $R_S$ and spontaneous curvature $C_s$ under the action of a rigid polymer loop of radius $R_0$ binding along its equator. The deformed vesicles are characterized by the ratio $r_0=R_0/R_S$: if $r_0 < 1$ ($r_0 >1$) the ring constricts (stretches) the vesicle along its equator. Both deformations have biophysical relevance: the constriction of the membrane is an important step in the fission process induced by a protein complex, whereas the stretching of the membrane might represent the situation when outward traction is applied to the membrane or the confinement of a loop within the vesicle, scenario that contrasts with the very different limit studied previously in which a semiflexible polymer is confined within a rigid sphere \cite{Spakowitz2003, Stoop2011, GuvenVazquez2012}. Although this model greatly simplifies the biophysics, it provides several pointers to a better understanding of the different stages of such cellular processes, in particular, it permits us to quantify the force deforming the vesicle.
\\
We determine the configurations of the membrane employing the spontaneous curvature model subject to the local constraint that the equator conforms to the ring, and to the global constraints that either the total area or the total volume of the membrane is fixed. The Euler-Lagrange (EL) equation of a fluid membrane, whose solutions describe the equilibrium configurations, can be expressed in terms of the conservation law of the stress tensor, which is also determined by the geometry of the membrane \cite{EvansSkalak1980, Jenkins1977, Steigman1999, CapoGuven2002, Guven2004, Lomholt2006, Deserno2015, Guven2018}. In Sec.\ \ref{Sect:ELeq} we present how the constraint imposed by the ring can be implemented in the variational principle by means of a vectorial Lagrange multiplier, which modifies the boundary terms.
\\
For the sake of simplicity, we consider deformations preserving the axial symmetry, as well as the mirror symmetry with respect to the equatorial plane. The axial symmetry implies the existence of a first integral of the EL equation: the integration of the projection of conservation law of the stress tensor along the axis of symmetry provides a direct derivation \cite{CapoGuven2002}, which we present in Sec.\ \ref{Sect:FirsInt} for the spontaneous curvature model. The presence of the ring at the equator sets the boundary condition for the radial coordinate of the membrane. Furthermore, the condition of finite energy demands the continuity of the tangent planes across the equator and at the poles, such that kinks do not occur. To examine the response of the membrane to the force exerted by the ring, the first integral is solved numerically with these boundary conditions in the two source-free regions into which the membrane is partitioned.
\\
The force exerted by the ring induces a distribution of stresses on the membrane. If the force is negative (positive) the ring exerts an inward (outward) radial force. In Sec.\ \ref{Sect:Equatorialforce} we examine the local and global effects of the force exerted by the ring on the surface geometry. First, from the balance of stresses at the equator we find that the external force originates a discontinuity in the arclength derivative of the meridian curvature. Also, from the scale invariance of the energy we obtain the external force in terms of the global quantities of the membrane and the Lagrange multipliers fixing area and volume.
\\
The length scale is given by the radius of the spherical vesicle $R_S$. In Sec.\ \ref{Sect:nondimquant} we define the geometric and physical quantities rescaled with appropriate powers of $R_S$ that we employ in the discussion of the deformed configurations of the membranes.
\\
We examine the equilibrium shapes of membranes with different values of spontaneous curvature and with their area or volume fixed in Secs. \ref{Sec:Aconst} and \ref{Sec:Vconst} respectively. For membranes with spontaneous curvature lower than certain values we obtain a unique configuration for each value of the equatorial radius.
\\
The behavior of constricted membranes ($0< r_0 <1$), illustrated in Fig.\ \ref{fig:1}, is similar for fixed area or volume, differing only in their size. Slightly constricted membranes adopt a prolate shape (Fig.\ \ref{fig:1}(a)), followed by a transition to a nonconvex dumbbell geometry (Figs.\ \ref{fig:1}(b)-(c)). Upon further constriction, a second transition occurs associated with the conversion to a pair of quasispherical vesicles connected by an infinitesimal neck (Figs.\ \ref{fig:1}(d)-(e)). The constriction involves a finite inward force, but its dependence on the equatorial radius is not monotonic, it varies between a local minimum and local maximum values before decreasing towards a finite value in the limit of the very small neck. Usually, to analyze such necks it is assumed that they can be described by a section of an unduloid (a constant mean curvature surface) and the corresponding results are compared with numerical results \cite{Fourcade1994, Agudo2016, Lipowsky2022, Lipowsky2024}. By looking at the numerical solutions the neck apparently adopts a catenoidal shape in this limit (shown in Fig.\ \ref{fig:1}(f)). However, since the catenoid is a minimal surface it satisfies the source-free EL equation, so it cannot represent a constricted small neck, for which a finite force persists. A more careful analysis employing perturbation theory reveals that the geometry of the small neck is well approximated by a first order perturbation of the catenoid, which modifies its mean curvature such that it alternates between the spontaneous curvature of the membrane and the mean curvature of the quasispherical vesicles, making it compatible with a finite equatorial force. In addition, the perturbative analysis provides the force exerted on these small necks, which is proportional to the difference of between the mean curvature of the quasispherical vesicles and the spontaneous curvature of the membrane.
\vskip0pc \noindent
The sequences of stretched vesicles ($r_0>1$) with fixed area and fixed volume are illustrated in Figs.\ \ref{fig:2} and \ref{fig:3} respectively. Slightly stretched membranes adopt an oblate shape for both constraints (Figs.\ \ref{fig:2}(a)-(b) and \ref{fig:3}(a)-(b)), but as the equatorial radius increases the shapes of vesicles with fixed area and fixed volume become very dissimilar. If the area is fixed, the membrane flattens more and more (Fig.\ \ref{fig:2}(c)) until it consists of two flat discs joined along their perimeter (Fig.\ \ref{fig:2}(d)). The force increases monotonically as the membrane is stretched and diverges in the limit of the disclike shape. If the volume is fixed, the poles of the membrane get closer and closer (Figs.\ \ref{fig:3}(c)-(e)) until they touch, whereas the exterior remains round, so the membrane adopts a shape similar to a discocyte (Fig.\ \ref{fig:3}(f)), for which the force is finite.
\begin{figure}[htb]
\begin{center}
 \begin{tabular}{ccc}
  $\vcenter{\hbox{\includegraphics[scale=0.4]{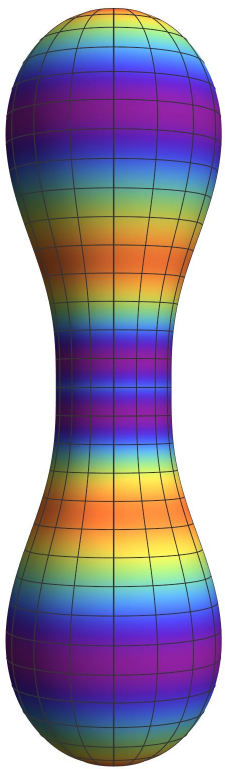}}}$ &
  $\vcenter{\hbox{\includegraphics[scale=0.4]{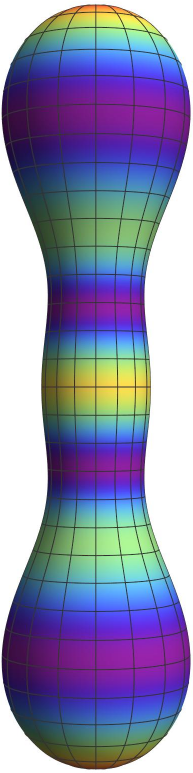}}}$ &
  $\vcenter{\hbox{\includegraphics[scale=0.4]{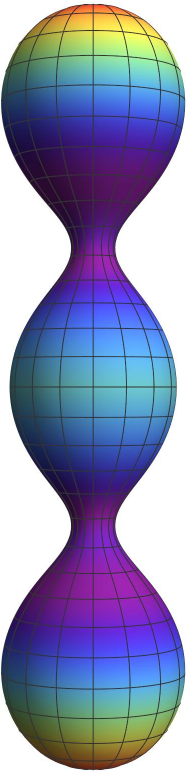}}}$ \\
{\small (a) $r_0=0.313$} & {\small (b) $r_0=0.325$} & {\small (c) $r_0=0.5$}
  \end{tabular}
\end{center}
\begin{center}
 \begin{tabular}{cc}
$\vcenter{\hbox{\includegraphics[scale=0.4]{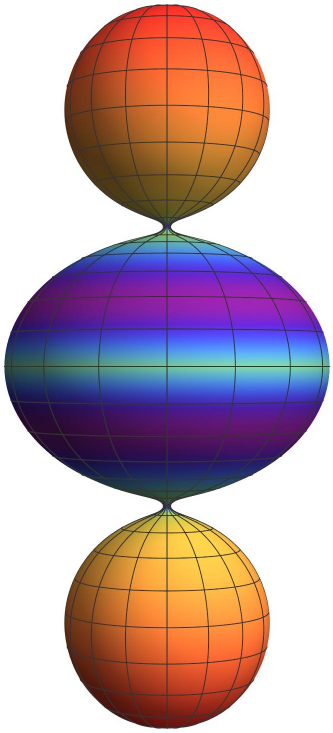}}}$ &
  $\vcenter{\hbox{\includegraphics[scale=0.325]{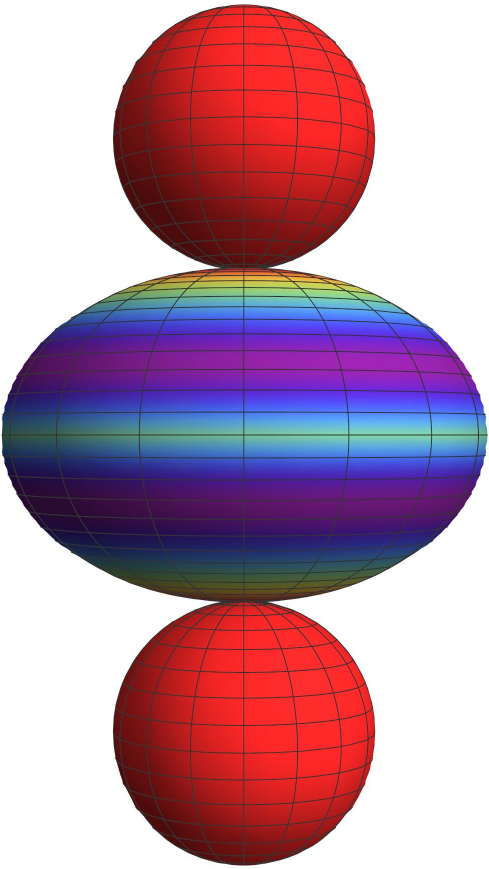}}}$ \\
{\small (d) $r_0=0.78$} & {\small (e) $r_0=0.85$}
   \end{tabular}
\end{center}
  \includegraphics[scale=0.6]{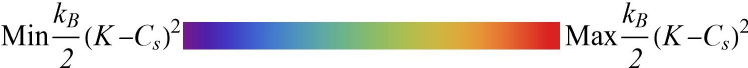}
\caption{(a)-(d) Sequence of vesicles corresponding to a branch of a bifurcation with fixed area and $C_s=2\sqrt{2}/R_S$ and (e) $C_s=3/R_S$. The bending energy density is color coded.}
\label{fig:4}
\end{figure}
\vskip0pc \noindent
For both constraints the increase of the spontaneous curvature to values close to the mean curvature of the quasispherical vesicles, obtained in the limit of maximum constriction, gives rise to bifurcations and two new branches in the solutions of the EL equation. For either constraint both branches begin with different constricted vesicles and for the case of fixed area they end with a common stretched vesicle, whereas for the case of fixed volume they end with different stretched configurations. In both cases, below a certain value of the equatorial radius the configurations belonging to one of the branches possess lower total bending energy than the configurations of the original sequence of constricted vesicles with the same equatorial radius. The sequence of configurations corresponding to those branches is illustrated in Fig.\ \ref{fig:4} for fixed area (the sequence for fixed volume is similar). It begins with a dumbbell shape (Fig.\ \ref{fig:4}(a)), as the vesicle is stretched its central region bulges, resembling an onduloid (Figs.\ \ref{fig:4}(b)-(c)), which morphs into a central oblate shape connected by small necks to two quasispherical vesicles above and below (Fig.\ \ref{fig:4}(d)). Upon further stretch and increase of spontaneous curvature the necks become vanishingly small (Fig.\ \ref{fig:4}(e)), but which are free of external forces, similar to the remote constriction of a vesicle \cite{Bozic2014}.
\\
The conclusions and discussion are presented in Sec.\ \ref{sectdiscussion}. The parametrization and geometric quantities of an axisymmetric surface are reviewed in Appendix \ref{App:axisym}. The derivations of the first integral of the EL equation and of the external force adapted to axisymmetric surfaces are presented in Appendix \ref{App:HamDer}.

\section{Bending energy and Euler-Lagrange equation} \label{Sect:ELeq}

The membrane is represented by its midsurface, parametrized by coordinates $u^a$, $a=1,2$, and embedded in  three-dimensional space by the mapping $\Sigma : (u^1,u^2) \mapsto \bfX(u^1,u^2)$. The two tangent vectors adapted to this parametrization are $\bfe_a=\partial_a \bfX$; $\bfn$ is the unit outward normal; $g_{ab} = \bfe_a\cdot \bfe_b$ is the metric tensor and $g^{ab}$ its inverse; $K_{ab} = \bfe_a\cdot \partial_b \bfn$ is the extrinsic curvature tensor. The two invariants of the shape operator $K^{a}_{\phantom{a}b} = g^{ac} K_{cb}$ are (twice) the mean curvature, $K = \mathrm{tr} K^{a}_{\phantom{a}b} = C_1 + C_2$ and the Gaussian curvature, $K_G = \mathrm{det}K^a_{\phantom{a}b} = C_1 C_2$, where $C_1$ and $C_2$ are the two principal curvatures \cite{DoCarmoBook}. The boundary of $\Sigma$, denoted by $\partial \Sigma$, is parametrized by its arc length $s$, so its embedding functions are given by the composition of maps $\partial \Sigma: s \mapsto \bfX_0(U^a(s))$, where $U^a$ are the surface coordinates along the boundary. The Darboux frame adapted to $\partial \Sigma$ is given by the right-handed basis defined by the unit tangent vector $\bfT := d\bfX_0/ds = \rmT^a \bfe_a$, where $\rmT^a= dU^a/d s$, the surface unit normal $\bfn$ and the unit outward conormal $\bfL := \bfT \times \bfn = \rmL^a \bfe_a$.
\\
On mesoscopic scales, the free energy of a fluid membrane is given by the Canham-Helfrich  energy, given by the sum of the bending energy, quadratic in the mean curvature, and the Gaussian energy \cite{Canham1970, Helfrich1973, Willmore}
\begin{subequations} \label{eq:Hbending}
\begin{eqnarray}
 H_{CH}[{\bf X}] &=& H_B[\bfX] +H_G[\bfX] \,,\\
 H_B[\bfX] &:=& \frac{\rmk_B}{2}\, \int dA\, K_D^2  \,, \\
 H_G[\bfX]&:=& \rmk_G  \int dA\, K_G \,,
\end{eqnarray}
\end{subequations}
where we have defined $K_D = K - C_s$, the difference between the mean curvature and the spontaneous curvature $C_s$; $dA$ is the area element; $\rmk_B$ and $\rmk_G$ are the bending and Gaussian moduli with units of energy \cite{SeifertAP, Deserno2015, LipowskyBook}. An important consequence of the fact that the energy $H_{CH}$ depends only on the geometrical degrees of freedom of the membrane is that the stresses are also determined completely by the membrane geometry.
\\
The constraints of fixed area and volume, which are of global character, as well as the restriction on the surface to conform to the ring, which is enforced locally, are introduced explicitly in the variational principle. Fixing the area $A$ and volume $V$ of the membrane to given values $A_0$ and $V_0$ involves the introduction of constant Lagrange multipliers $\sigma$ and $\rmP$. The former contributes to the tangential stresses and the latter represents the difference between the internal and external pressure, $\rmP =  \rmP_{int} - \rmP_{ext}$. The ring will be represented by a space curve $\bfY(s)$, which is also parametrized by the boundary arc length $s$. The constraint $\bfY (s) = \bfX_0(U^a(s))$, is implemented by introducing a vectorial Lagrange multiplier $\bm{\lambda}$ \cite{GuvenVazquez2012, GuvValVaz2014}. Thus the constrained functional to be minimized is
\begin{eqnarray} \label{eq:Hconst}
 H &=& H_{CH}[\bfX] + \sigma (A[\bfX] - A_0) - \rmP (V[\bfX]-V_0) \nonumber \\
 &&+ \int ds  \bm{\lambda} (s) \cdot \left[ \bfX_0(U^a (s)) - \bfY(s) \right] \,,
 \end{eqnarray}
where $ds$ is the line element of the ring. The Lagrange multiplier $\bm{\lambda}(s)$ quantifies the change of the bending energy under a deformation of the ring
\begin{equation}
 \bm{\lambda} =- \frac{\delta  H_{B}}{\delta \bfX_0} = \frac{\delta  H_{B}}{\delta \bfY}  \,,
\end{equation}
so it can be identified as the linear force density exerted by the ring on the membrane boundary.
\\
From the variation of the coordinates $U^a$ we get that $\bm{\lambda} \cdot \bfe_a = 0$, thus, the force density along the boundary is normal to the membrane, \cite{GuvenVazquez2012}
\begin{equation} \label{lambdavn}
\bm{\lambda} = \lambda_n \bfn\,.
\end{equation}
The total variation of the constrained functional, given in Eq.\ (\ref{eq:Hconst}), has two terms representing the changes in the energies of the bulk and of the boundary \cite{CapoGuven2002, Guven2004, Guven2018}
\begin{subequations} \label{Eq:deltaH}
\begin{eqnarray}
\delta H &=& \delta H_\Sigma + \delta H_{\partial \Sigma} \,,\\
\delta H_\Sigma &=& \int_\Sigma dA \left(\nabla_a {\bf f}^a - \rmP \bfn \right) \cdot \delta \bfX \,, \\
\delta H_{\partial \Sigma} &=& \int_{\partial \Sigma} ds \, \left(\rmL_a \delta Q^a - \bm{\lambda} \cdot \delta \bfY \right)\,. \label{Eq:deltaHpartialSigma}
\end{eqnarray}
\end{subequations}
In the bulk term $\delta H_\Sigma$, $\nabla_a$ is the covariant derivative compatible with $g_{ab}$;
$\bff^a$ is the stress tensor, which expressed in the adapted basis reads
\begin{subequations} \label{eq:stressdef}
\begin{eqnarray}
\bff^a &=& f^{ab} \bfe_b + f^a \bfn\,, \\
f^{ab} &=& \rmk_B\, K_D \left( K^{ab}- \frac{1}{2} K_D g^{ab} \right) - \sigma g^{ab}\,, \label{eq:fabcomp}\\
f^a &=& -\rmk_B \nabla^a K\,. \label{eq:facomp}
\end{eqnarray}
\end{subequations}
The normal component $f^a$ vanishes for constant mean curvature surfaces, and in particular for minimal surfaces with vanishing mean curvature.
\\
In the boundary term $\delta H_{\partial \Sigma}$, $\rmL_a$ are the covariant components  of the outward conormal along the equator, $\bfL$, and we have defined
\begin{subequations} \label{Eq:deltaQadef}
 \begin{eqnarray}
\delta Q^a &=& -\bff^a \cdot \delta \bfX_0+ H^{ab} \bfe_b  \cdot \delta \bfn \,, \\
H^{ab} &=& \rmk_B K_D g^{ab} + \rmk_G(K g^{ab}-K^{ab}) \,.
\end{eqnarray}
\end{subequations}
In equilibrium, the stresses within the membrane bulk, given by the divergence of the surface stress tensor, are balanced with the pressure difference, which acts in the normal direction\footnote{In the presence of an external normal force per unit area across the surface, $\Phi(u^a)$, the right-hand side of Eq.\ (\ref{eq:fcons}) is replaced by $(\Phi + \rmP ) \bfn$.}
\begin{equation} \label{eq:fcons}
\nabla_a \bff^a = \rmP \bfn\,.
\end{equation}
If $\rmP=0$, as in the case of open membranes, $\bff^a$ is conserved. Equation (\ref{eq:fcons}) captures the fluid character of the bilayer lipid membranes: since $\nabla_a \bff^a \cdot \bfe_b=0$, there is no cost associated with deformations along its tangential plane, only normal deformations are penalized. The Euler-Lagrange (EL) equation is given by the normal projection of the divergence of the stress tensor, $\nabla_a \bff^a \cdot {\bf n} = \rmP$, which in full reads \cite{CapoGuven2002, Guven2004, Guven2018}
\begin{equation} \label{Eq:ELfree}
(-\Delta + \bar{\sigma}) K + K_D \left(2 K_G - \frac{K}{2}\, (K + C_s)\right) = \bar{\rmP}\,,
\end{equation}
where $\Delta = g^{ab} \nabla _a \nabla_ b$ is the Laplace-Beltrami operator on the surface and from now on parameters with an overbar will be scaled with the inverse of $\rmk_B$: $\bar{\sigma} = \sigma/\rmk_B$ and $\bar{\rmP} = \rmP/\rmk_B$, which have dimensions of inverse area and inverse volume respectively. These two unknown parameters are not independent, as shown below in Sec.\ \ref{Sect:Scaling}. There is a scaling relation connecting them, so even if both of them are nonvanishing, there is actually only one independent parameter, which is determined by imposing the area or volume constraint.
\\
The Gauss-Bonnet theorem implies that the Gaussian energy is given by the sum of a topological invariant and the line integrals of the geodesic curvature along the boundaries \cite{DoCarmoBook}. In consequence, neither the bulk membrane stresses nor the EL equation depend on the Gaussian rigidity $\rmk_G$, it only enters the change in the boundary energy. Stationarity of the energy also entails the vanishing  of $\delta H_{\partial \Sigma}$, given in Eq.\ (\ref{Eq:deltaH}), which by expressing the surface tangent vectors $\bfe_a$ in the tangent basis adapted to the boundary $\partial \Sigma$ and taking into account that $\delta \bfY=\delta \bfX_0$, can be recast as
\begin{eqnarray} \label{Eq:balstrbdry}
 &-&\int ds \left( \bm{\lambda} + \bff_\perp \right) \cdot \delta \bfX_0 \nonumber \\
 &+&
 \int ds \left(\rmk_G \tau_g \bfT  + \left(\rmk_B K_D + \rmk_G \kappa_n \right) \bfL \right) \cdot \delta \bfn = 0\,, \qquad
\end{eqnarray}
where we have defined the projection of the stress tensor along the outward conormal by
\begin{subequations} \label{Eq:Lafadef}
\begin{eqnarray}
\bff_\perp &:=& \rmL_a \bff^a = f_{\perp \perp} \bfL +  f_{\perp \parallel} \bfT + f_\perp \bfn \,, \\
f_{\perp \parallel} &:=& \rmL_a \rmT_b f^{ab}=\rmk_B K_D \tau_g\,,  \label{Eq:fprppar} \\
f_{\perp \perp} &:=& \rmL_a \rmL_b f^{ab}=\frac{\rmk_B}{2} K_D \left(\kappa_{n \perp} -\kappa_n +C_s \right) -\sigma , \qquad \label{Eq:fprpprp} \\
f_\perp &:=& \rmL_a f^a = -\rmk_B \nabla_\perp K  \,, \label{Eq:fprp}
\end{eqnarray}
\end{subequations}
where $\tau_g = \rmT^a \rmL^b K_{ab}$ is the geodesic torsion; $\kappa_n = \rmT^a \rmT^b K_{ab}$ and $\kappa_{n\perp} = \rmL^a \rmL^b K_{ab}$ are the normal curvatures in the directions along the boundary and orthogonal to it \cite{DoCarmoBook}; $\nabla_\perp = \rmL^a \nabla_a$ is the projected covariant derivative along $\bfL$. The vector $\bff_\perp$ is the force per unit length exerted by a region of the membrane on its neighboring region separated by a line element $ds$ \cite{Guven2018}. In Sec. \ref{Sect:stressbal}, we specialize this equation to axisymmetric surfaces in order to analyze the external normal force deforming the equatorial radius of the membrane.

\section{Axial symmetry and first integral of the EL equation} \label{Sect:FirsInt}

\begin{figure}[t]
\begin{center}
\includegraphics[scale=0.75]{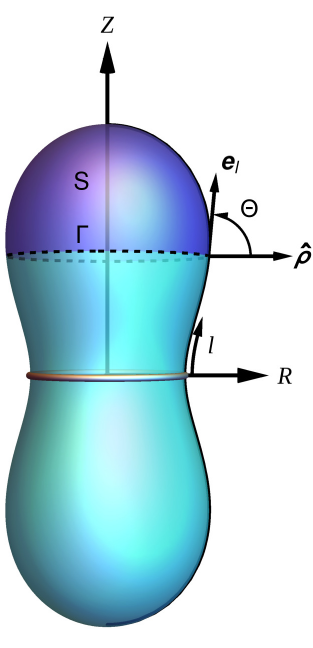}
\end{center}
\caption{Axisymmetric membrane constrained by a ring at the equator. The meridian is parametrized by its arclength $l$, and $\Theta$ is the angle that the tangent vector $\bfe_l$ makes with the radial direction $\hat{\bm{\rho}}$. The integration contour $\Gamma$ on the cap $\mathrm{S}$ of the surface is shown with a dashed line.}
\label{fig:5}
\end{figure}

We assume that the membrane remains axially symmetric under a radial deformation along one of its parallels caused by a circular ring  of radius $R_0$. Moreover, for simplicity we consider that the ring is located at equator, so the deformed membrane also has mirror symmetry with respect to the equatorial plane (Fig.\ \ref{fig:5}).
\vskip0pc \noindent
Therefore, the membrane can be parametrized by arclength $l$ along the meridian (measured from the ring) and the azimuthal angle $\varphi$. The arclength $l$ is measured from the ring, so $l=0$ at the equator and  at the pole it will be denoted by $l_p$. In these coordinates the line element assumes the form $ds^2 = dl^2 + R(l)^2 d\varphi^2$, where $R$ is the radius of the parallel circles. As discussed in Appendix \ref{App:axisym}, the derivatives of the radial and height coordinates of the meridian can also be described using the angle $\Theta$ that its tangent vector makes with the radial direction $\hat{\bm{\rho}}$ (Fig.\  \ref{fig:5}),
\begin{equation} \label{def:RpZp}
R'(l) = \cos \Theta \,, \quad Z'(l) = \sin \Theta\,,
\end{equation}
where the prime stands for differentiation with respect to $l$, $'=d/dl$. The principal curvatures of an axisymmetric surface occur along the meridians and parallels, which in terms of $\Theta$ are given by
\begin{equation} \label{Axscurvs0}
C_\perp = \Theta'\,, \qquad C_\parallel = \frac{\sin \Theta}{R}\,.
\end{equation}
Thus, the mean and Gaussian curvatures are given by $K = C_\perp + C_\parallel$ and $K_G = C_\perp C_\parallel$.
\\
The rotational symmetry of the membrane geometry implies the existence of a first integral of the EL equation (\ref{Eq:ELfree}) \cite{CapoGuven2002}. Integrating Eq.\ (\ref{eq:fcons}) within the cap $\mathrm{S}$ bounded by parallel $\Gamma$ of radius $R$ above the equator (see Fig.\ \ref{fig:1}), and using the divergence theorem to recast the left-hand side as a closed line integral along $\Gamma$, we get
\begin{equation} \label{eq:oint}
\oint\limits_\Gamma ds\, \bff_\perp = \rmP \int\limits_\Sigma dA \, {\bf n}\,,
\end{equation}
where $ds = R d \varphi$ is the arclength. Thus, the total force exerted on $\mathrm{S}$, given by the line integral of the projection of stress tensor onto the unit outward conormal along $\Gamma$, is proportional to the integrated normal vector over the same region. The equator is not included in the region of integration, so Eq.\ (\ref{eq:oint}) does not capture the force exerted by the ring, which will be calculated in Sec.\ \ref{Sect:stressbal}.
\\
In this case the normal curvatures along the parallel and orthogonal to it are $\kappa_n = C_\parallel$ and $\kappa_{n \perp} = C_\perp$, whereas the geodesic torsion vanishes, $\tau_g =0$, (see Appendix \ref{App:axisym}), so the projected stress tensor and its tangential and normal components, given in Eq.\ (\ref{Eq:Lafadef}), simplify to
\begin{subequations} \label{fproys}
\begin{eqnarray}
\bff_\perp &=& f_{\perp \perp} \bfL + f_\perp \bfn \,, \\
f_{\perp\perp} &=& \frac{\rmk_B}{2} K_D \left(C_\perp - C_\parallel + C_s  \right) - \sigma \,,  
\label{fperpperpctrsprn} \\
f_\perp &=&\rmk_B K'  = k_B \left(C'_\perp + C'_\parallel \right) \,, \label{fperpctrsprn}
\end{eqnarray}
\end{subequations}
where the derivatives of the curvatures are
\begin{equation} \label{eq:dCprpdCpar}
C_\perp' = \Theta''\,, \quad C_\parallel'=\frac{\cos \Theta}{R} (C_\perp-C_\parallel)\,.
\end{equation}
The only nontrivial projection of Eq.\ (\ref{eq:oint}) is along the axis of symmetry, $\hat{\bf z}$, and it is independent of $s$. Using the result (\ref{fperpzproj}) of Appendix \ref{App:axisym},
we find that the vertical projection of the left-hand side of Eq.\ (\ref{eq:oint}) reads
\begin{equation} \label{eq:lhsint}
 \oint\limits_\Gamma ds\, \bff_\perp \cdot \hat{\bfz} \\
= - 2 \pi R \left( \cos\Theta \, f_\perp + \sin\Theta  \, f_{\perp\perp} \right)\,.
 \end{equation}
Using Eq.\ (\ref{eq:naxsym}) of Appendix \ref{App:axisym} for the normal vector to an axisymmetric surface along with Eq.\ (\ref{def:RpZp}) for $R'$, we have ${\bf n}\cdot \hat{\bf z}= -\cos\Theta =-R'$. Thus, integrating the right-hand side of Eq.\ (\ref{eq:oint}) from a parallel with arclength $l$ to the pole where $l=l_p$ and $R(l_p)=0$, we get
\begin{eqnarray} \label{eq:rhsint}
\frac{1}{2 \pi} \int\limits_\Sigma dA \, {\bf n} \cdot \hat{\bfz} = -\int_{l}^{l_p} d\mathrm{l} R R' 
= \int_0^R d \mathrm{R} \mathrm{R} = \frac{R^2}{2}\,. \quad
\end{eqnarray}
Substituting Eqs.\ (\ref{eq:lhsint}) and (\ref{eq:rhsint}) in Eq.\ (\ref{eq:oint}) we obtain the first integral of the EL equation
\begin{equation} \label{eq:firsint}
\cos\Theta \, f_\perp + \sin\Theta  \, f_{\perp\perp} + \frac{\rmP}{2} \, R =0\,,
\end{equation}
which estates the balance of normal and tangential stresses with the pressure difference. By substituting Eqs.\ (\ref{fproys}), in full this equation reads
\begin{align} \label{eq:shapeaxi}
\cos\Theta  \left(C'_\perp +C'_\parallel \right) & + \sin \Theta \left(\frac{1}{2} \left( C^2_\perp - \left(C_\parallel - C_s \right)^2 \right) - \bar{\sigma} \right) \nonumber \\
& + \frac{\bar{\rmP}}{2} R = 0 \,,
\end{align}
which is to be solved along with Eqs.\ (\ref{def:RpZp}). This axially symmetric shape equation is a second order differential equation for $\Theta$. In practice, numerical precision and stability are improved by casting it in the more familiar equivalent Hamiltonian form (as described in Appendix \ref{App:HamDer}) in terms of $\Theta$ and its conjugate momenta $P_\Theta = R K_D$ (which replaces the curvature $C_\perp =\Theta'$). Alternatively, instead of working with $P_\Theta$, it proves useful to work with $K_D$ as a dependent variable along with $R$ and $\Theta$, so we solve the system of first order differential equations given by Eqs.\ (\ref{def:RpZp}) and
\begin{subequations} \label{Eqs:systThetaKD}
\begin{eqnarray}
\Theta'- K_D &+& C_\parallel - C_S = 0 \,, \label{Eq:Theta} \\
 \cos\Theta \, K_D' &+& \sin \Theta \left(K_D \left( \frac{K_D}{2} - C_\parallel + C_s \right) - \bar{\sigma} \right) \nonumber \\
 &+& \frac{\bar{\rmP}}{2} R = 0 \,. \label{Eq:ELKD}
\end{eqnarray}
\end{subequations}
To solve this system of four differential equations in order to determine the shape of the deformed equilibrium states, it is required to identify the four boundary conditions on the geometry along the equator that are consistent with the constraint imposed by the ring.

\subsection{Boundary conditions}\label{Sect:bcs}

We consider deformed vesicles with mirror symmetry with respect to the equatorial plane, so the system of equations (\ref{def:RpZp}) and (\ref{Eqs:systThetaKD}) are solved for one half of the generating curve with total arclength $l_p$. Setting the equator at the plane $Z=0$, the boundary conditions of the radial and height coordinates are given by
\begin{equation}
R_0:=R(0)=R_{ring} \,, \quad Z_0:=Z(0)=0\,.
\end{equation}
Finite bending energy demands the surface to be continuous and smooth across the equator and at the poles, otherwise the curvatures (and the energy) would diverge. The existence of tangent planes along the equator and at the poles impose the boundary conditions (BCs) on the angle $\Theta$,
\begin{equation} \label{BCsTheta}
\Theta_0:=\Theta(0) = \frac{\pi}{2} \,, \quad \Theta_p := \Theta(l_p) = \pi \,.
\end{equation}
On account of Eqs.\ (\ref{def:RpZp}), these BCs also imply that the equator and the poles correspond to extrema of the radial and the height coordinates respectively. It should be noted that, since Eq.\ (\ref{Eq:ELKD}) does not hold at the equator, where the force is applied by the ring, it is not legitimate to evaluate the first integral at $\Theta_0=\pi/2$. Doing so, the coefficient of the first term vanishes and it reduces to an equation relating the Lagrange multiplier $\sigma$ and $\rmP$, which in general is inconsistent with the constraints of fixed area and volume or with the condition of smoothness at the pole, so in practice we evaluate at a very close value, such as $\Theta_0=(1 \pm 10^{-6}) \pi/2$. The second BC replaces the specification of $K_D(0)$, which involves $\Theta'(0)$, value that is not fixed by the requisite of continuity along the equator. Furthermore, since $l_p$ is not known a priori, this is a free boundary problem. The shooting method is used to solve this boundary value problem: the initial value $K_D(0)$ (or $\Theta'(0)$) is fine-tuned so that the second BC in Eq.\ (\ref{BCsTheta}) is fulfilled and the geometry closes smoothly at the pole.
\\
The poles with $\Theta=0, \pi$ and $R=0$ are regular singular points where the curvature $C_\parallel$ seems to be undetermined. However, taking the limit $l \rightarrow l_p$ and using the expression of $R'$, given in Eq.\ (\ref{def:RpZp}), we find that the poles are umbilical points, i. e. both curvatures are equal \cite{Svetina1989},
\begin{equation} \label{Umbilicpoles}
C_{\parallel p}:=\lim_{l \rightarrow l_p}C_\parallel = \frac{\cos \Theta \Theta'}{R'} \Big|_{l_p} = C_{\perp p}\,.
\end{equation}
Thus, at the poles we have $K_p = 2 C_{\perp p}=2 \Theta'_p$.
\\
From Eq.\ (\ref{eq:dCprpdCpar}), we have that $C_\parallel \propto (C_\perp - C_\parallel)/R$, so on account of Eq.\ (\ref{Umbilicpoles}) and also that $R(l_p)=0$, apparently also this derivative is undefined at the poles, but by taking the limit $l \rightarrow l_p$, we get
\begin{equation} \label{eq:dCperpCparp}
\lim_{l \rightarrow l_p} C'_\parallel = \cos \Theta \left(\frac{ C'_\perp-C'_\parallel}{R'}\right)\Bigg|_{l_p} =(C'_\perp-C'_\parallel)\Big|_{l_p} \,.
\end{equation}
Therefore at the poles we have $ C_{\perp p}' = 2 C_{\parallel p}'$. Furthermore, from  Eq.\ (\ref{eq:shapeaxi}) we have that at the poles $K_p'=C'_{\perp p}+C'_{\parallel p}=0$, so  the derivatives of the principal curvatures vanish at the poles, $C_{\parallel p}'= 0$ and $C_{\perp p}'=\Theta''_p=0$.

\section{Force exerted by the ring} \label{Sect:Equatorialforce}

In this section we determine the external force in two ways, using the balance of forces established at the equator and by means of the scale invariance of the energy.

\subsection{Equatorial force as a balance of stresses} \label{Sect:stressbal}

We can consider the vesicle as composed by two halves bounded by the equator, where they are in contact with the ring. Taking into account the contributions of each half, the upper one denoted by a $+$ and the bottom one with a $-$, we have from  Eq.\ (\ref{Eq:deltaHpartialSigma}) that the change of the energy along the equator is given by
\begin{equation}
\delta H_{\partial \Sigma} = \int ds \left( \rmL_{a+} \delta Q_{+}^a + \rmL_{a-} \delta Q_{-}^a - \bm{\lambda} \cdot \delta \bfY \right) \,.
 \end{equation}
Stationarity of the energy implies the vanishing of $\delta H_{\partial \Sigma}$. Taking into account that the deformations of the embedding functions and the normal $\delta \bfY$ and $\delta \bfn$ are the same for both boundary parallels, and that the equator is also a principal curve, so $\tau_g=0$, we have in this case that Eq.\ (\ref{Eq:balstrbdry}) reads
\begin{eqnarray}
\int ds \bm{\lambda} \cdot \delta \bfY &=& -\int ds \left( \bff_{\perp +} +  \bff_{\perp -} \right) \cdot \delta \bfY \nonumber \\
&+& \int ds \big[ \left( \rmk_B K_{D+} + \rmk_G C_{\parallel +} \right) \bfL_+ \nonumber \\
&& +\left( \rmk_B K_{D -} + \rmk_G C_{\parallel -} \right) \bfL_- \big] \cdot \delta \bfn \,.
\end{eqnarray}
The avoidance of kinks at the equator implies that the angle $\Theta$ ought to be continuous and smooth, rendering the parallel curvature $C_\parallel$ to be also continuous and smooth across the equator. However its counterpart along the meridian, $C_\perp=\Theta'$ also should be continuous, but not necessarily smooth, for the presence of external forces will be reflected in its derivatives as explained below. Thus, $C_{\parallel +} = C_{\parallel -}$ and $C_{\perp +} = C_{\perp -}$. Furthermore, the tangent vectors along the two boundaries have opposite directions $\bfT_+ = -\bfT_-=\bfT=\hat{\bm{\varphi}}$, and since the normal is the same along both of them, the conormals also have opposite directions $\bfL_+ = -\bfL_-=\bfL =-\bfe_l$. Taking this into account, we have that the vector multiplying $\delta \bfn$ vanishes,\footnote{Besides, we are interested in the particular case of equatorial deformation in which the normal does not change, so $\delta \bfn= \mathbf{0}$.} so the Gaussian rigidity does not enter. Moreover, since $\delta \bfY$ is arbitrary, we have that the stress transmitted to the membrane boundary is
\begin{equation} \label{lambdaaxysym}
\bm{\lambda} = -\bff_{\perp +} - \bff_{\perp -} \,.
\end{equation}
Thus, $\bm{\lambda}$ is identified as the external force exerted by the ring on the membrane
boundary (recall $\bff_\perp$ is the force per unit length exerted by the membrane). Decomposing the
force vector in the basis adapted to the boundary
\begin{equation}
  \bff_{\perp \pm} = f_{\perp \parallel \pm} \bfT_{\pm} +  f_{\perp \perp \pm} \bfL_{\pm} + f_{\perp \pm} \bfn \,.
\end{equation}
Since the parallels are principal lines of curvature, the tangential projections vanish $f_{\perp \parallel \pm}=0$. The projection of Eq.\ (\ref{lambdaaxysym}) onto the conormal reads
\begin{equation}
 \lambda_L := \bm{\lambda} \cdot \bfL = - f_{\perp \perp +} + f_{\perp \perp -}  \,.
\end{equation}
Since the curvatures are continuous, we have $f_{\perp \perp +} = f_{\perp \perp -}$ and in consequence the tangential projection vanishes, so $\lambda_L=0$, consistent with Eq.\ (\ref{lambdavn}). The projection onto the normal reads
\begin{equation} \label{lambdadotn}
 \lambda_n := \bm{\lambda} \cdot \bfn =  -f_{\perp +} - f_{\perp -} \,.
\end{equation}
From Eq.\ (\ref{Eq:fprp}) we have that the normal components of the stress tensor on each side are given by the derivatives of the mean curvature along the conormal ($\rmL_+ = -1$ and $\rmL_- = 1$),  so $f_{\perp +} =\rmk_B K_+'$ and $f_{\perp -}= -\rmk_B K_-'$. Substituting in Eq.\ (\ref{lambdadotn}) it reads
\begin{equation} \label{eq:difdK}
 \bar{\lambda}_n = - K_+' + K_-'\,,
\end{equation}
where $K' = C_\perp' + C_\parallel'$, and $C_\perp'$ and $C'_\parallel$ were defined in Eq.\ (\ref{eq:dCprpdCpar}). Due to the continuity of $\Theta$ and $\Theta'$, we have $C_{\parallel +}'=C_{\parallel -}'$, so Eq.\ (\ref{eq:difdK}) reduces to
\begin{equation}
 \bar{\lambda}_n = - C_{\perp +}' + C_{\perp -}'= - \Theta_{+}'' + \Theta_{-}''\,.
\end{equation}
Thus, the linear force density on the equator involves only the difference of the derivatives of the meridian curvatures. This contrasts with the case of lipid membrane domains where the force is due to the line tension and is given in terms of the difference of the curvatures on both sides, rather than of their derivatives, \cite{Julicher1996}. In particular, if the vesicle has mirror symmetry with respect to the equatorial plane, $\Theta(-l)=\pi - \Theta(l)$, so the derivative of $\Theta$ is an odd function of arc length, $\Theta_+''=-\Theta_-''=\Theta''_0$, and the linear force density is proportional to the second derivative of $\Theta$ at the equator
\begin{equation} \label{def:lambdan}
 \bar{\lambda}_n = -2 \Theta_0''\,.
\end{equation}
The total normal force on the equator is obtained by integrating this linear force density
\begin{equation} \label{def:Fext}
 \bar{F} := \int ds \bar{\lambda}_n = -4 \pi R_0 \Theta_0''\,.
\end{equation}
These expressions for the linear force density and the total force can be derived also by integrating the EL equation over a region centered at the equator \cite{Guven2018}. 

\subsection{Equatorial force from the scale invariance of the bending energy} \label{Sect:Scaling}

In the absence of external forces, the scale invariance of bending energy, Eq.\ (\ref{eq:Hbending}), provides a useful relation between the spontaneous curvature $C_s$, the Lagrange multipliers $\sigma$ and $P$ and the global quantities of the membrane
\begin{equation} \label{scalenof}
(C_s ^2 + 2\bar{\sigma}) A - 3 \bar{\rmP} V - C_s M = 0  \,,
\end{equation}
where we have defined the total mean curvature
\begin{equation}
M = \int dA K \,.
\end{equation}
If $C_s = 0$ the two Lagrange multipliers are related, $2\sigma A = 3 \rmP V$.
\\
Let us apply the scaling argument for a surface constrained by a ring to see how this relation generalizes. To this end, we consider the effect of a membrane rescaling ${\bf X}\to \Lambda {\bf X}$ on the constrained energy $H$, given by Eq.\ (\ref{eq:Hconst}). Under a rescaling the line element of the boundary changes as $s \rightarrow \Lambda s$, whereas the area, volume and mean curvature of the membrane scale as  $A \rightarrow \Lambda^2 A$, $V \rightarrow \Lambda^3 V$ and $K \rightarrow \Lambda^{-1} K$, respectively, so the energy changes as
 \begin{eqnarray}
H[\Lambda {\bf X}] &=& \frac{\rmk_B}{2} \int dA (K - \Lambda C_s)^2 \nonumber \\
&&+ \sigma \, (\Lambda^2 \, A - A_0) - \rmP\, (\Lambda^3\, V-V_0) \nonumber \\
&&+ \int  ds \Lambda \, \bm{\lambda}(s) \, \cdot (\Lambda \bfX_0 (s) -  \bfY(s)) \,.
 \end{eqnarray}
While the bending energy itself is scale invariant, the spontaneous curvature breaks this invariance. Stationarity of the energy under a rescaling imposes the condition
\begin{equation} \label{fmult}
 \frac{d H}{d\Lambda}\Big{|}_{\Lambda =1} = 0 \,,
\end{equation}
which implies that
 \begin{equation} \label{ScalingdeltaHmemb}
 \int ds\, \bm{\lambda} \cdot \bfX_0 = -(2\sigma + k_B C_s^2) A_0 + 3 \rmP V_0 + \rmk_B C_s M \,,
 \end{equation}
where we have used that $A=A_0$ and $V=V_0$. This quantity is proportional to the change of the energy of the membrane boundary, $\delta H_{\partial \Sigma} = \int ds \bm{\lambda} \cdot \delta \bfX_0$, under an infinitesimal dilation, $\delta \bfX_0 = \delta \Lambda \bfX_0$. Since the change of the energy of the membrane boundary is the negative of the change of the energy of the ring,  $\delta H_{ring} = -\delta H_b$, we get that under an infinitesimal dilation the change in the energy of the ring is proportional to minus the right-hand side of Eq.\ (\ref{ScalingdeltaHmemb}). In particular, for a circular ring  acting on an axially symmetric vesicle, we have on account of the rotational symmetry $\bfY (s) = R_0 \hat{\bm{\rho}}$ and $\bm{\lambda} = \lambda_n \hat{\bm{\rho}} $, so the magnitude of the total force exerted by the ring is
\begin{equation} \label{eq:FpHpR}
F = \frac{1}{R_0} \left( ( 2\sigma + \rmk_B C_s^2) A_0 - 3 \rmP V_0 - \rmk_B C_s M \right).
 \end{equation}
If $C_s = 0$ the constraining force is proportional to $\sigma$ for fixed area ($\rmP=0$), $\bar{F}= 2 \sigma A_0/ R_0$, whereas it is proportional to $\rmP$ for fixed volume ($\sigma = 0$) $\bar{F}= -3 \rmP V_0  / R_0$, which is less than evident at the level of the EL equation. In each case, in the limit of maximum constriction, with $R_0 \to 0$, the force is proportional to the derivatives of $\sigma$ and $\rmP$ with respect to $R_0$, $\bar{F} = 2 A_0 \partial \sigma /\partial R_0 $ for fixed area and $\bar{F} = -3 V_0 \partial \rmP /\partial R_0 $ for fixed volume. The introduction of a new scale associated with the constraint on the ring is consistent with a nonvanishing value for $\sigma$ or $\rmP$. Only in the limit $R_0\to 0$, the scale invariance is restored.
\\
Combining Eqs.\ (\ref{def:Fext}) and (\ref{eq:FpHpR}) we can express the second derivative of the angle $\Theta$ as
\begin{equation} \label{ThetappsigmaPM}
 4 \pi R^2_0 \Theta''_0 = - (2\bar{\sigma}+C_s^2) A_0 + 3 \bar{\rmP} V_0 + C_s M\,.
\end{equation}
Since this equation involves the total mean curvature $M$, which is a global quantity, it is not useful in the determination of $\sigma$, $\rmP$ or $\Theta''_0$. However, as shown below it is useful in the limit of maximum constriction, where it permits us to determine $\sigma$ if the area is fixed ($\rmP=0$) or to determine $\rmP$ if the volume is fixed ($\sigma=0$).

\section{Nondimensional quantities} \label{Sect:nondimquant}

The relevant length scale is set by the radius of the spherical vesicle. In the following we consider nondimensional quantities obtained by a rescaling with appropriate powers of $R_S$. The scaled arclength along the meridian, measured from the equator is
\begin{equation}
\ell : = \frac{l}{R_S}\,.
\end{equation}
The scaled arclength at the poles is $\ell_p =l_p /R_S$. The derivative with respect to the nondimensional arclength will be denoted by a dot, $\dot{} =  d/d \ell = R_S d/dl$. The reduced coordinates are defined by
\begin{equation} \label{redcoords} r(\ell) := \frac{R}{R_S} \,, \quad  z(\ell):=\frac{Z}{R_S}\,.
 \end{equation}
Thus, the nondimensional embedding functions are $\bfx(\ell) := \bfX/R_S = r(\ell) \hat{\bm{\rho}}+z(\ell)\hat{\bfz}$. The radius of the ring located at the equator with $\ell=0$ is $r_0:=R_0/R_s$.
\\
The principal curvatures and the spontaneous curvature are scaled with $R_S$,
\begin{subequations}
\begin{eqnarray}
c_\perp &:=& R_S C_\perp=\dot{\Theta} \,, \\
c_\parallel &:=& R_S C_\parallel = \frac{\sin \Theta}{r}\,, \\
c_s&:=& R_S C_s\,.
\end{eqnarray}
\end{subequations}
The nondimensional mean curvature and the mean curvature difference are
\begin{subequations}
\begin{eqnarray}
k&:=&R_S K=c_\perp + c_\parallel \,, \\
\kappa&:=& R_S K_D = k - c_s\,.
\end{eqnarray}
\end{subequations}
We define the reduced area, volume and total mean curvature by
\begin{equation}
a:=\frac{A}{A_S}\,, \quad v:=\frac{V}{V_S} \,, \quad m := \frac{M}{M_S} \,,
\end{equation}
where $A_S=4\pi R_S^2$, $V_S = 4 \pi R_S^3/3$ and $M_S = 8 \pi R_S$ are the area, volume and total mean curvature of a sphere of radius $R_S$. For surfaces with equatorial mirror symmetry we have
\begin{subequations}
\begin{eqnarray}
a &=& \int_0^{\ell_p} d \ell r\,, \\
v &=& \frac{3}{2} \int_0^{\ell_p} d \ell \sin  \Theta \, r^2 \,, \\
m &=& \frac{1}{2} \int_0^{\ell_p} d \ell \, k \, r\,.
\end{eqnarray}
\end{subequations}
The bending energy of a spherical vesicle with spontaneous curvature is $H_{BS}(c_s)= 2 \pi \rmk_B (2-c_s)^ 2$. We rescale the total bending energy of the vesicles with the energy of a spherical vesicle with null spontaneous curvature, $H_{BS}(0)=8 \pi \rmk_B$. Thus, for a spherical vesicle we have
\begin{equation} \label{hBsphere}
h_B:=\frac{H_B}{H_{BS}(0)}= \left(1 -\frac{c_s}{2} \right)^2 \,,
\end{equation}
which vanishes if $c_s=2$. The scaled total enegy of symmetric membranes is given by
\begin{equation}
 h_B = \frac{1}{4} \int_0^{\ell_p} d \ell \kappa^2 r\,.
\end{equation}
Since $\bar{\sigma}$ and $\bar{P}$ have dimensions of inverse area and inverse volume, the nondimensional intrinsic tension and nondimensional pressure are defined by
\begin{equation}
\mu := R_S^2 \, \bar{\sigma}\,, \quad \mrmp := R_S^3 \, \bar{\rmP} \,.
\end{equation}
We define the scaled linear force density and the scaled total force on the equator (given by Eqs.\ (\ref{eq:FpHpR}))
\begin{subequations} \label{eq:phif0}
 \begin{eqnarray}
\phi_0&:=& R_S^2 \bar{\lambda}_n = -2 \ddot{\Theta}_0\,, \\
f_0&:=&R_S \bar{F} = -4 \pi r_0 \ddot{\Theta}_0 \nonumber \\
&=& \frac{4 \pi}{r_0} \left((c_s^2+ 2 \mu) a_0 - \mrmp \, v_0 - 2 c_s m \right)\,.
\end{eqnarray}
\end{subequations}
So the analog of Eq.\ (\ref{ThetappsigmaPM}) reads
\begin{equation} \label{scalrelsimp}
r_0^2 \ddot{\Theta}_{0} = -\left(2 \mu + c_s^2 \right) a_0 + \mrmp \, v_0 + 2 c_s m  \,.
\end{equation}
Expressing Eqs.\ (\ref{def:RpZp}) and (\ref{Eqs:systThetaKD}) in terms of these nondimensional coordinates and parameters, we have ($\,\dot{}:=d/d \ell$)
\begin{subequations} \label{nondimhamsystem}
 \begin{eqnarray}
\dot{r} &=& \cos \Theta\,, \\
\dot{z} &=& \sin \Theta\,, \\
\dot{\Theta} &=& \kappa -\frac{\sin \Theta }{r} + c_s\,, \\
\cos \Theta \, \dot{\kappa} &+& \sin \Theta \left( \kappa \left( \frac{\kappa}{2} -\frac{\sin \Theta}{r}+ c_s \right) - \mu  \right) \nonumber \\
&+& \frac{\mrmp}{2} r = 0\,. \label{EL:kappa}
\end{eqnarray}
\end{subequations}
The second derivative of $\Theta$ is given by
\begin{equation}
\ddot{\Theta} = \dot{\kappa} - \frac{\cos \Theta}{r} \left(\kappa - \frac{2\sin \Theta}{r} +c_s\right) \,.
\end{equation}
The initial values of the reduced mean curvature difference and its derivative can be expressed in terms of the initial values of the angle and its derivatives as
\begin{subequations} \label{Eq:k0dk0Theta0dTheta0}
\begin{eqnarray}
\kappa_0&=&\dot{\Theta}_0+\frac{\sin \Theta_0}{r_0} - c_s\,, \\
\dot{\kappa}_0 &=& \ddot{\Theta}_0 + \frac{\cos \Theta_0}{r_0} \left(\dot{\Theta}_0 - \frac{\sin \Theta_0}{r_0} \right)\,.
\end{eqnarray}
\end{subequations}
The reduced coordinates, angle and mean curvature difference of a sphere of radius $R_\mathfrak{s}$, whose equator is at a height $Z_e$ and with arc length $l_e$ (so the reduced height and arc length of the equator are $z_e=Z_e/R_S$ and $\ell_e=l_e/R_S$), are given by 
\begin{subequations} \label{sphericalsol}
 \begin{align}
 r(\ell) &= r_\mathfrak{s} \cos \theta \,, & z(\ell)&=r_\mathfrak{s} \sin  \theta +z_e\,, \\
 \Theta(\ell) &= \theta + \frac{\pi}{2}  \,, & \kappa(\ell) &= \frac{2}{r_\mathfrak{s}} - c_s \, ,
 \end{align}
\end{subequations}
where we have defined the reduced radius and the angle $\theta$ by 
\begin{equation} \label{eq:Sphericalsolrel}
r_\mathfrak{s}=\frac{R _\mathfrak{s}}{R_S} \,,  \quad \theta := \frac{\ell-\ell_e}{r_\mathfrak{s}}\,.
\end{equation}
The sphere is a solution of the first integral (\ref{EL:kappa}) if the parameters satisfy the equation 
\cite{Zhong-Can1989}
\begin{equation} \label{eq:spheresclrl}
r_\mathfrak{s} \mu - \frac{\mrmp}{2} r_\mathfrak{s}^2 + c_s \left( r_\mathfrak{s} \frac{c_s}{2}-1 \right)=0 \,.
\end{equation}
This agrees with Eq.\ (\ref{scalrelsimp}) for $a_0=1$, $v_0=1$, $m=1$, and $\ddot{\Theta}_0=0$, so the sphere is a solution free of external forces. If $c_s=0,2/r_\mathfrak{s}$ the Lagrange multipliers enforcing the area and volume are proportional, $\mrmp r_\mathfrak{s} = 2\mu$. For the initial spherical configuration $R_\mathfrak{s}=R_S$, so the scaled radius is unit, $r_\mathfrak{s}=1$, and we set the equator on the plane $Z=0$, so $\ell_e=0$ and $z_e=0$.

\section{Equilibrium shapes with fixed area} \label{Sec:Aconst}

\begin{figure*}
\begin{center}
\begin{tabular}{cccccc}
 $\vcenter{\hbox{\includegraphics[scale=0.35]{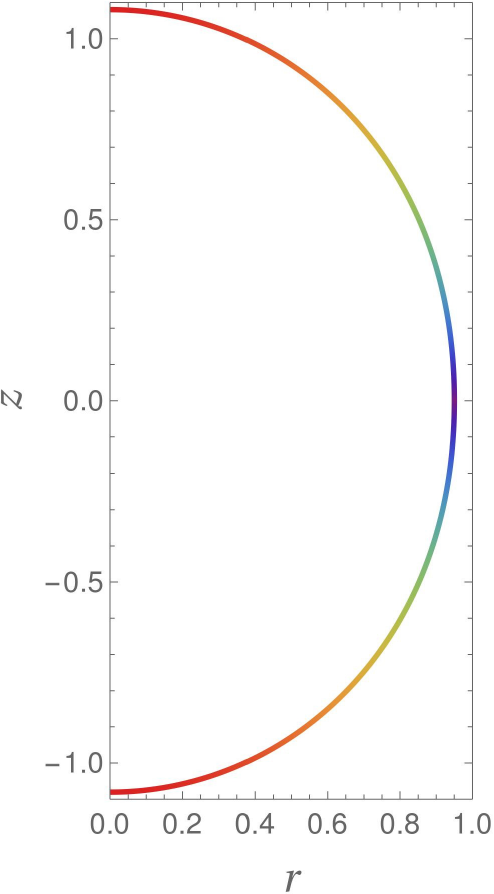}}}$ &
  $\vcenter{\hbox{\includegraphics[scale=0.375]{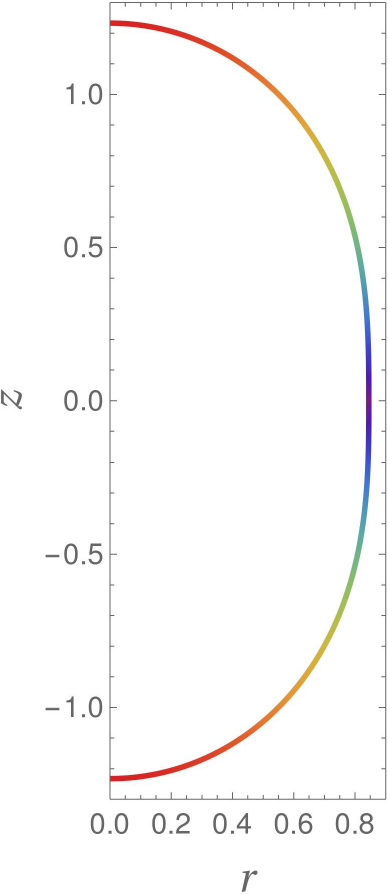}}}$ &
  $\vcenter{\hbox{\includegraphics[scale=0.375]{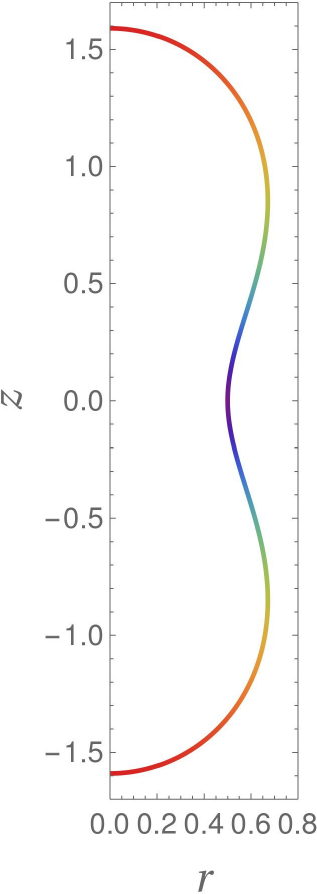}}}$ &
   $\vcenter{\hbox{\includegraphics[scale=0.375]{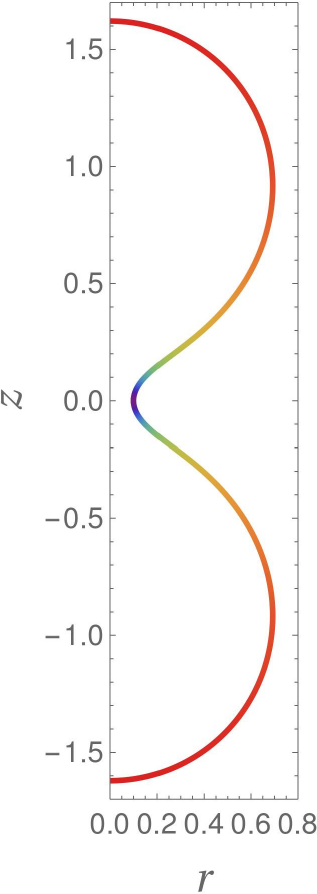}}}$ &
  $\vcenter{\hbox{\includegraphics[scale=0.375]{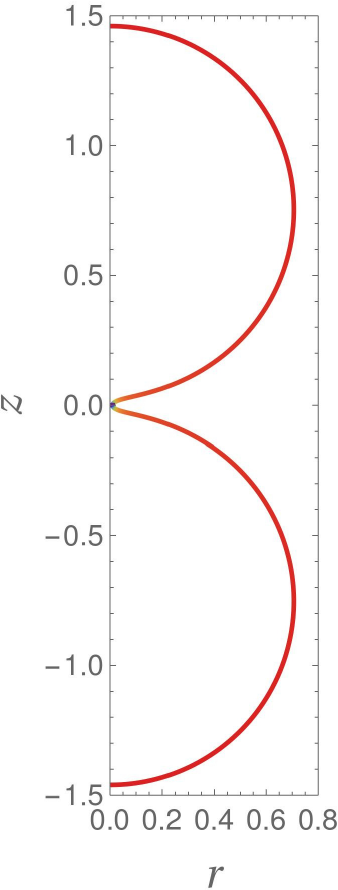}}}$ &
  $\vcenter{\hbox{\includegraphics[scale=0.275]{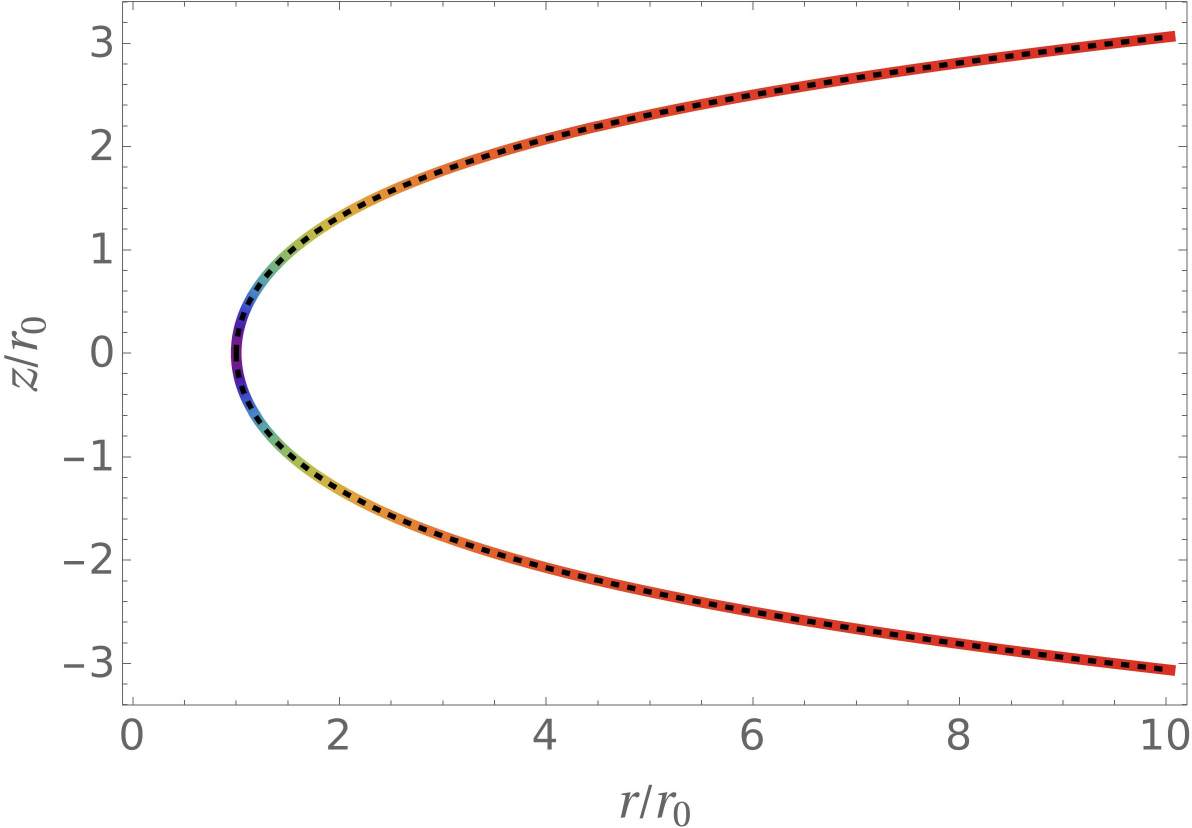}}}$ \\ \\
{\small (a) $r_0 = 0.95$} & {\small (b) $r_0 = 0.845$}& {\small (c) $r_0 = 0.5$} & {\small (d) $r_0 = 10^{-1}$} & {\small (e) $r_0=10^{-2}$} & {\small (f) $r_0 = 10^{-3}$} 
\end{tabular}
 \end{center}
\includegraphics[scale=0.6]{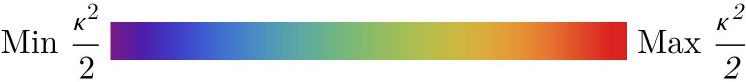}
\caption{(a)-(e) Sequence of constricted vesicles with $a=1$ and $c_s=0$ as $r_0$ is decreased. (e) For a very small radius, the vesicle is constituted by two quasispherical vesicles connected by a small catenoid-like neck. (f) Magnification of the profile curve of the small neck (the solid line corresponds to the numerical solution and the black dashed line to the catenoid). The scaled bending energy density is color coded.}
\label{fig:6}
\end{figure*}

\begin{figure*}
\begin{center}
\begin{tabular}{cccc}
$\vcenter{\hbox{\includegraphics[scale=0.35]{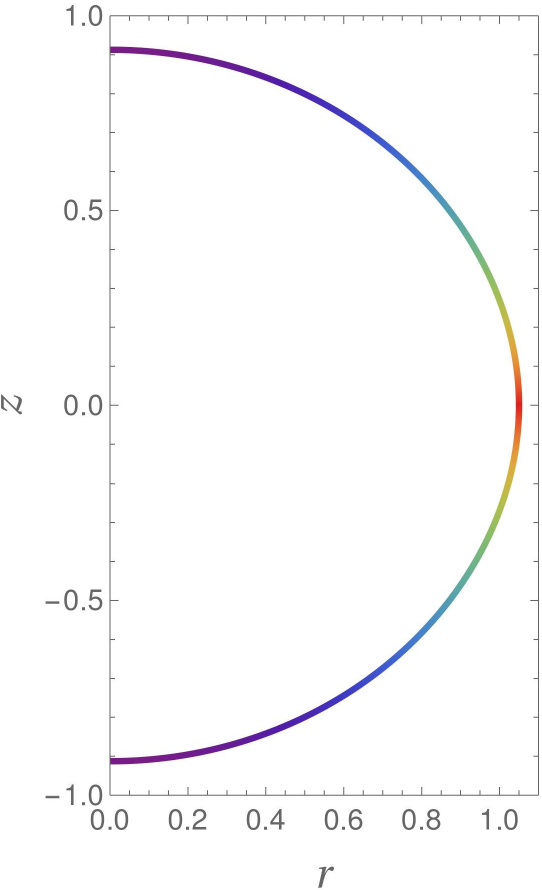}}}$ &
$\vcenter{\hbox{\includegraphics[scale=0.35]{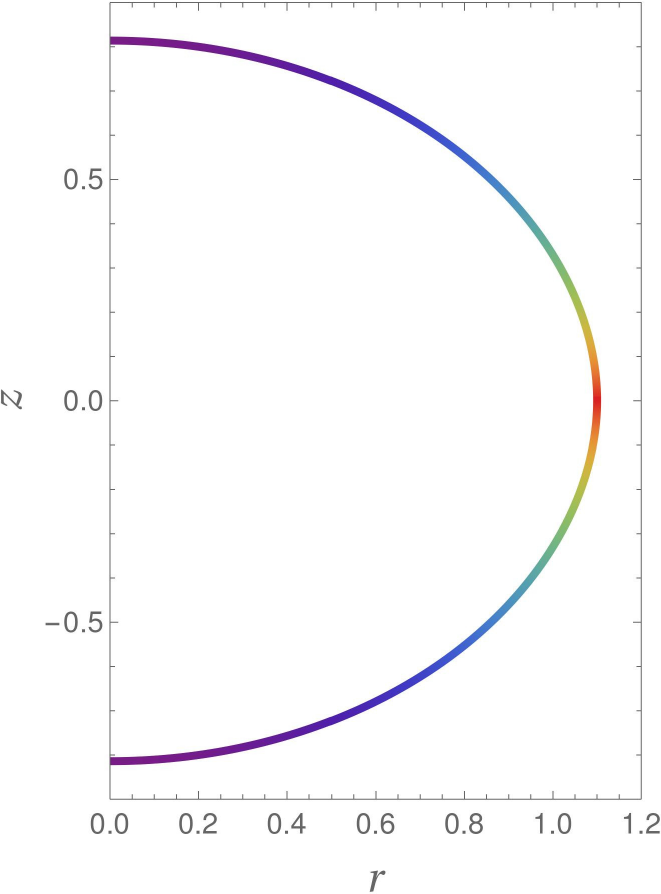}}}$ &
$\vcenter{\hbox{\includegraphics[scale=0.35]{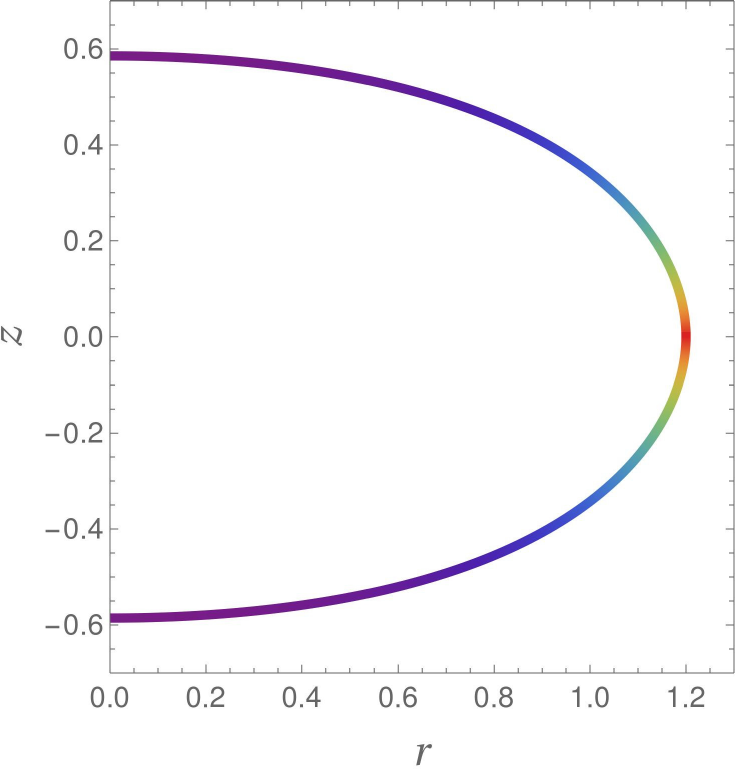}}}$ &
$\vcenter{\hbox{\includegraphics[scale=0.35]{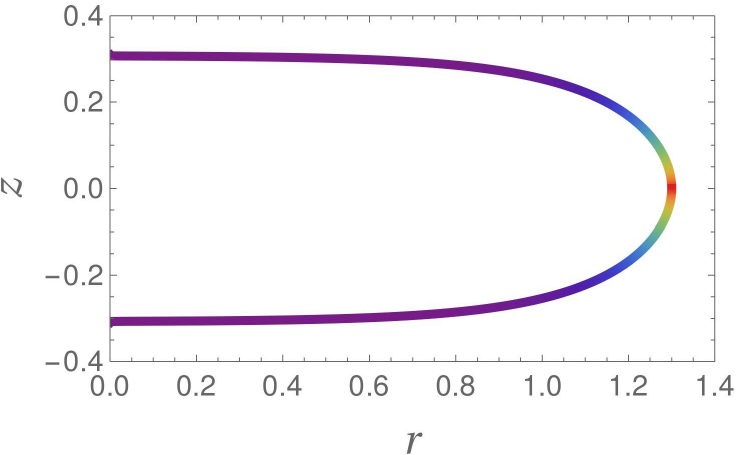}}}$ \\ \\
{\small (a) $r_0 = 1.05$} & {\small (b) $r_0 = 1.1$} & {\small (c) $r_0 = 1.2$} & {\small (d) $r_0 = 1.3$}
  \end{tabular}
 \end{center}
 \begin{center}
 \begin{tabular}{c}
 $\vcenter{\hbox{\includegraphics[scale=0.45]{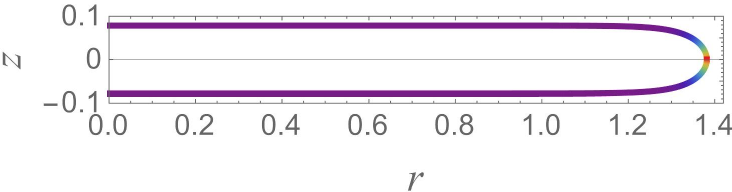}}}$ \\
 {\small (e) $r_0 = 1.38$} 
  \end{tabular}
 \end{center}
  \includegraphics[scale=0.6]{fig6g} 
\caption{(a)-(e) Sequence of stretched vesicles with $a=1$ and $c_s=0$ as $r_0$ is increased. The scaled bending energy density is color coded.}
\label{fig:7}
\end{figure*}

\begin{figure*}
\begin{center}
\begin{tabular}{cccccc}
$\vcenter{\hbox{\includegraphics[scale=0.475]{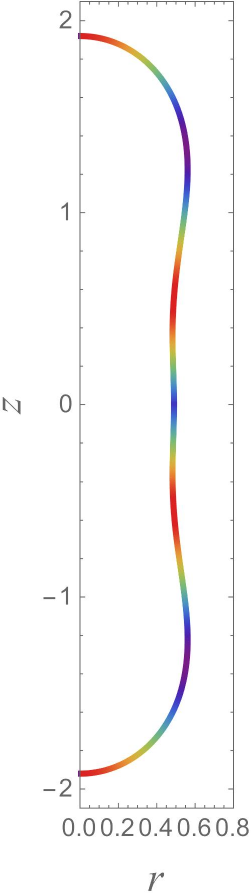}}}$ &
$\vcenter{\hbox{\includegraphics[scale=0.475]{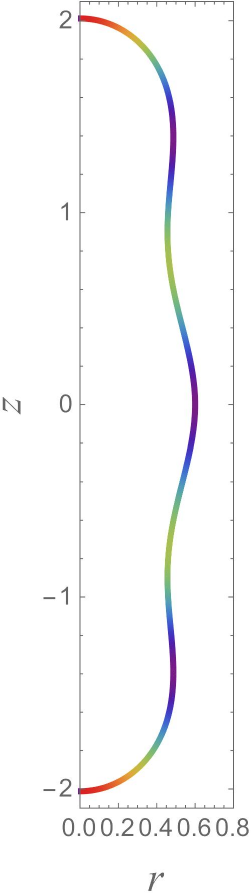}}}$ &
$\vcenter{\hbox{\includegraphics[scale=0.475]{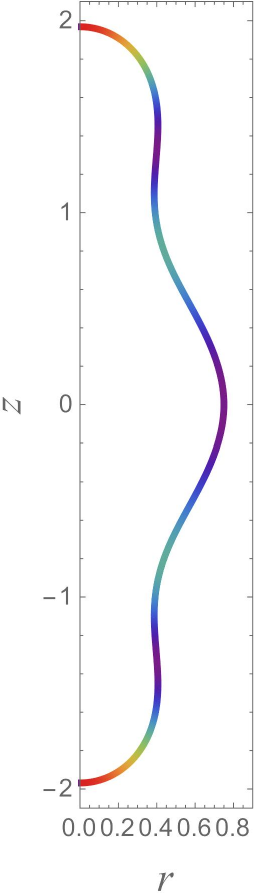}}}$ &
$\vcenter{\hbox{\includegraphics[scale=0.475]{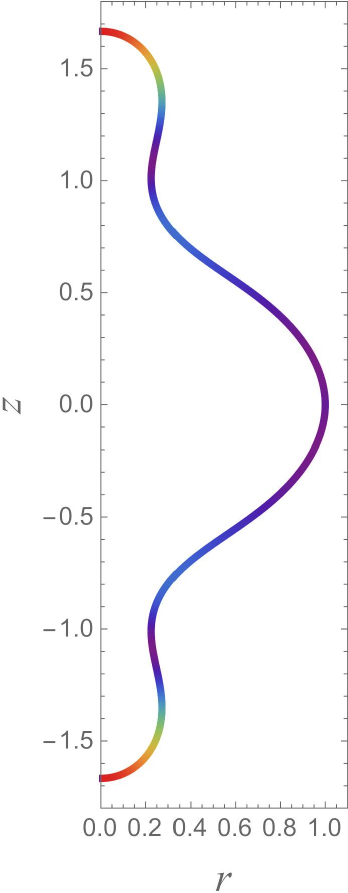}}}$ &
$\vcenter{\hbox{\includegraphics[scale=0.485]{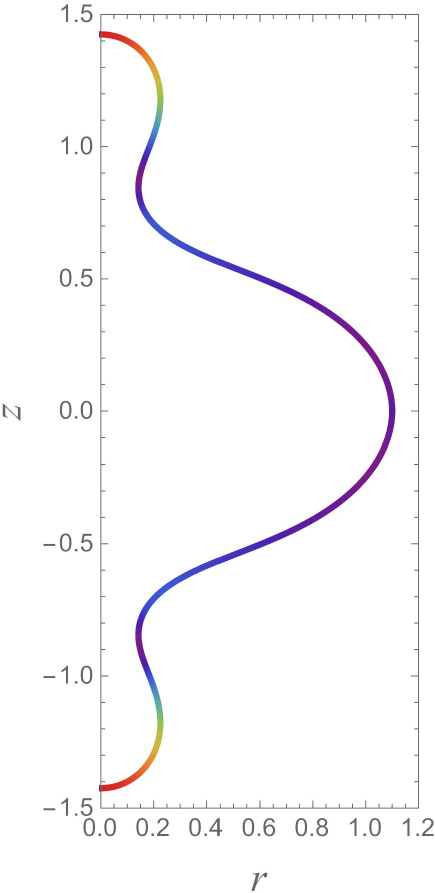}}}$ &
$\vcenter{\hbox{\includegraphics[scale=0.475]{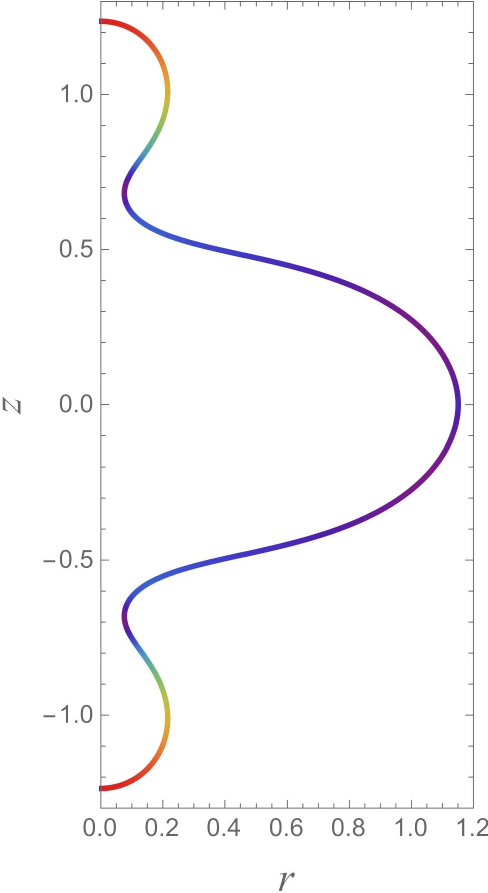}}}$ \\ \\
  {\small (a) $\medbullet$ $r^a_{01}=0.49$} & {\small (b) $r_0=0.6$} & {\small (c) $r_0=0.75$}  & {\small (d) $r_0=1$} & {\small (e) $r_0=1.1$} & {\small (f) $\blacksquare$ $r^a_{02}=1.15$}
  \end{tabular}
 \end{center}
  \includegraphics[scale=0.6]{fig6g}
\caption{(a)-(f) Sequence of vesicles with $a=1$ and $c_s=2\sqrt{2}$ corresponding to the first branch ($\medbullet$ $r^a_{01} < r_0 < r_{02}^a$ $\blacksquare$) as $r_0$ is increased. The scaled bending energy density is color coded.}
\label{fig:8}
\end{figure*}

\begin{figure*}
\begin{center}
\begin{tabular}{ccccc}
$\vcenter{\hbox{\includegraphics[scale=0.475]{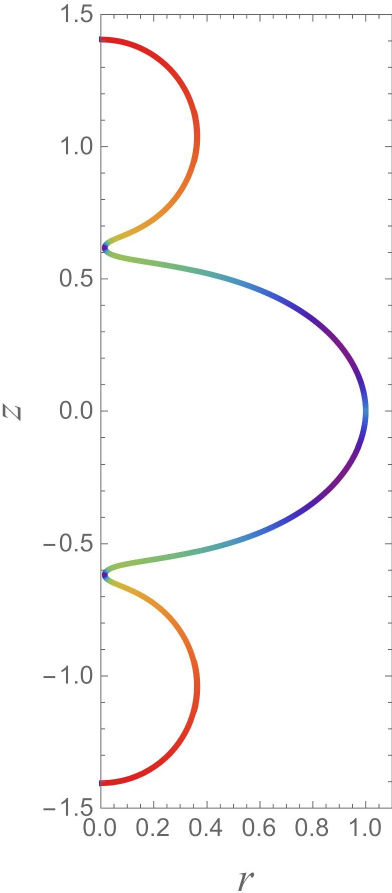}}}$ &
$\vcenter{\hbox{\includegraphics[scale=0.475]{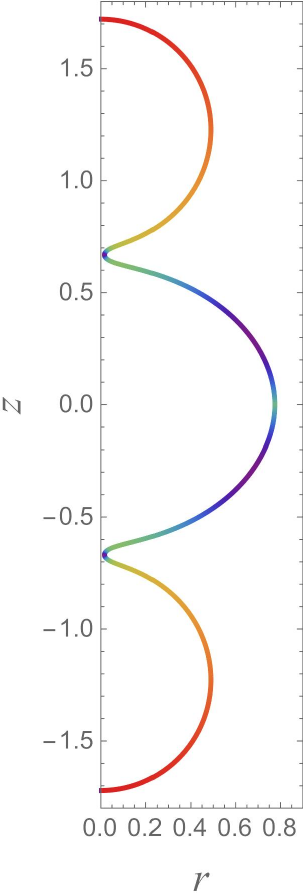}}}$ &
$\vcenter{\hbox{\includegraphics[scale=0.475]{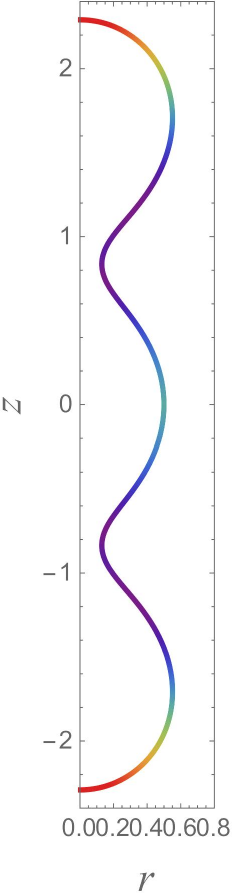}}}$ &
$\vcenter{\hbox{\includegraphics[scale=0.475]{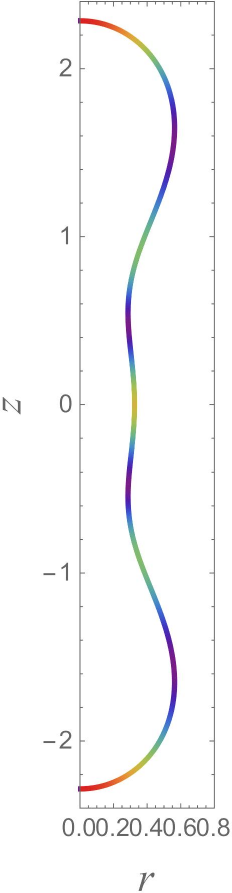}}}$ &
$\vcenter{\hbox{\includegraphics[scale=0.475]{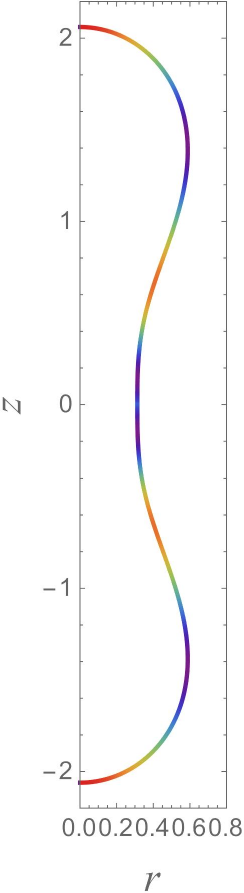}}}$ \\ \\
  {\small (a) $r_0=1$} &  {\small (b) $r_0=0.78$} & {\small (c) $r_0=0.5$} & {\small (d) $r_0=0.325$} &  {\small (e) $\blacktriangle$ $r^a_{03}=0.31$} 
  \end{tabular}
 \end{center}
  \includegraphics[scale=0.6]{fig6g}
\caption{(a)-(e) Sequence of vesicles with $a=1$ and $c_s=2\sqrt{2}$ corresponding to the second branch ($\blacktriangle$ $r^a_{03} < r_0 < r_{02}^a$ $\blacksquare$) as $r_0$ is decreased. The scaled bending energy density is color coded.}
\label{fig:9}
\end{figure*}

\begin{figure*}
\begin{center}
\begin{tabular}{ccc}
$\vcenter{\hbox{\includegraphics[scale=0.475]{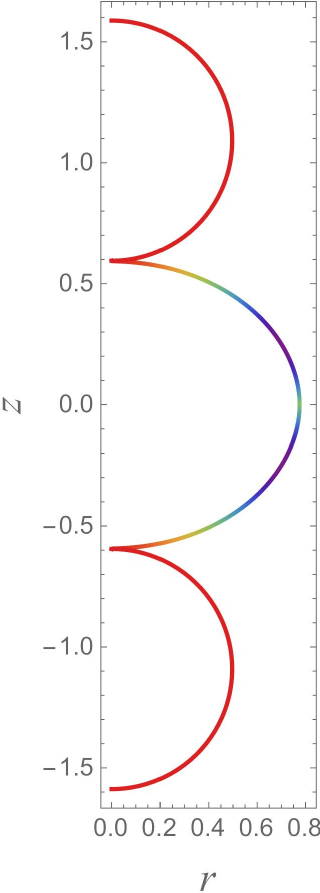}}}$ &
$\vcenter{\hbox{\includegraphics[scale=0.475]{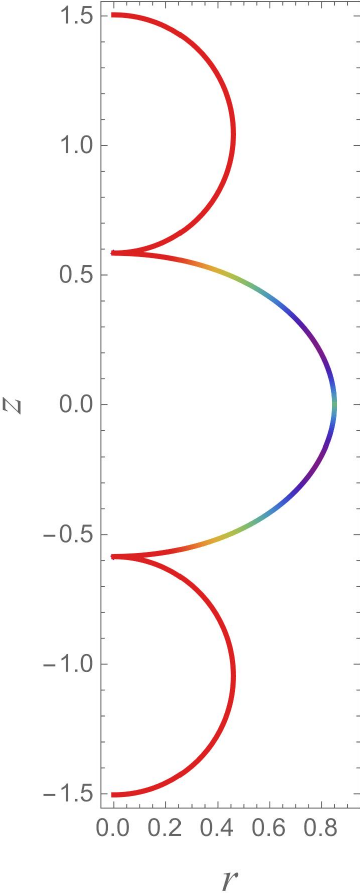}}}$ &
$\vcenter{\hbox{\includegraphics[scale=0.475]{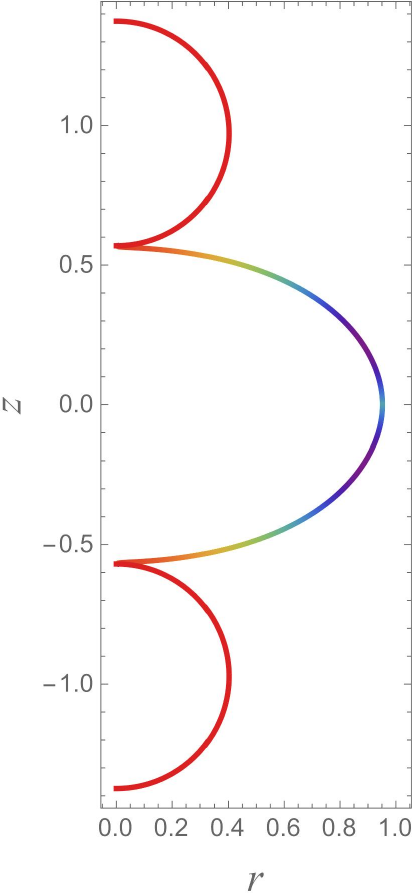}}}$ \\ \\
  {\small (a) $r_0=0.77$} & {\small (b) $r_0=0.85$} & {\small (c) $r_0=0.95$}
  \end{tabular}
 \end{center}
  \includegraphics[scale=0.6]{fig6g}
\caption{(a)-(c) Vesicles with $a=1$ and $c_s=3$ corresponding to the second branch and with vanishingly small necks. The scaled bending energy density is color coded.}
\label{fig:10}
\end{figure*}

\begin{figure}[htb]
 \centering
\includegraphics[width=0.475\textwidth]{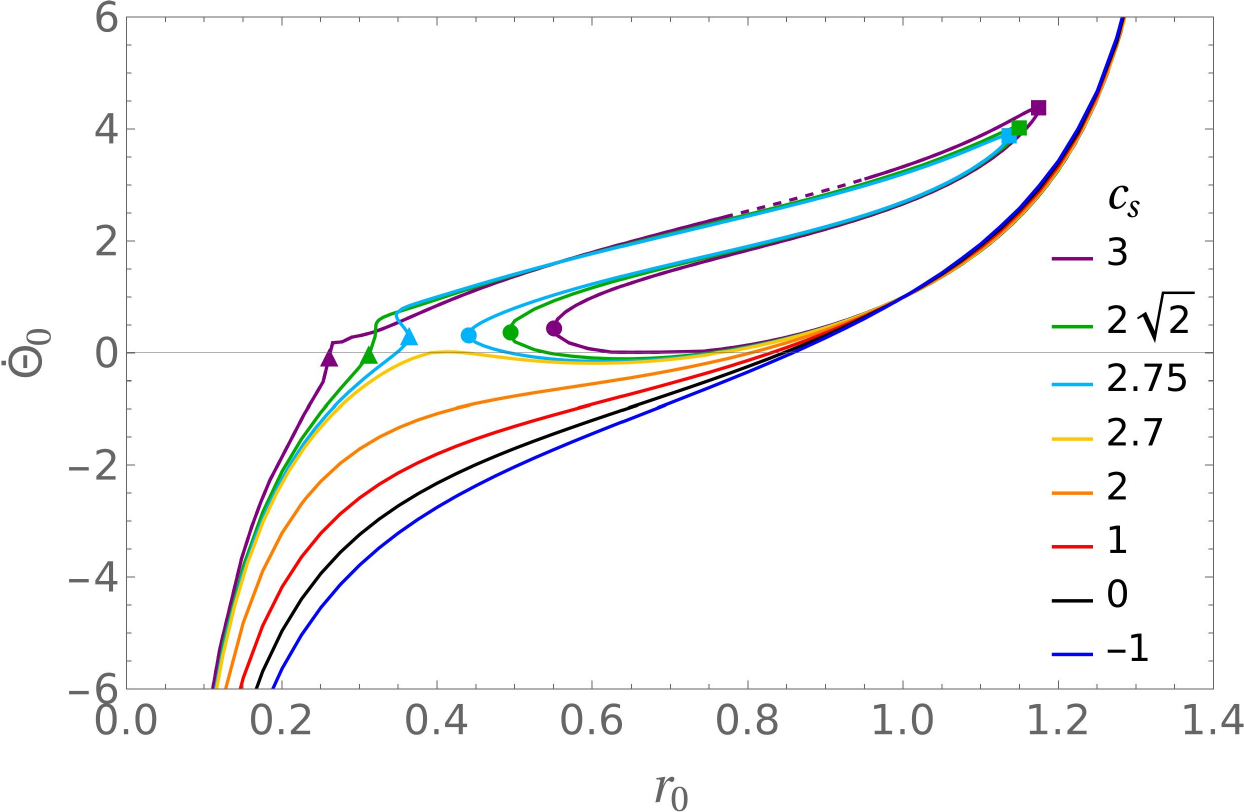}
\includegraphics[width=0.475\textwidth]{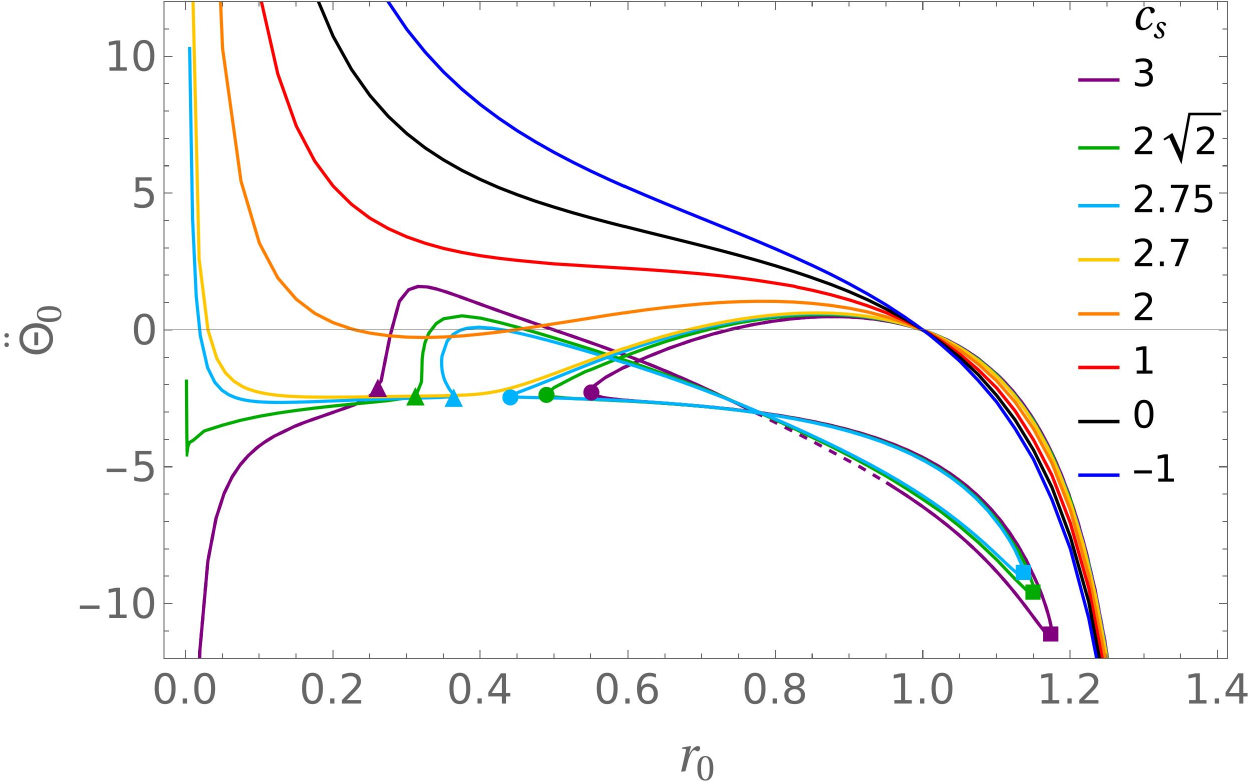}
\caption{Dependence of $\dot{\Theta}_0$ and $\ddot{\Theta}_0$ on $r_0$ for membranes with $a=1$. The first and second branches are demarcated with the symbols $(\medbullet,\blacksquare)$ and $(\blacktriangle,\blacksquare)$ respectively. The dashed line represents configurations with very small necks.}
\label{fig:11}
\end{figure}

\begin{figure}[htb]
 \centering
\includegraphics[width=0.475\textwidth]{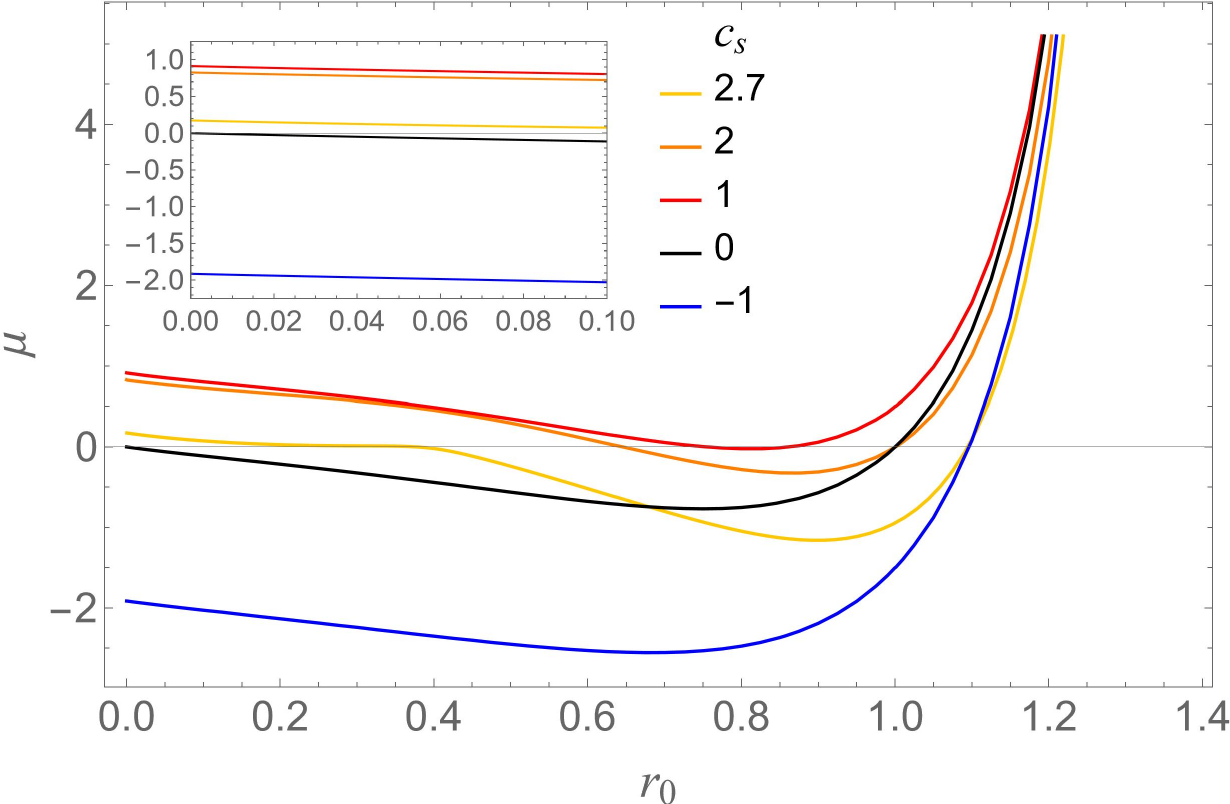}
\includegraphics[width=0.475\textwidth]{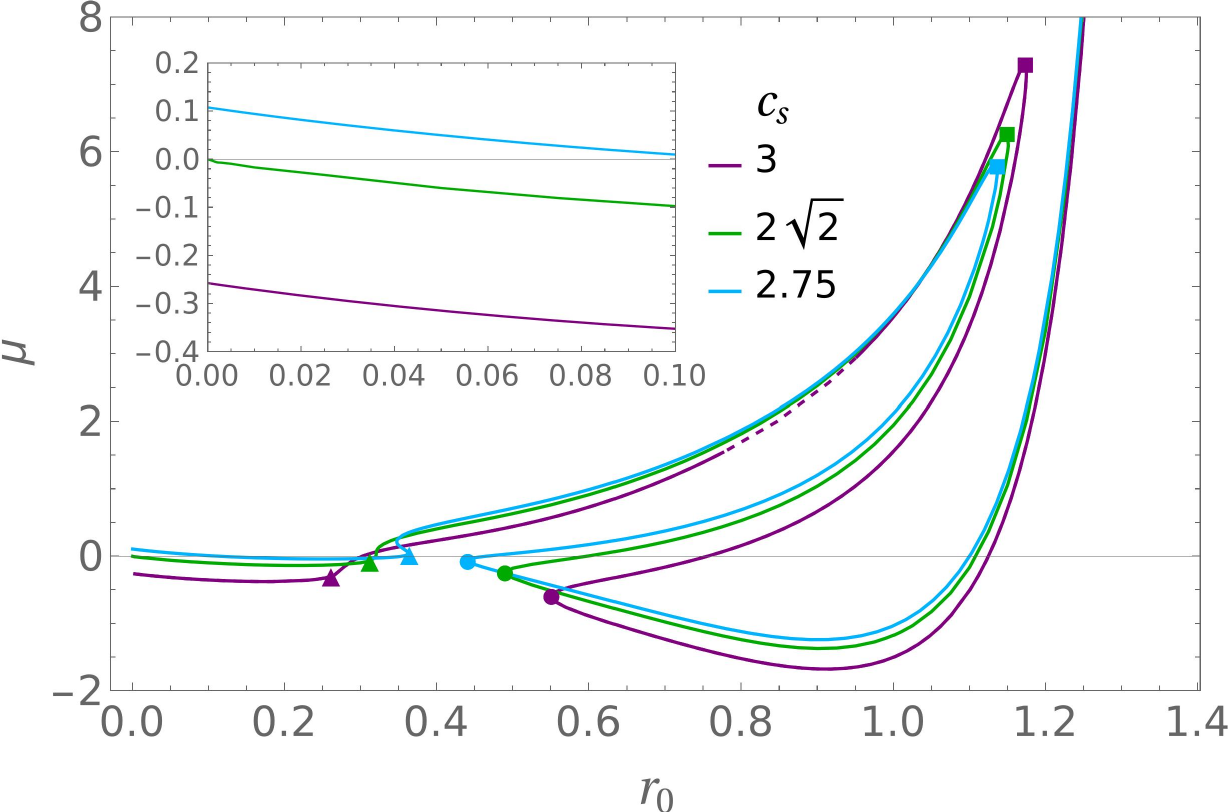}
\caption{Lagrange multiplier $\mu$ as a function of $r_0$ for membranes with $a=1$. The insets show values close to the limit of maximum constriction. The first and second branches are demarcated with the symbols $(\medbullet,\blacksquare)$ and $(\blacktriangle,\blacksquare)$, respectively. The dashed line represents configurations with very small necks.}
\label{fig:12}
\end{figure}

\begin{figure}[htb]
\includegraphics[width=0.475\textwidth]{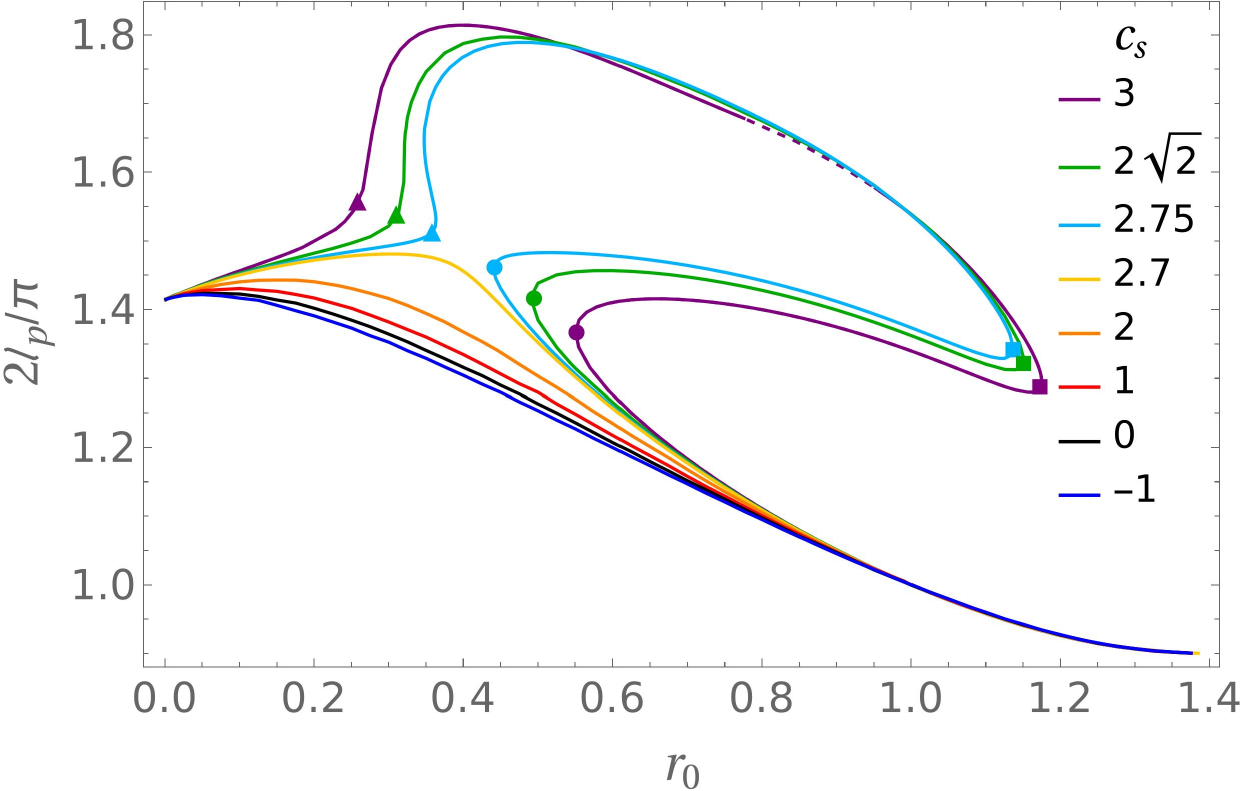}
\includegraphics[width=0.475\textwidth]{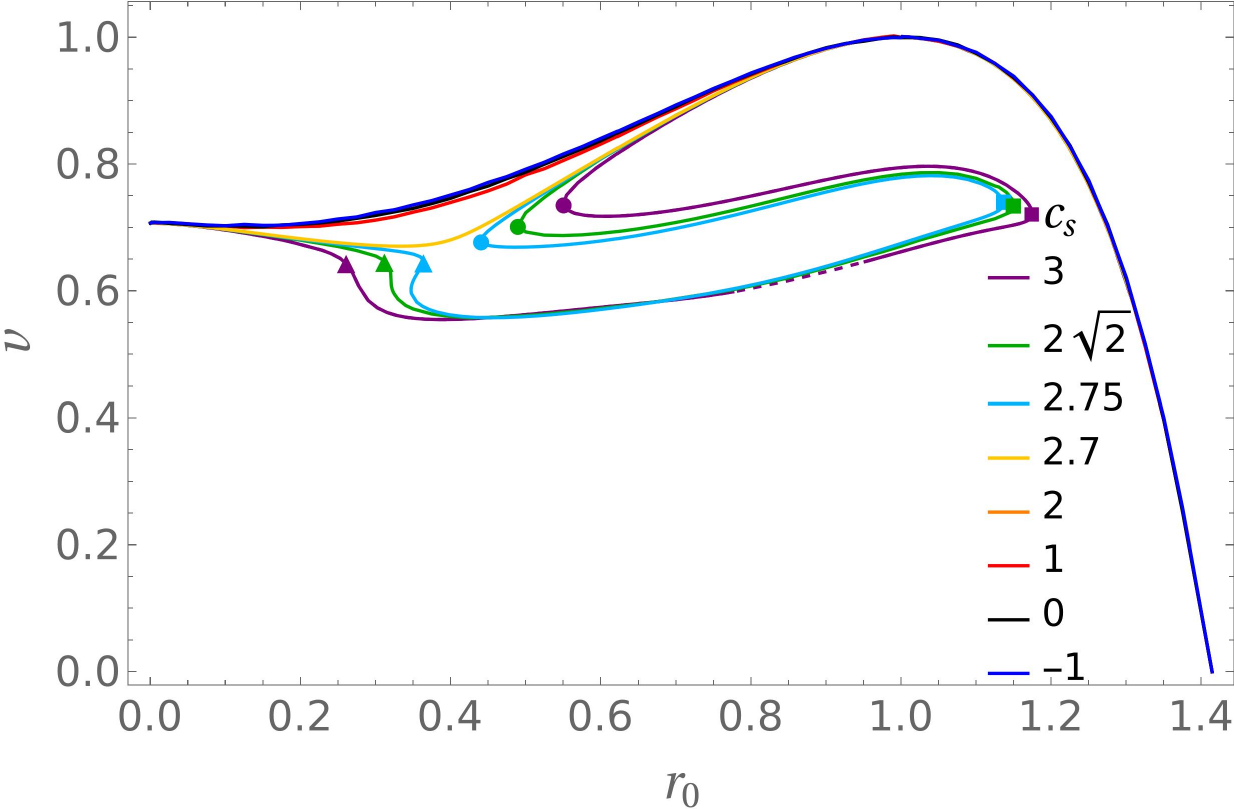}
\caption{Total length of the generating curve $2 \ell_p$ and volume $v$ as functions of $r_0$ for membranes with $a=1$. The first and second branches are demarcated with the symbols $(\medbullet,\blacksquare)$ and $(\blacktriangle,\blacksquare)$, respectively. The dashed lines represent configurations with very small necks.
}
\label{fig:13}
\end{figure}

\begin{figure}[htb]
 \centering
\includegraphics[width=0.48\textwidth]{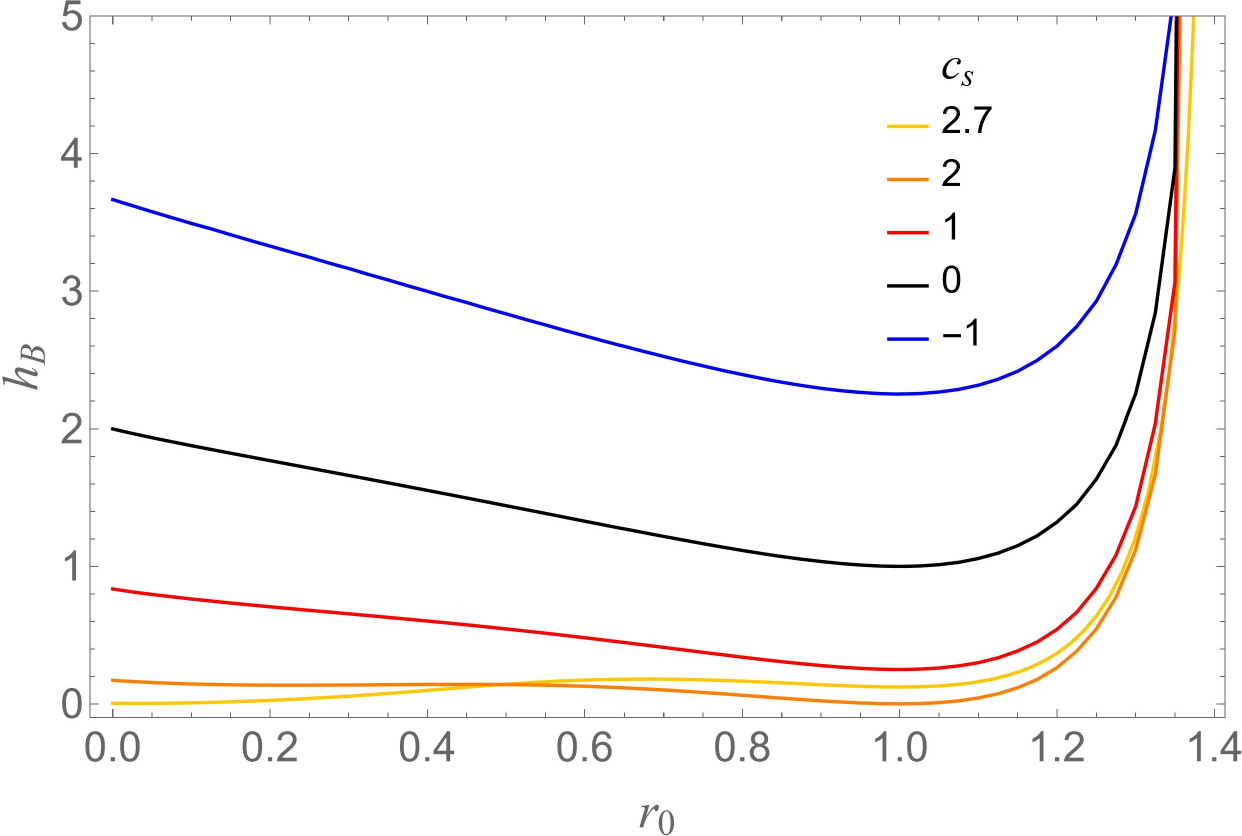}
\includegraphics[width=0.48\textwidth]{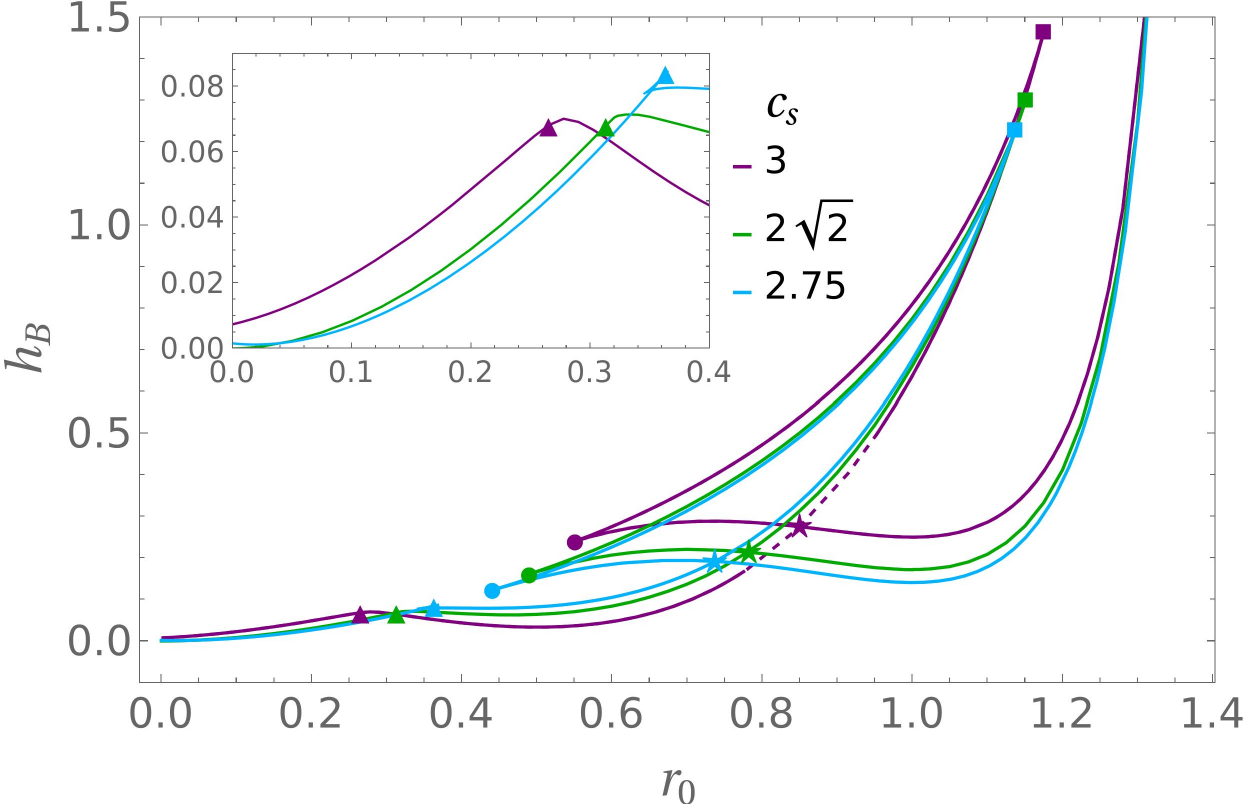}
\caption{Scaled total energy $h_B$ as a function of $r_0$ for membranes with $a=1$. The first and second branches are demarcated with the symbols $(\medbullet,\blacksquare)$ and $(\blacktriangle,\blacksquare)$, respectively. The dashed line represents configurations with very small necks.
}
\label{fig:14}
\end{figure}

\begin{figure}[htb]
 \centering
\includegraphics[width=0.48\textwidth]{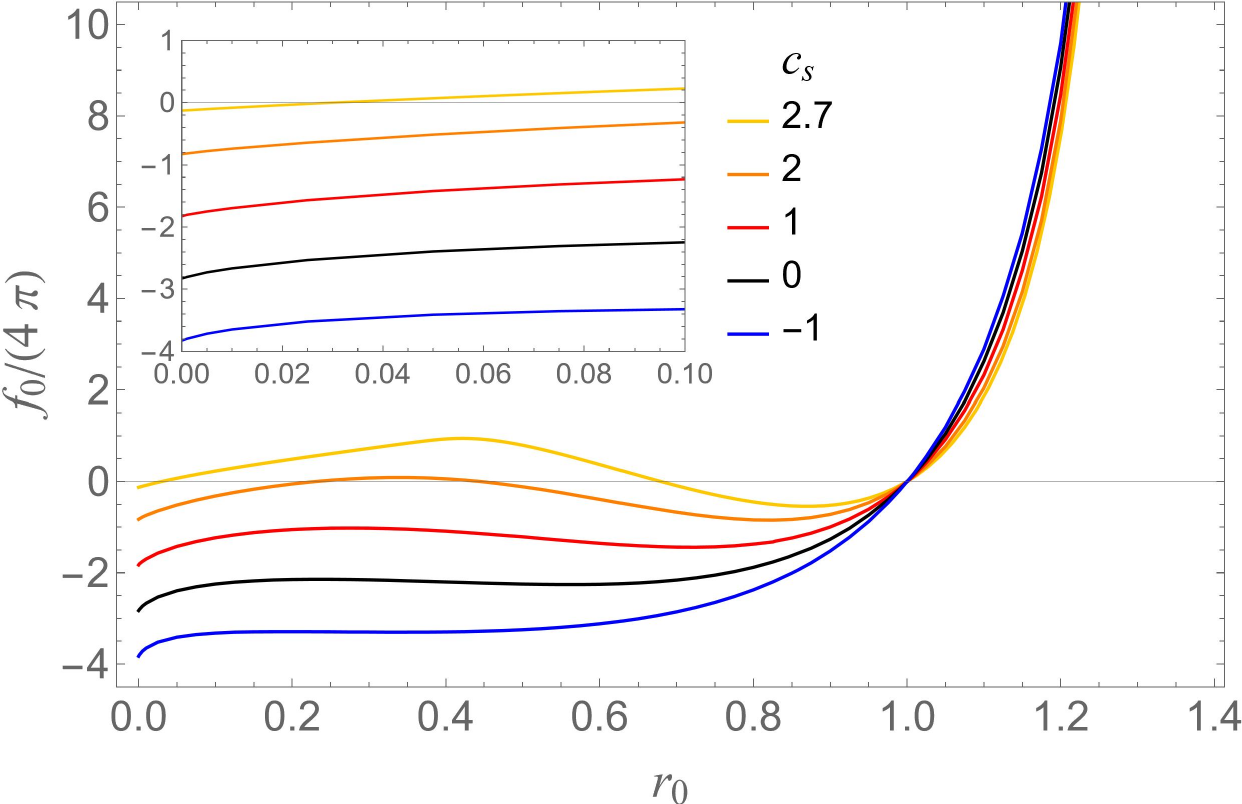}
\includegraphics[width=0.48\textwidth]{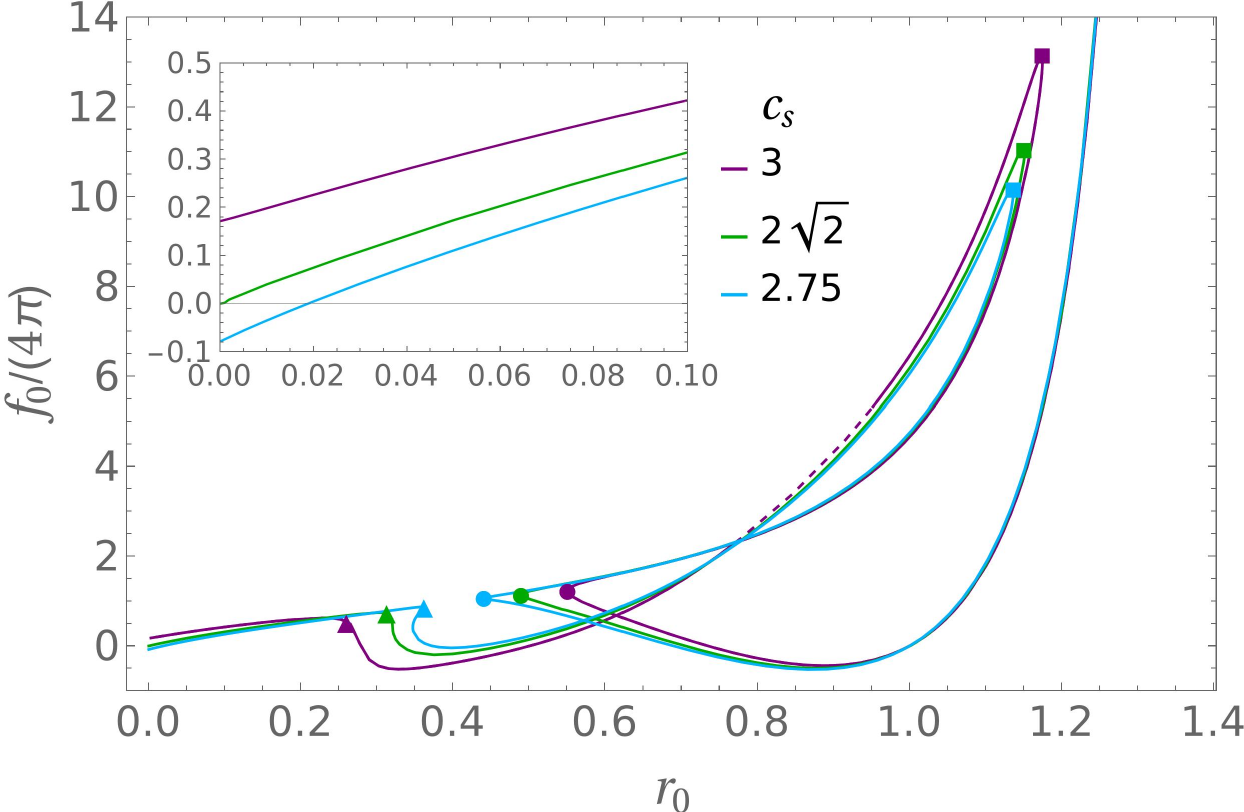}
\caption{Scaled total force $f_0$ as a function of $r_0$ for membranes with $a=1$. The first and second branches are demarcated with the symbols $(\medbullet,\blacksquare)$ and $(\blacktriangle,\blacksquare)$, respectively. The dashed line represents configurations with very small necks.
}
\label{fig:15}
\end{figure}

We first consider deformed vesicles whose area is equal to the area of the spherical vesicle, $A=A_S$, so their reduced area is unit, $a = 1$. The reduced volume of the vesicle $v$ is variable, so $\mrmp=0$. This case could represent the situation in which the control parameter is the pressure difference across the membrane and with a fixed temperature, so that the enclosed volume varies. In this case Eq.\ (\ref{EL:kappa}) reduces to
\begin{equation} \label{EL:kappaAfix}
\cos \Theta \dot{\kappa} + \sin \Theta \left( \kappa \left( \frac{\kappa}{2} -\frac{\sin
\Theta}{r}+ c_s \right) - \mu  \right) = 0 \,.
\end{equation}
Evaluating Eq.\ (\ref{EL:kappaAfix}) at a parallel with $\Theta_0$ and $r_0$, and using Eqs.\ (\ref{Eq:k0dk0Theta0dTheta0}), we can determine $\mu$ in terms of the initial values of the angle and its derivatives
\begin{eqnarray}
 \mu &=&\cot \Theta_0 \left(\ddot{\Theta}_0+\frac{\cos \Theta_0}{r_0} \left(\dot{\Theta}_0-\frac{\sin \Theta_0}{r_0}\right)\right) \nonumber \\
 &&+ \frac{1}{2} \left(\dot{\Theta}_0^2-\left(\frac{\sin \Theta_0}{r_0}-c_s\right)^2\right)\,.
\end{eqnarray}
We solved the system of differential equations (\ref{nondimhamsystem}) for different values of $c_s$. Starting with values close to $c_s=0$ and up to $c_s=2.7$ we found that there is a single sequence of configurations characterized by the equatorial radius $r_0$, which are illustrated in Fig.\ \ref{fig:6} for constriction and in Fig.\ \ref{fig:7} for dilation. However, for higher values of $c_s$ two bifurcations in the solutions arise, which comprise constricted and dilated vesicles, shown in Figs.\ \ref{fig:8} - \ref{fig:10}.
\vskip0pc \noindent
The equatorial radius of a constricted vesicle is in the interval $0 < r_0 <1$. As $r_0$ is decreased, the vesicle adopts a prolate shape (Fig.\ \ref{fig:6}(a)) and elongates along the symmetry axis. Then the equatorial region becomes cylindrical ($\dot{\Theta}_0=0$), setting the beginning of the transition from a prolate  shape (Fig.\ \ref{fig:6}(b)) to a dumbbell shape with a waistline, (Fig \ref{fig:6}(c)). As $r_0$ is further decreased the waist decreases rapidly and the vesicle is mainly composed of two increasing quasi spherical vesicles (Fig.\ \ref{fig:6}(d)). In the limit of maximum constriction, $r_0 \rightarrow 0$, the vesicle consists of two quasispherical vesicles of radius $r_\rma=1/\sqrt{2}$ (so $\ell_p=\pi/\sqrt{2}$ and $k_\rma=2 \sqrt{2}$) connected by a vanishingly small catenoid-like neck (Figs.\ \ref{fig:6}(e)-(f)).
\\
The equatorial radius of a stretched vesicle is in the interval $1< r_0 < \sqrt{2}$. As $r_0$ is increased the vesicle becomes oblate and the distance between the poles decreases (Figs.\ \ref{fig:7}(a)-(c)), then the polar regions flatten more and more, (Fig.\ \ref{fig:7}(d)), up to the limit of maximum stretching with $r_0\to \sqrt{2}$, where the vesicle is composed of two discs joined along their boundary, (Fig.\ \ref{fig:7}(e)), so $\ell_p \rightarrow \sqrt{2}$.
\begin{table}
 \begin{tabular}{||c|c|c|c||}
 \hline \hline
$c_s$ & $2.75$ & $2 \sqrt{2}$ & $3$ \\
 \hline \hline
$\medbullet$ $r^a_{0 1}$ & 0.44 & 0.49 & 0.55 \\
 \hline
$\blacksquare$ $r^a_{0 2}$ & 1.14 & 1.15 & 1.17 \\
\hline
$\blacktriangle$ $r^a_{0 3}$ & 0.36 & 0.31 & 0.26 \\
\hline
$\bigstar$ $r^a_{0 4}$ & 0.74 & 0.78 & 0.85 \\
\hline \hline
 \end{tabular}
 \caption{Numerical values of $r_0$ for vesicles with $a=1$ corresponding to the bifurcations and their branches, which are indicated with the respective symbols in Figs. \ref{fig:8}, \ref{fig:9} and \ref{fig:11}-\ref{fig:15}.} \label{TableI}
\end{table}
\vskip0pc \noindent
As the spontaneous curvature is increased above $c_s=2.7$ the sequence of configurations is not simple, because bifurcations occur in the solutions of the system of differential equations. Some relevant values of the equatorial radius pertaining to these bifurcations and their branches, referred to as $r^a_{0i}$, $i=1,2,3,4$ in the following discussion, and indicated with markers in the plots of the vesicle quantities below, are presented in Table \ref{TableI}. The first bifurcation occurs in the sequence of constricted vesicles, as $r_0$ is decreased there are no further configurations for $r_0<r^a_{01}$, and the sequence of the configurations continues for higher values of $r_0$ along a branch in the interval $r^a_{01} < r_0 < r^a_{02}$, whose corresponding configurations are illustrated in Fig.\ \ref{fig:8}. At $r^a_{01}$ the vesicle is in the transition from prolate to dumbbell shapes (Fig.\ \ref{fig:8}(a)). As $r_0$ is increased the equatorial region begins to bulge (Figs.\ \ref{fig:8}(b)-(c)), and as $r_0$ is further increased it swells while the regions close to the poles become rounded and develop necks (Figs.\ \ref{fig:8}(d)-(e)). At $r^a_{02}$ the central region reaches a maximum stretching (Fig.\ \ref{fig:8}(f)) and another bifurcation occurs: there are no further solutions with higher $r_0$ and the sequence of configurations continues for lower values of $r_0$ along another branch in the interval $r^a_{03} < r_0 < r^a_{02}$, whose corresponding configurations are shown in Fig.\ \ref{fig:9}. As $r_0$ is reduced below $r^a_{02}$ the central region deflates and the necks shrink, while the top and bottom round regions become quasispherical, Figs.\ \ref{fig:9}(a)-(b). Then, the central region keeps deflating, but the radius of the necks increases and the vesicle adopts an onduloidlike shape Fig.\ \ref{fig:9}(c). Upon further reduction of $r_0$ the necks conflate with the central region while the extremes are kept rounded (Fig.\ \ref{fig:9}(d)), until the vesicle adopts a dumbbell shape Fig.\ \ref{fig:9}(e) at $r^a_{03}$, where the second branch merges with the original sequence of configurations, after which the constriction of the vesicle proceeds just as discussed above.
\\
The radius of the necks of the configurations of the second branch can be rendered very small by increasing $c_s$, for instance for $c_s=3$ in the interval $0.77 < r_0 < 0.95$ the configurations consist of oblate membranes connected to quasispherical vesicles above and below by an infinitesimal neck (Fig.\ \ref{fig:10}). This situation could be regarded as the remote constriction of the vesicle, similar to the process presented in Ref.\ \cite{Bozic2014}. The quasispherical vesicles satisfy Eq.\ (\ref{eq:spheresclrl}) with $\mrmp=0$, so they are not subject to external forces and they have a radius given by
\begin{equation}
r_\mathbf{s} = \frac{2 c_s}{c_s^2+2 \mu}\,. 
\end{equation}
The plots of the parameters of the vesicle are shown in Figs.\ \ref{fig:11}-\ref{fig:15}, where the values corresponding to configurations delimiting the first and second branches (shown in Figs.\ \ref{fig:8} and \ref{fig:9}) are indicated with the symbols $(\medbullet,\blacksquare)$ and $(\blacktriangle,\blacksquare)$,  respectively (cf. Table \ref{TableI}), whereas the values corresponding to the configurations with small necks are shown with a dashed line.
\\
The values of $\dot{\Theta}_0$ and $\ddot{\Theta}_0$ as functions of $r_0$ are plotted in Fig.\ \ref{fig:11}. For the initial spherical state with $r_0=1$ their values are $\dot{\Theta}_0=1$ and $\ddot{\Theta}_0 =0$. For values close to $c_s=0$ and constriction, $\dot{\Theta}_0$ decreases ($\ddot{\Theta}_0$ increases) monotonically, tending to $-\infty$ ($+\infty$) as $r_0 \rightarrow 0$. As $c_s$ is increased $\dot{\Theta}_0$ ($\ddot{\Theta}_0$) takes lower (higher) values until it develops a maximum (minimum) and for $c_s>2\sqrt{2} $, $\ddot{\Theta}_0 \rightarrow -\infty$ as $r_0 \rightarrow 0$, which has the consequence that the total force changes of sign. Above $c_s=2.7$, $\dot{\Theta}_0$ ($\ddot{\Theta}_0$) increases (decreases) along the first branch, whereas along the second one it decreases (first increases and then decreases). For dilation $\dot{\Theta}_0>1$ ($\ddot{\Theta}_0<0$) and it increases (decreases) monotonically, tending to $+\infty$ ($-\infty$) as $r_0 \rightarrow \sqrt{2}$.
\begin{table}
 \begin{tabular}{||c|c|c|c|c|c|c|c|c||}
 \hline \hline
$c_s$ & -1 & 0 & 1 & 2 & 2.7 & 2.75 & $2\sqrt{2}$ & 3 \\
 \hline \hline
$\mu^a_{mc}$ & -1.91 & 0 & 0.91 & 0.83 & 0.17 &0.11 & 0 & -0.26\\
\hline
$h^a_{B mc}$ & 3.66 & 2 & 0.84 & 0.17 & $0.004$ & 0.0015 & 0 & $0.0074$ \\
\hline
$\frac{1}{4\pi} f^a_{0 mc}$ & -3.83 & -2.83 & -1.83 & -0.83 & -0.13 & -0.08 & 0 &0.17\\
\hline \hline
 \end{tabular}
 \caption{Numerical values of quantities of vesicles with $a=1$ and different values of $c_s$ in the limit of maximum constriction.} \label{TableII}
\end{table}
\vskip0pc \noindent
The dependence of the Lagrange multiplier $\mu$ on $r_0$ is plotted in Fig \ref{fig:12}. For the spherical membrane with $r_0=1$ it has a value $\mu=c_s(1-c_s/2)$. For values close to $c_s=0$, as $r_0$ is decreased $\mu$ first decreases reaching a global minimum and then increases up to the value $\mu^a_{mc}$ (given in Table \ref{TableII}) in the limit of maximum constriction. For stretched membranes $\mu$ increases with $r_0$, diverging in the limit of maximum stretching, $\mu \rightarrow +\infty$ as $r_0 \rightarrow \sqrt{2}$. For $c_s>2.7$, $\mu$ increases along the first branch and decreases along the second one.
\vskip0pc \noindent
The scaled total length of the profile curve $2 \ell_p$ and the reduced volume $v$ are plotted in Fig.\ \ref{fig:13}. The scaled total length begins at the value $\pi$ and for values close to $c_s=0$ it increases as $r_0$ is decreased, reaching a maximum and then decreases reaching the value $\sqrt{2}$ in the limit of maximum constriction. For dilation it decreases monotonically reaching the value $2\sqrt{2}$ in the maximum stretching limit. For $c_s>2.7$, along both branches $2 \ell_p$ first increases and then decreases.
\\
The behavior of the reduced volume is practically the same for $c_s < 2.7$, it begins with the value $v=1$ at $r_0=1$ and for constriction it decreases with $r_0$, after it reaches a minimum it increases up to the value $v=1/\sqrt{2}$ at the limit of maximum constriction. For dilation $v$ increases with $r_0$ reaching a maximum and then it decreases towards $0$ in the limit of maximum stretching. For $c_s > 2.7$, along the first branch $v$ increases reaching a maximum, then it decreases and along the second branch it keeps decreasing,
it reaches a minimum and then it increases.
\vskip0pc \noindent
The scaled total energy $h_B$ is plotted in Fig.\ \ref{fig:14}. The total energy of the initial spherical configuration with $r_0=1$ is given in Eq.\ (\ref{hBsphere}). For values close to $c_s=0$, as the equator is reduced, $h_B$ increases monotonically, reaching the value $h^a_{B mc}$ (given in Table \ref{TableII}) for the totally constricted membranes, which agrees with the exact value 
\begin{equation}
h^a_{B mc} = (\sqrt{2}-c_s/2)^2 \,,
\end{equation}
corresponding to the scaled total energy of two quasispherical vesicles joined by an infinitesimal neck, which vanishes if $c_s = 2\sqrt{2}$. For stretched membranes $h_B$ increases with $r_0$, diverging to $+\infty$ as $r_0\to \sqrt{2}$. For $c_s > 2.7$, $h_B$ changes twice in a nonsmooth way at the bifurcations, first at the transition from the original sequence of constricted configurations to the first branch, where it changes from decreasing to increasing and at the transition to the second branch, where it becomes decreasing again and intersects the original sequence at $r_0=r^a_{04}$ (indicated in Fig. \ref{fig:14} with the symbol $\bigstar$), corresponding to two different configurations with the same total energy. For $r_0< r^a_{04}$ the configurations of the second bifurcation have the lowest energy, so it is conceivable that the original constricted configurations tend to decay to the configurations of the second bifurcation.
\\
The scaled total force $f_0$ is plotted in Fig.\ \ref{fig:15}. The data of the numerical solutions confirm that the equatorial force is given by the derivative of the bending energy with respect to the equatorial radius, $F=\partial H_B/\partial R_0$ or in terms of scaled quantities, $f_0=8 \pi \partial h_b/\partial r_0$.
The initial spherical membrane is free of external forces, $f_0 = 0$. For values close to $c_s = 0$ the total force on the constricted membranes is negative (so it is constrictive), and displays an oscillating behavior: as $r_0$ is decreased $f_0$ also decreases to a minimum, then it increases to a maximum, after which it decreases again towards the limit of maximum constriction, where it reaches a finite value $f^a_{0 \, mc}$ (given in Table \ref{TableII}). Let us examine the behavior of the total force in the limit of maximum constriction a little more closely. Although the linear force density diverges as $r_0 \rightarrow 0$, $\phi_0 = - 2\ddot{\Theta}_0 \rightarrow \infty$ ($-\infty$ for $c_s>2\sqrt{2}$), the product $r_0 \ddot{\Theta}_0$ converges to the finite value given above, so $f_0$ is finite. As we will see below, this is possible because in this limit the geometry of the neck does not correspond to a surface of constant mean curvature, in particular it is not the central segment of a catenoid, which could not sustain an equatorial force. For stretched vesicles the force increases monotonically as $r_0$ is increased, diverging in the limit of two flat discs connected along their rim, $f_0 \rightarrow \infty$ as $r_0 \to \sqrt{2}$. Since it is not possible to stretch the membrane into such disclike geometry by applying a finite force, one would expect the membrane to rupture beyond some critical force, which would occur through pore formation \cite{Farago2005, Illya2008}. As $c_s$ is increased $f_0$ takes more positive values and for $c_s>2$ it becomes positive (so it is dilative) for some configurations. For $c_s>2.7$, $f_0$ is positive and increasing along the first branch, whereas along the second one it decreases towards negative values, reaches a minimum and then increases to positive values where it merges with the original sequence of constricted vesicles.

\subsection{Limit of maximum constriction: Quasicatenoidal neck}

\begin{figure}[htb]
 \centering
\includegraphics[width=0.475\textwidth]{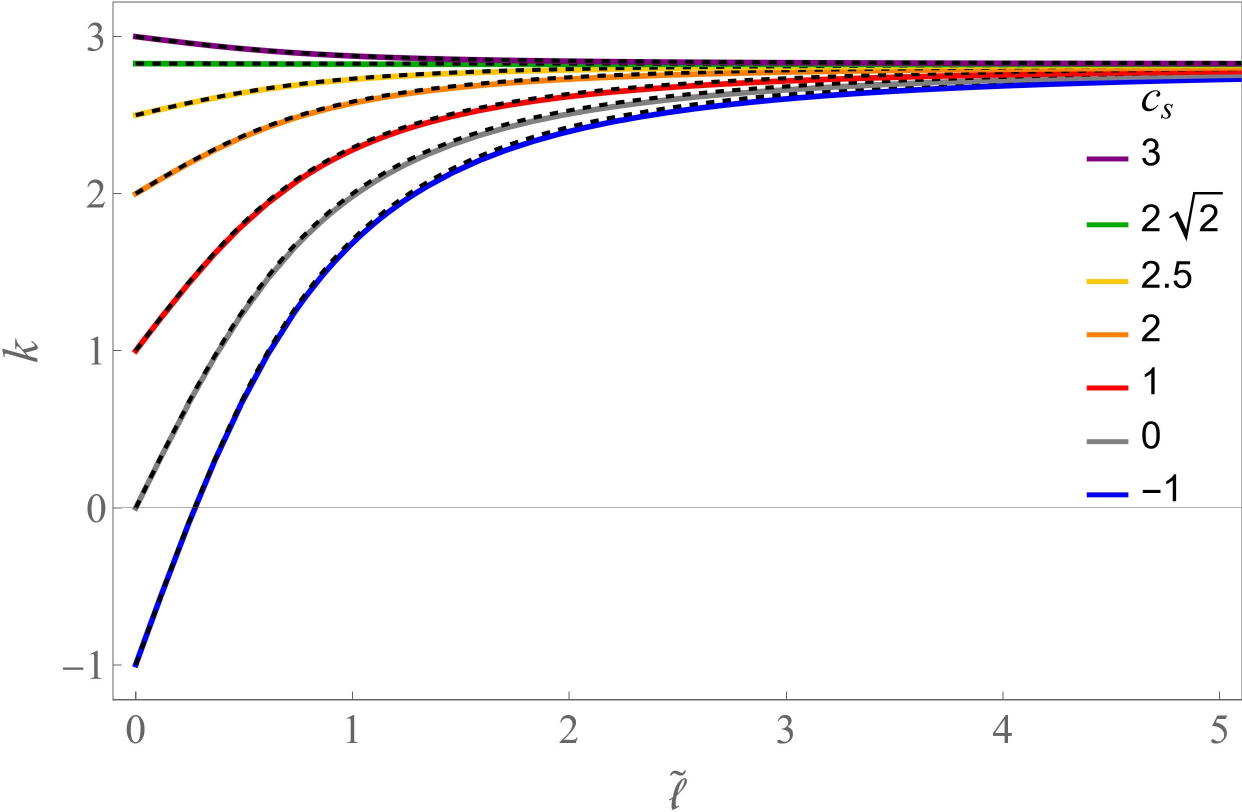}
\caption{Mean curvature as a function of the arclength of configurations with $a=1$ in the region close to the neck. At the equator its value is given by spontaneous curvature and away from the equator its value tends to the mean curvature of the quasispherical membrane, $k_a=2\sqrt{2}$. The solid lines correspond to the numerical data and the dashed lines to plots of the first order correction $\tilde{k}_{(1)}/r_0$.} 
\label{fig:16}
\end{figure}

\begin{figure}[hbt]
\begin{center}
  \includegraphics[width=0.475\textwidth]{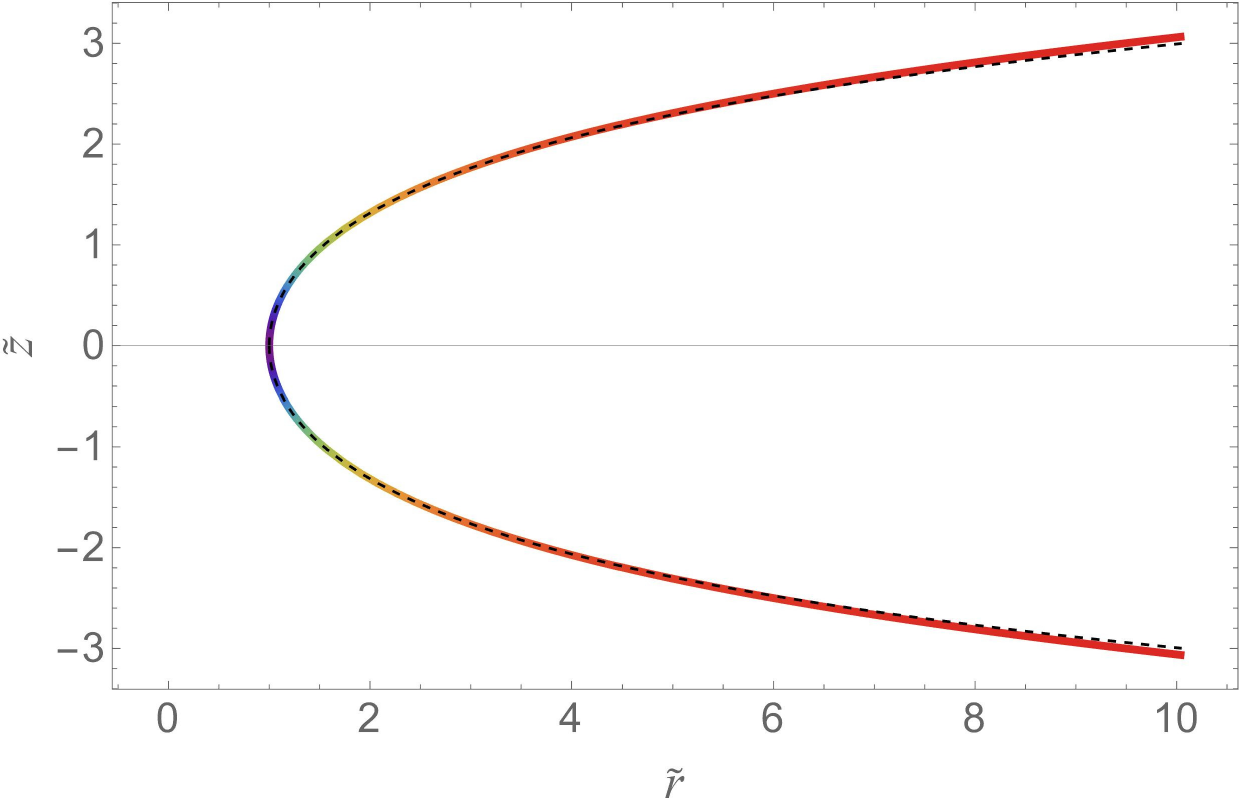}
  \vskip1pc
  \includegraphics[scale=0.55]{fig6g}
\end{center}
\caption{Profile curve of the narrow neck of the vesicle in the limit of maximum constriction: the solid line corresponds to the first-order deformed catenoid and the black dashed line to the original catenoid. The magnitude of the perturbation has been augmented for illustration purposes. The bending energy density is color coded.}
\label{fig:17}
\end{figure}

At first glance, from the numerical solutions it appears that the shape of the neck is well approximated by the central section of a catenoid (its generating curve, a catenary, is shown with a dashed line in Figs.\ \ref{fig:1}(f) and \ref{fig:6}(f)), regardless of the value of $c_s$. However, such comparison is deceptive, although the catenoid is a solution of Eq.\ (\ref{EL:kappa}) for vanishing $\mu$, $\mathrm{p}$ and $c_s$, since it is a minimal surface it cannot withstand a force, in stark contrast with the fact that a finite force is required to hold the membrane at the maximum constriction. Furthermore, we see in Figs.\ \ref{fig:1}(f) and \ref{fig:6}(f) that the mean curvature of the neck varies along it, so it cannot be part of constant mean curvature surface (in particular of a minimal surface). To resolve this apparent contradiction  we employ a perturbative analysis about the catenoid. To this end, we consider a catenoidal neck whose radius $R_0$ is much smaller than the radius of the vesicle $R_S$, so the scaled neck radius $r_0 = R_0/R_S \ll 1$ sets the length scale in this regime. Since in the neighborhood of the neck we have $\ell \ll 1$ ($l \ll R_S$), we consider the rescaled arclength
\begin{equation}
 \tell=\frac{\ell}{r_0}=\frac{l}{R_0}\,.
\end{equation}
Like for the sphere, we consider the deformation of just one half of the catenoid, say the one with $\tell > 0$. The rescaled arc length derivative is given by $d/d \tell = r_0 d/d\ell$. Likewise, since $r \ll 1$ ($R \ll R_S$) and $|z| \ll 1$ ($Z \ll R_S$) we consider the rescaled coordinates
\begin{equation}
\tilde{r}(\tilde{\ell})=\frac{r}{r_0}=\frac{R}{R_0}\,, \quad  \tilde{z}(\tilde{\ell})=\frac{z}{r_0}=\frac{Z}{R_0}\,.
\end{equation}
Also, the magnitudes of the curvatures are large
\begin{equation}
c_\parallel = -c_\perp = \frac{1}{r_0 \left(1 + \tilde{\ell}^2 \right)} \gg 1\,,
\end{equation}
so we consider the rescaled curvatures $\tilde{c}_\parallel = r_0 c_\parallel$ and $\tilde{c}_\perp = r_0 c_\perp$, as well as the rescaled mean curvature difference $\tilde{\kappa} = r_0 \kappa=\tilde{c}_\perp+\tilde{c}_\parallel-\tilde{c}_s$, where $\tilde{c}_s=r_0 c_s$.
We expand the coordinates employing $r_0$ as the small parameter
\begin{subequations} \label{Eq:catneckexpansion}
\begin{eqnarray}
\tilde{r}(\tell) &=& \tilde{r}_{(0)}(\tell) + \tilde{r}_{(1)}(\tell) + \dots \,, \\
\tilde{z}(\tell) &=& \tilde{z}_{(0)}(\tell) + z_{(1)}(\tell) + \dots \,, \\
\Theta (\tell) &=& \Theta_{(0)}(\tell) + \Theta_{(1)}(\tell)  \dots \,,  \\
\tilde{\kappa}(\tell) &=& \tilde{\kappa}_{(0)}(\tell) + \tilde{\kappa}_{(1)}(\tell) + \dots \;,
\end{eqnarray}
\end{subequations}
where in the mean curvature difference $\tilde{\kappa}_{(0)}(\tell)= \tilde{c}_{\perp(0)} + \tilde{c}_{\parallel(0)}$ and $\tilde{\kappa}_{(1)}(\tell)= \tilde{c}_{\perp(1)} + \tilde{c}_{\parallel(1)} -\tilde{c_s}$.
The deformation should keep the equator fixed, so the following BCs are to be fulfilled by the first order perturbations
\begin{equation} \label{Cat:BCs}
\tilde{r}_{(1)}(0)=0 \,, \quad \tilde{z}_{(1)}(0)=0 \,, \quad \Theta_{(1)}(0)=0 \,.
\end{equation}
The zeroth-order terms are given by the functions corresponding to the catenoid\footnote{$\tilde{\kappa}_{(0)}=0$ because the mean curvature of the catenoid is zero and it is only a solution of the EL for null spontaneous curvature.}
\begin{subequations} \label{catsol}
\begin{eqnarray}
\tilde{r}_{(0)}(\tell) &=& \sqrt{1+\tell^2}\,, \\
\tilde{z}_{(0)}(\tell) &=& \arcsinh \tell \,, \\
\Theta_{(0)}(\tell) &=& \arccot \tell \,,  \\
\tilde{\kappa}_{(0)}(\tell) &=& 0 \,.
\end{eqnarray}
\end{subequations}
The derivatives of the radial and height coordinates are
\begin{subequations}
\begin{eqnarray}
\frac{d \tilde{r}_{(0)}}{d \tell} &=& \cos \Theta_{(0)} = \frac{\tell}{\tilde{r}_{(0)}}\,, \\
\frac{d\tilde{z}_{(0)}}{d \tell} &=& \sin \Theta_{(0)} = \frac{1}{\tilde{r}_{(0)}}\,.
\end{eqnarray}
\end{subequations}
The curvatures are given by
\begin{subequations}
\begin{eqnarray}
\tilde{c}_{\parallel(0)} &=& \frac{\sin \Theta_{(0)}}{\tilde{r}_{(0)}} = \frac{1}{\tilde{r}_{(0)}^2} \,,
\\
\tilde{c}_{\perp(0)} &=& \frac{d \Theta_{(0)}}{d \tell} = - \frac{1}{\tilde{r}_{(0)}^2} \,.
\end{eqnarray}
\end{subequations}
Thus, the scaled mean curvature $\tilde{k}_{(0)}=\tilde{c}_{\parallel (0)} + \tilde{c}_{\perp (0)}$ vanishes as it is a minimal surface. Moreover, the derivative of the meridian curvature is proportional to the scaled arclength
\begin{equation}
 \frac{d \tilde{c}_{\perp(0)}}{d \tell} = \frac{d^2\Theta_{(0)}}{d \tell^2}= \frac{2 \tell}{\tilde{r}_{(0)}^4}\,,
\end{equation}
so it vanishes at the equator where $\tell = 0$. In consequence the force vanishes as well
\begin{equation}
f_0 = - 4 \pi r_0 \ddot{\Theta}_{(0)}(0) = - \frac{4 \pi}{r_0} \frac{d^2{\Theta}_{(0)}}{d\tell^2}(0)= 0 \,,
\end{equation}
consistent with the fact that the catenoid is a minimal surface, which satisfies the source-free EL
equation when the parameters vanish, i.e.
\begin{equation} \label{eq:catenoidcts}
\mu_{(0)}=0\,, \quad \mathrm{p}_{(0)}=0 \,, \quad  c_{s(0)}=0 \,.
\end{equation}
The system of differential equations, given in Eqs.\ (\ref{def:RpZp}) and (\ref{Eqs:systThetaKD}), read
\begin{subequations} \label{Eq:catnecksystEqs}
 \begin{eqnarray}
 \frac{d \tilde{r}}{d\tell} &=&\cos \Theta \,, \\
 \frac{d \tilde{z}}{d \tell} &=&\sin \Theta \,,\\
 \frac{d \Theta}{d \tell} &=& \tilde{\kappa} - \frac{\sin \Theta}{\tilde{r}} + \tilde{c}_s\,, \\
 \cos \Theta \frac{d \tilde{\kappa}}{d \tell} &=& - \sin \Theta \left( \tilde{\kappa} \left( \frac{\tilde{\kappa}}{2} - \frac{\sin \Theta}{\tilde{r}} + r_0 c_s\right) - \tilde{\mu} \right) \nonumber \\
 &&-\frac{\tilde{\mathrm{p}}}{2} \tilde{r} \,,
  \end{eqnarray}
\end{subequations}
where
\begin{equation}
 \tilde{\mu}= r_0^2 \mu \,, \quad \tilde{\mathrm{p}}=r_0^3 \mathrm{p} \,.
\end{equation}
Since the scaled spontaneous curvature and the Lagrange multipliers are proportional to powers of $r_0$, they enter only at orders of the expansion higher than one.
\\
Substituting Eqs.\ (\ref{Eq:catneckexpansion}) we get that at first order the system of Eqs.\ (\ref{Eq:catnecksystEqs}) is given by
\begin{subequations}
\begin{eqnarray}
\frac{d \tilde{r}_{(1)}}{d \tell} + \frac{\Theta_{(1)}}{\tilde{r}_{(0)}}  &=&0\,, \label{Cat:Eqdr}\\
\frac{d\tilde{z}_{(1)}}{d\tell} -\frac{\tell}{\tilde{r}_{(0)}} \Theta_{(1)}&=&0 \,, \label{Cat:Eqdz}\\
\frac{d\Theta_{(1)}}{d \tell} + \frac{\tell}{\tilde{r}_{(0)}^2} \Theta_{(1)} -\frac{1}{\tilde{r}_{(0)}^3} \tilde{r}_{(1)} -\tilde{\kappa}_{(1)} &=&0 \,, \label{Cat:EqdTheta} \\
\tell \, \frac{d\tilde{\kappa}_{(1)}}{d \tell} - \frac{1}{\tilde{r}_{(0)}^2} \tilde{\kappa}_{(1)} &=&0 \label{Cat:Eqdkappa}\,.
\end{eqnarray}
\end{subequations}
The solution to Eq.\ (\ref{Cat:Eqdkappa}) is\footnote{In general, the first-order correction to the mean curvature difference is $\kappa_{(1)} = C_{(1)} \sqrt{r^2-1}/r$ where $C_{(1)}$ is a constant \cite{Fourcade1994, Yang2017}. So by substituting the catenoidal radial coordinate $r=\sqrt{1+\ell^2}$ we obtain the same result.
}
\begin{equation} \label{Cat:k1sol}
\tilde{\kappa}_{(1)}(\tell) = \tilde{\kappa}_{(1)A} \frac{\tell}{\tilde{r}_{(0)}} \,,
\end{equation}
where $\tilde{\kappa}_{(1)A}$ is a constant of integration. This first-order correction to the mean curvature difference vanishes at the equator and tends asymptotically to $\tilde{\kappa}_{(1)A}$ for $\tell \gg 1$. The asymptotic value $\tilde{\kappa}_{(1)A}$ ought to match the value of the mean curvature difference (scaled by $r_0$), $\kappa_\rma= k_\rma-c_s$, where $k_\rma= 2 \sqrt{2}$ is the mean curvature of the quasispherical vesicles, so we get
\begin{equation}
\tilde{\kappa}_{(1)A}  = r_0 \kappa_\rma = r_0\left(2 \sqrt{2} -c_s\right)\,.
\end{equation}
The first-order correction to the mean curvature is $\tilde{k}_{(1)}=\tilde{\kappa}_{(1)}+\tilde{c}_s$.
Since the lowest order mean curvature vanishes, the mean curvature of the neck is given by $k^a_{mc}=\tilde{k}_{(1)}/r_0$, which reads
\begin{equation}
 k^a_{mc}=  2\sqrt{2} \frac{\tell}{\tilde{r}_{(0)}} + c_s \left(1-\frac{\tell}{\tilde{r}_{(0)}} \right) \,.
\end{equation}
Therefore, the mean curvature of the neck starts at the value of $c_s$ at the equator, feature known as kissing condition \cite{Fourcade1994, Julicher1996, Yang2017}, and far from it reaches the value $k_\rma$. Thus, up to first order, we see that only for $c_s = 2 \sqrt{2}$ the neck is represented by a constant mean curvature surface. The mean curvature of the neck of the numerical solutions and this first-order correction are plotted in Fig.\ \ref{fig:16}, showing a good agreement.
\vskip0pc \noindent
Using Eq.\ (\ref{Cat:Eqdr}) in Eq.\ (\ref{Cat:EqdTheta}) to replace $\Theta_{(1)}$ in favor of $\tilde{r}_{(1)}$, substituting Eq.\ (\ref{Cat:k1sol}) and integrating with the BCs (\ref{Cat:BCs})
we determine $\Theta_{(1)}$ in terms of $r_{(1)}$
\begin{equation} \label{Cat:Theta1r1}
 \Theta_{(1)} =\frac{\tell}{\tilde{r}_{(0)}} \tilde{r}_{(1)} + \tilde{\kappa}_{(1)A}
(\tilde{r}_{(0)}-1) \,.
\end{equation}
Substituting this result in Eq.\ (\ref{Cat:Eqdr}) we obtain an ODE for $\tilde{r}_{(1)}$, whose solution also determines $\Theta_{(1)}$ on account of Eq.\ (\ref{Cat:Theta1r1}), and which in turn allows for the integration of (\ref{Cat:Eqdz}) to determine $\tilde{z}_{(1)}$. These solutions, with the BCs (\ref{Cat:BCs}), are given by
\begin{subequations}
\begin{eqnarray}
\frac{\tilde{r}_{(1)}}{\tilde{\kappa}_{(1)A}}&=& \left(\frac{1}{\tilde{r}_{(0)}} -
\frac{1}{2}\right) \tell - \frac{\Upsilon(\tell)}{2 \tilde{r}_{(0)}}  \,,\\
\frac{\tilde{z}_{(1)}}{\tilde{\kappa}_{(1)A}}&=& \frac{1}{\tilde{r}_{(0)}}\left(1+\frac{\tell
\Upsilon(\tell)}{2} \right) \nonumber \\
&+& \frac{1}{4} \left(\tell^2 - \Upsilon(\tell)^2 \right)-1\,,  \\
2 \frac{\Theta_{(1)}}{\tilde{\kappa}_{(1)A}} &=& \tilde{r}_{(0)}+\frac{1}{\tilde{r}_{(0)}}
-\frac{1}{\tilde{r}_{(0)}^2} \left(\tell \Upsilon(\tell)+2\right)\,, \\
\Upsilon(\tell)&:=&\arctanh \left(\frac{\tell}{\tilde{r}_{(0)}}\right)\,.
 \end{eqnarray}
\end{subequations}
These first-order deformations make the generating curve of the neck slightly wider in comparison with the catenoid, as shown in Fig.\ \ref{fig:17}.
\vskip0pc \noindent
The derivatives of $\Theta_{(1)}$ are
\begin{subequations}
\begin{eqnarray}
 \frac{1}{\tilde{\kappa}_{(1)A}}\frac{d\Theta_{(1)}}{d \tell}&=&
\left(\frac{1}{2}-\frac{1}{\tilde{r}^2_{(0)}}\right) \left(\frac{\tell}{\tilde{r}_{(0)}} +
\frac{\Upsilon(\tell)}{\tilde{r}^2_{(0)}} \right) \nonumber \\
 &+& \frac{2 \tell}{\tilde{r}^4_{(0)}} \,, \\
 \frac{1}{\tilde{\kappa}_{(1)A}} \frac{d^2\Theta_{(1)}}{d \tell^2}&=&
\left(1-\frac{2}{\tilde{r}_{(0)}}\right)\left(\frac{3}{\tilde{r}^3_{(0)}}
-\frac{4}{\tilde{r}^5_{(0)}}\right) \nonumber \\
 &+&\left(\frac{4}{\tilde{r}^2_{(0)}}-1\right) \frac{\tell}{\tilde{r}^4_{(0)}}\Upsilon(s)
 \end{eqnarray}
\end{subequations}
Although the first-order derivative vanishes at the equator ($\tell=0$), the second-order derivative is nonvanishing
\begin{equation}
  \frac{d^2\Theta_{(1)}}{d\tell^2}(0)=\tilde{\kappa}_{(1)A}=r_0 \left(2 \sqrt{2}-c_s\right)\,.
 \end{equation}
Since $r_0^2 \ddot{\Theta}_{(1)}= d^2\Theta_{(1)}/d\tell^2$, we have
 \begin{equation}
  \ddot{\Theta}_{(1)}(0)=\frac{2\sqrt{2}-c_s}{r_0}\,.
\end{equation}
Thus, even though the scaled equatorial force density $\phi_0=-2\ddot{\Theta}_0$ diverges as $r_0 \rightarrow 0 $, the total scaled force converges to a finite value
\begin{equation} \label{Eq:f0maxconst}
\frac{f^a_{0mc}}{4 \pi} = - r_0 \ddot{\Theta}_0 = -(2\sqrt{2} -c_s) \,.
\end{equation}
This result agrees with the numerical results of the scaled force, see Table \ref{TableII} and the insets of Fig.\ (\ref{fig:15}). In particular $f_0$ is negative (constrictive) for $c_s< 2 \sqrt{2}$, it vanishes for $c_s=2\sqrt{2}$ and becomes positive for $c_s> 2 \sqrt{2}$, so a stretching force is required to maintain the vesicle in equilibrium.
\\
The value of $\mu$ at the maximum constriction is determined from the scaling relation, Eq.\ (\ref{scalrelsimp}), relating the $\ddot{\Theta}_0$ and the global quantities of the surface, which in this case reads
\begin{equation} \label{Eq:escrelafix}
  -r_0^2 \ddot{\Theta}_{0} = 2 \mu + c_s (c_s - 2 m)\,.
\end{equation}
In the limit $r_0 \rightarrow 0$, $r_0^2 \ddot{\Theta}_0 \rightarrow 0$ and $m \rightarrow  k_\rma a_\rma = \sqrt{2}$, so we have that $\mu$ is quadratic in $c_s$
\begin{equation}
\mu^a_{mc}(c_s) = c_s \left(\sqrt{2} -\frac{c_s}{2}\right)\,.
\end{equation}
Thus, $\mu^a_{mc}$ is positive for $0<c_s < 2\sqrt{2}$ and it vanishes for $c_s=0,2\sqrt{2}$, which agrees with the numerical values presented in Table \ref{TableII}.

\subsection{Equatorial stretching: disc limit}

\begin{figure}[hbt]
\centering
  \includegraphics[width=0.475 \textwidth]{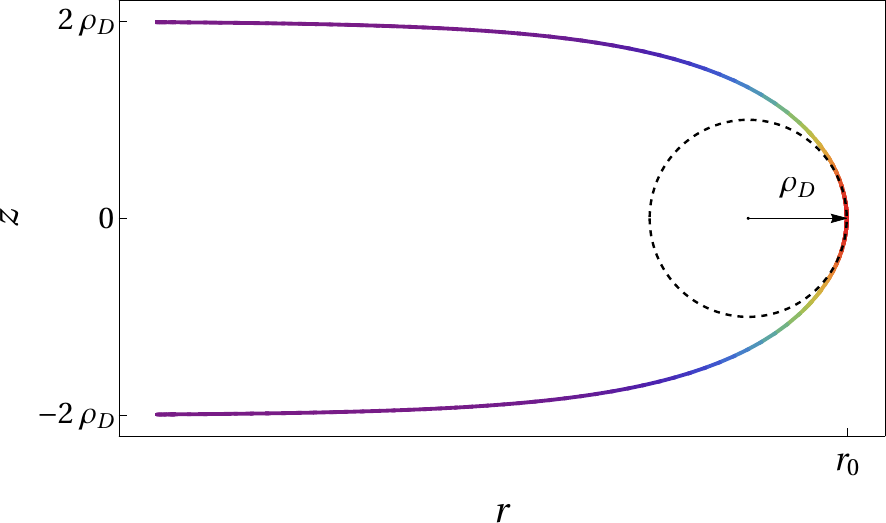}
  \vskip1pc
  \includegraphics[scale=0.55]{fig6g}
  \caption{Profile curve of the vesicle near the equatorial region in the disk limit. The osculating circle of radius $\rho_D$ is shown with a dashed line.  The scaled bending energy density is color coded.}
  \label{fig:18}
\end{figure}

In the limit of maximum stretching the curvature is concentrated near the equator, where the meridian curvature is large, $c_\perp \gg 1$. Since in this region $r_0 \approx 1$ and $|\sin \Theta| \leq 1$, we have $ c_\parallel \approx 1$, so $c_\perp \gg c_\parallel$. We also assume that the spontaneous curvature is finite,  $c_\perp \gg c_s$. Thus, to analyze this limit of a disclike surface, we can consider as a small parameter the radius of curvature given by the inverse of the value of $c_\perp$ at the equator,
\begin{equation}
\rho_D = \frac{1}{c_{\perp 0}} = \frac{1}{\dot{\Theta}_0} \ll 1 \,.
\end{equation}
Taking into account these considerations, the first integral, Eq.\ (\ref{EL:kappa}), simplifies to\footnote{Even if we considered the term proportional to $\mathrm{p}$ it would also be small compared with the other quantities, because in this limit $V \rightarrow 0$ so the pressure difference vanishes, $P \rightarrow 0$.}
\begin{equation} \label{eq:ELdisc}
 \cos \Theta \dot{c}_{\perp} + \sin \Theta \left( \frac{c_{\perp}^2}{2} - \mu \right) =0 \,.
\end{equation}
Evaluating at the equator we determine the constant
\begin{equation} \label{eq:mudisclim}
\mu =\frac{c_{\perp 0}^2}{2} = \frac{1}{2 \rho^2_D}\,.
\end{equation}
Thus, $\mu$ diverges with $1/\rho_D^2$ in the limit $r_0 \rightarrow \sqrt{2}$. Integrating Eq.\ (\ref{eq:ELdisc}) and imposing the asymptotic boundary condition that far from the equator the surface becomes planar, $\Theta \rightarrow \pi$ and $\dot{\Theta} \rightarrow 0$, yields
\begin{equation} \label{dThetadisc}
c_{\perp} =\dot{\Theta} =\frac{\sqrt{2}}{\rho_D} \, \cos \frac{\Theta}{2} \,. 
\end{equation}
Integrating once more, with the boundary condition $\Theta(0)=\pi/2$, we get
\begin{equation}
\sin \frac{\Theta}{2} = \tanh \Xi(\ell) \,, \quad  \Xi(\ell):= \frac{\ell - \ell_{0}}{\sqrt{2} \rho_D}\,,
\end{equation}
where $\ell_{0} = - \sqrt{2} \rho_D \mathrm{arctanh} (1/\sqrt{2})$. Since $\rho_D \ll 1$, $\Theta \approx \pi$ away from the equator, so in this planar region $c_\parallel \approx 0$. From this result, we can express the first order differential for the coordinates as
\begin{subequations}
\begin{eqnarray}
\dot{r} &=& \sech^2 \Xi(\ell) - \tanh^2 \Xi(\ell)  \,, \\
\dot{z} &=& 2 \sech \Xi(\ell) \tanh \Xi(\ell) \,.
\end{eqnarray}
\end{subequations}
Integrating these equations with the boundary conditions $r(0)=r_0$ and $z(0)=0$, we get
\begin{subequations}
\begin{eqnarray}
r &=& r_{0} -\ell+ 2 \rho_D \left( \sqrt{2} \tanh \Xi(\ell) - 1 \right)  \,, \\
z &=& 2 \rho_D \left(1 -\sqrt{2} \sech \Xi(\ell) \right) \,.
\end{eqnarray}
\end{subequations}
The scaled arclength at which the pole is reached ($r=0$) is $\ell_p \approx r_0 +2 (\sqrt{2}-1) \rho_D$. Calculating the scaled area, we have
\begin{eqnarray}
a = \int \displaylimits_0^{\ell_p} d \ell r &=& -\frac{\ell_p^2}{2} + \ell_p \left( r_0 - 2 \rho_D \right) \nonumber \\
&&+ 4 \rho _D^2 \ln \left[ \frac{1}{\sqrt{2}} \cosh \Xi(\ell_p)\right].\quad
\end{eqnarray}
Using the approximation $\cosh x \approx e^x/2$ for $x \gg 1$ and imposing the condition of unit scaled area, to first order in $\rho_D$ we have a quadratic equation for $r_0$
\begin{equation}
 \frac{r_0^2}{2} +2 \left(\sqrt{2}-1\right) r_0 \rho _D - 1 =0\,.
\end{equation}
Solving for $r_0$ and taking the positive solution we get
\begin{equation}
r_0 = \sqrt{2} - 2 (\sqrt{2}-1) \rho_D \,.
\end{equation}
Therefore, the small parameter is proportional to the difference of the maximum radius and the equatorial radius, $\rho_D \propto \sqrt{2} - r_0$, whereas the total arclength is $\ell_p \approx \sqrt{2}$.
In the limit $r_0 \to \sqrt{2}$, the geometry consists of two flat discs glued together along their perimeters (see Fig.\ \ref{fig:7}(e)). The profile of the equatorial region, which agrees very well with the numerical solution, is plotted in Fig.\ (\ref{fig:18}).
\vskip0pc \noindent
Differentiating Eq.\ (\ref{dThetadisc}) and using it to substitute $\dot{\Theta}$ in favor of $\Theta$, we obtain the second order derivative
\begin{equation}
 \ddot{\Theta} = - \frac{1}{2 \rho_D ^2} \sin \Theta\,.
\end{equation}
Evaluating at the equator we have that the scaled force is given by $f_0/(4 \pi) = - r_0 \ddot{\Theta}_0 = r_0/(\sqrt{2}\rho_D^2)$. Thus, it diverges as $1/\rho_D^2$ in the limit $r_0 \rightarrow \sqrt{2}$.
\\
Since in this limit $c_\perp \gg c_\parallel$, the scaled mean curvature is given only by the curvature along the meridian,
\begin{equation}
\kappa \approx c_\perp = \frac{\sqrt{2}}{\rho_D} \sech \Xi(\ell),
\end{equation}
so the total scaled bending energy reads
\begin{eqnarray}
 h_B &\approx& \int \displaylimits_{0}^{\sqrt{2}} d \ell c_\perp^2 r  \nonumber \\
 &=& 4 \Bigg[ \ln \left(\cosh \Xi(\ell) \right) -\sech^2\Xi(\ell) \nonumber \\
 &&- \left( \frac{\left(\sqrt{2} \ell-2\right)}{2 \rho_D} +2 \right) \tanh \Xi(\ell)  \Bigg]\Bigg|_{0}^{\sqrt{2}} .
\end{eqnarray}
Taking into account that $\rho_D \ll 1$, we get that the scaled total energy diverges with the inverse of $\rho_D$,
\begin{equation}
 h_B \approx \frac{2 (2-\sqrt{2})}{\rho_D} \,.
 \end{equation}
 These results agree with the behavior of the total force and enegy of the vesicles in the limit of maximum stretching obtained from the numerical results.
 
\section{Equilibrium shapes with fixed volume} \label{Sec:Vconst}

\begin{figure*}
\begin{center}
\begin{tabular}{ccccc}
  $\vcenter{\hbox{\includegraphics[scale=0.4]{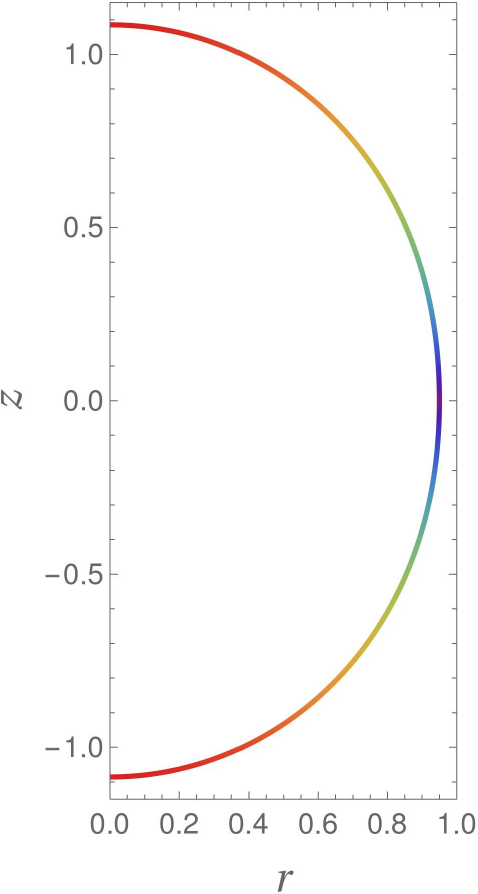}}}$ &
  $\vcenter{\hbox{\includegraphics[scale=0.4]{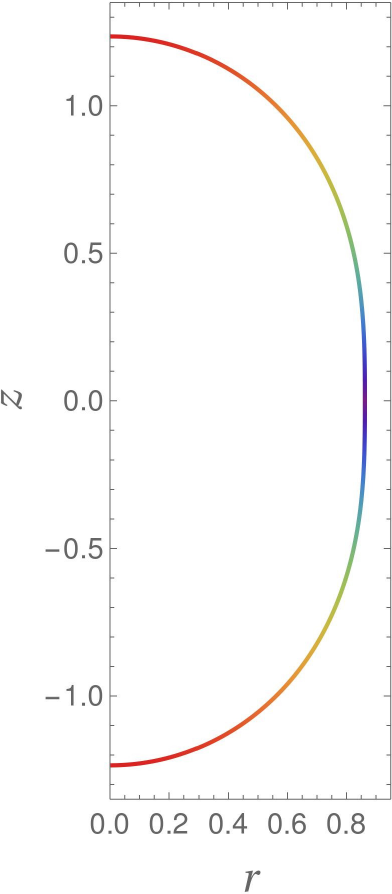}}}$ &
    $\vcenter{\hbox{\includegraphics[scale=0.4]{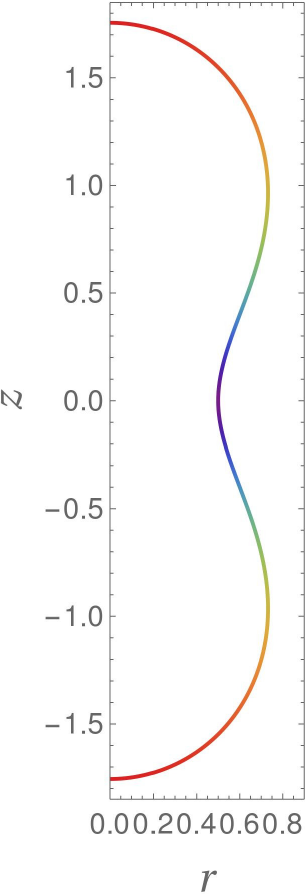}}}$ &
  $\vcenter{\hbox{\includegraphics[scale=0.4]{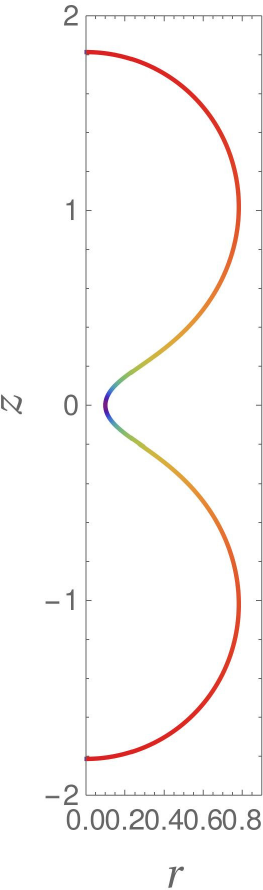}}}$ &
  $\vcenter{\hbox{\includegraphics[scale=0.4]{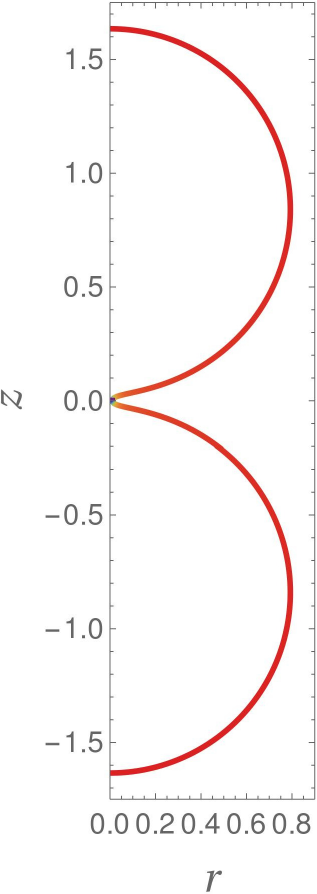}}}$ \\
{\small (a) $r_0=0.95$} & {\small b) $r_0=0.8625$} & {\small (c) $r_0=0.5$} & {\small d) $r_0=10^{-1}$} & {\small e) $r_0=10^{-2}$} 
  \end{tabular}
 \end{center}
  \includegraphics[scale=0.6]{fig6g}
\caption{(a)-(e) Sequence of vesicle configurations with $v=1$ and $c_s=0$ as $r_0$ is decreased. The scaled bending energy density is color coded.}
\label{fig:19}
\end{figure*}

\begin{figure*}
\begin{center}
\begin{tabular}{cccc}
$\vcenter{\hbox{\includegraphics[scale=0.35]{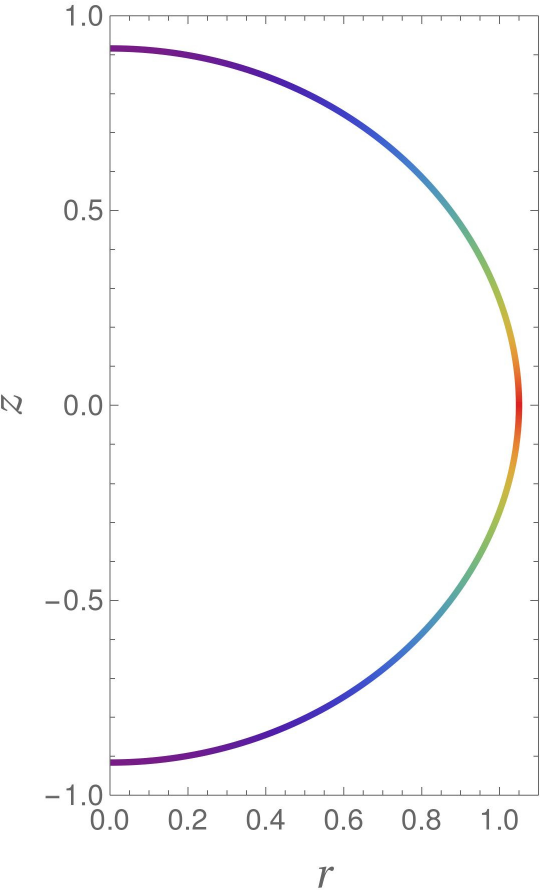}}}$ &
$\vcenter{\hbox{\includegraphics[scale=0.35]{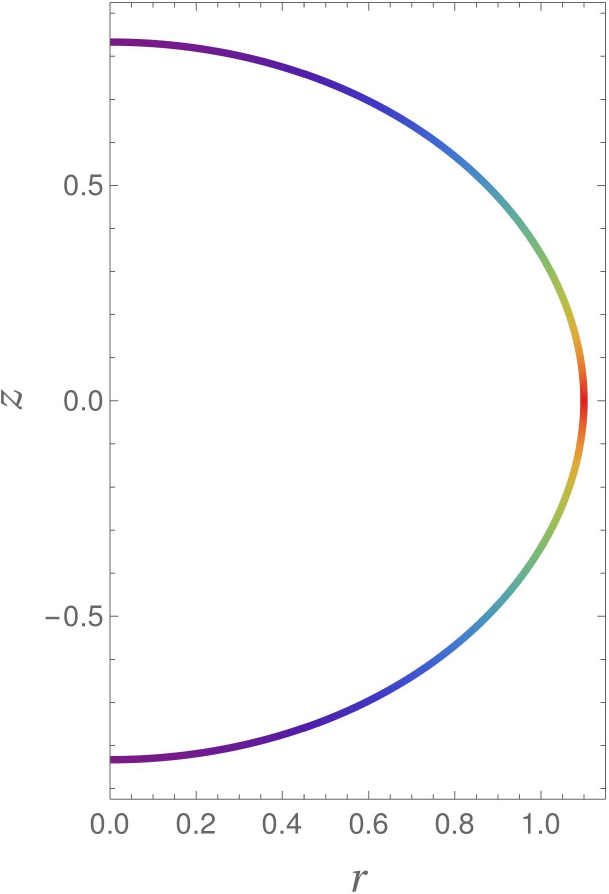}}}$ &
$\vcenter{\hbox{\includegraphics[scale=0.35]{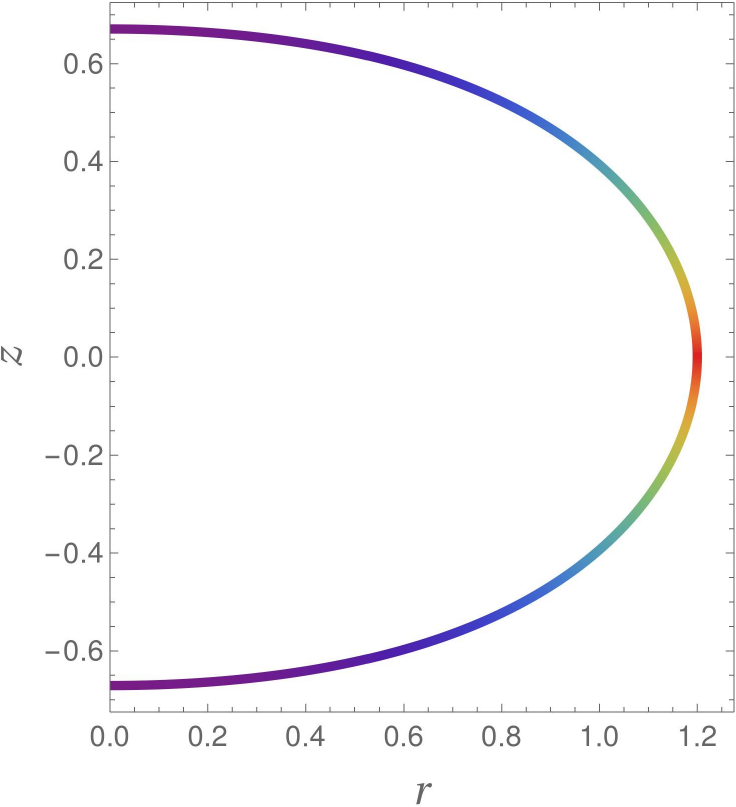}}}$ &
$\vcenter{\hbox{\includegraphics[scale=0.35]{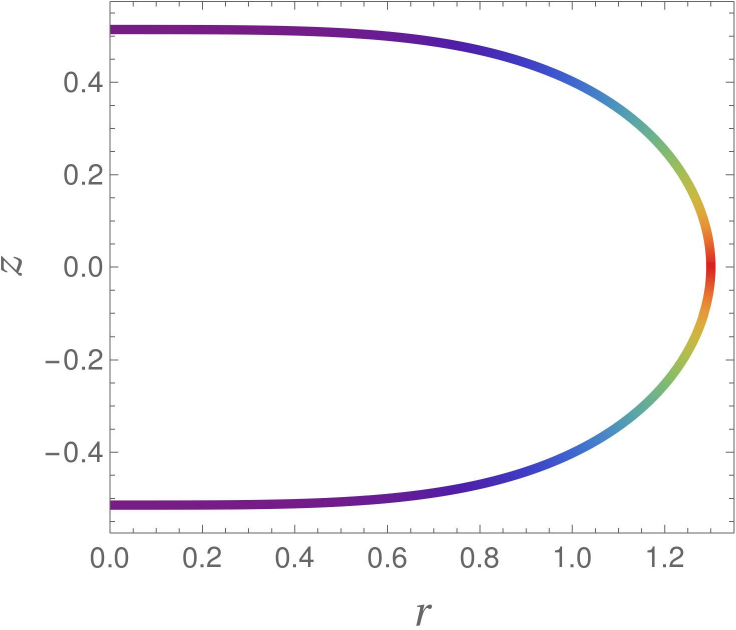}}}$ \\
{\small (a) $r_0=1.05$} & {\small (b) $r_0=1.1$} & {\small (c) $r_0=1.2$} & {\small (d) $r_0=1.3$}
\end{tabular}
\end{center}
\begin{center}
\begin{tabular}{ccc}
$\vcenter{\hbox{\includegraphics[scale=0.35]{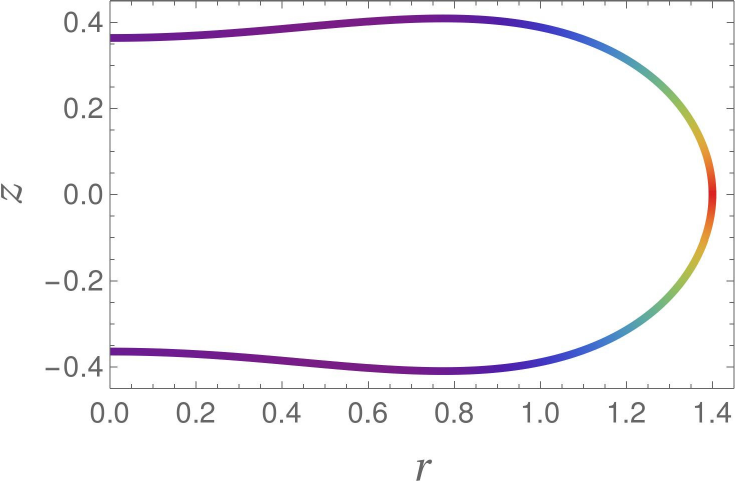}}}$ &
$\vcenter{\hbox{\includegraphics[scale=0.35]{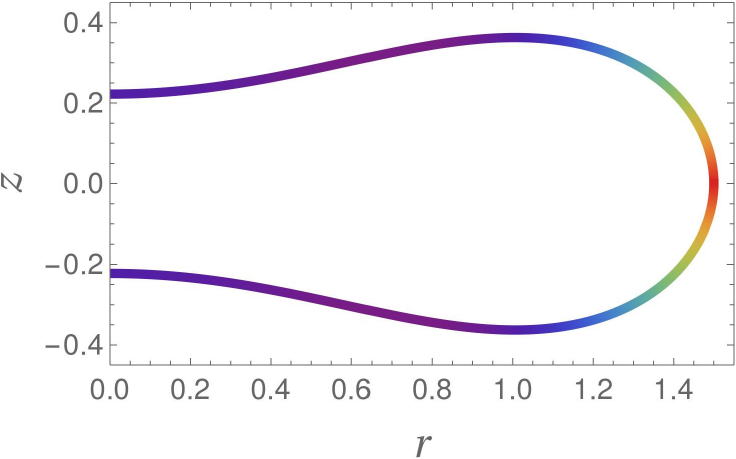}}}$ &
$\vcenter{\hbox{\includegraphics[scale=0.35]{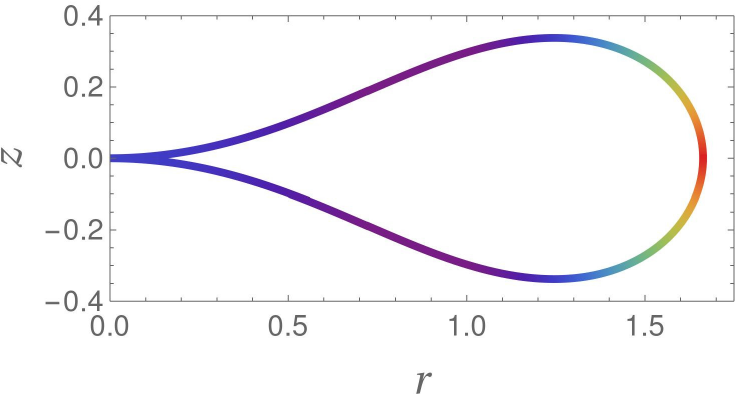}}}$\\
  {\small (e) $r_0=1.4$} & {\small (f) $r_0=1.5$} & {\small (g) $r_0=1.63$}
  \end{tabular}
\end{center}
  \includegraphics[scale=0.6]{fig6g}
\caption{(a)-(e) Sequence of vesicle configurations with $v=1$ and $c_s=0$ as $r_0$ is increased. The scaled bending energy density is color coded.}
\label{fig:20}
\end{figure*}

\begin{figure*}
\begin{center}
\begin{tabular}{ccccccc}
$\vcenter{\hbox{\includegraphics[scale=0.475]{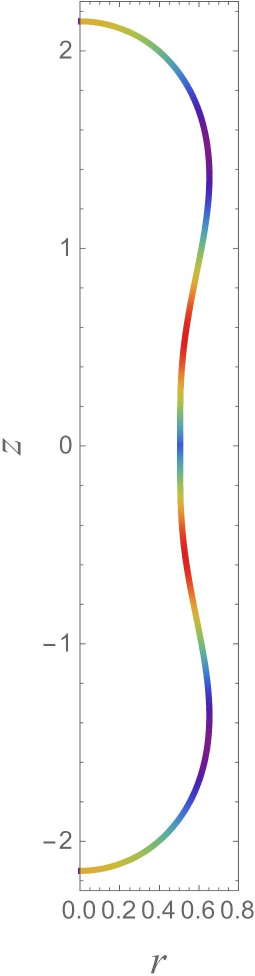}}}$ &
$\vcenter{\hbox{\includegraphics[scale=0.45]{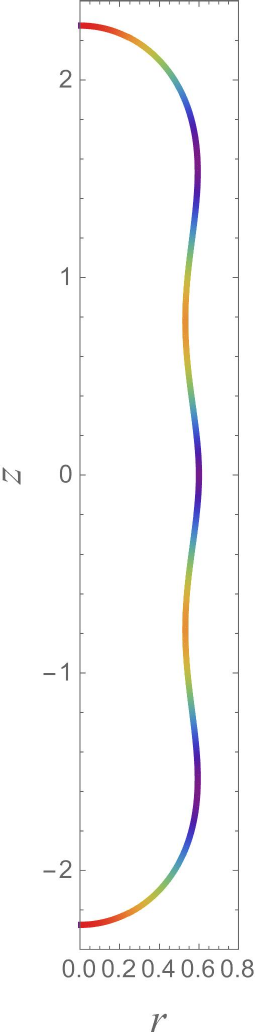}}}$ &
$\vcenter{\hbox{\includegraphics[scale=0.45]{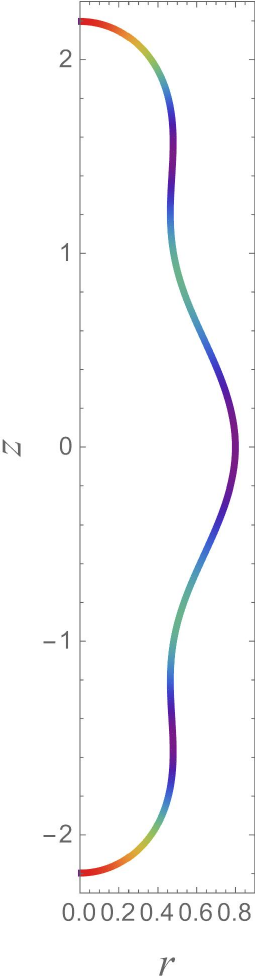}}}$ &
$\vcenter{\hbox{\includegraphics[scale=0.45]{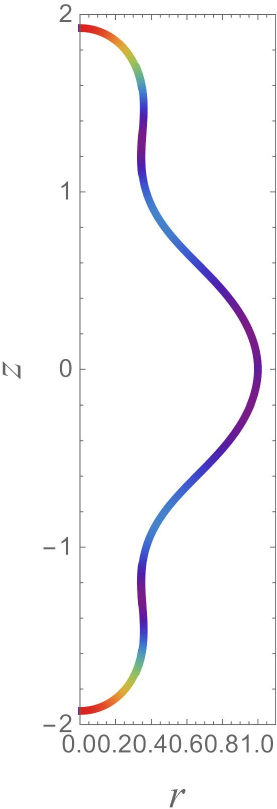}}}$ &
$\vcenter{\hbox{\includegraphics[scale=0.45]{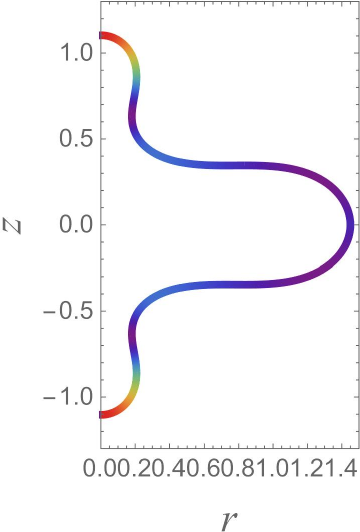}}}$ &
$\vcenter{\hbox{\includegraphics[scale=0.5]{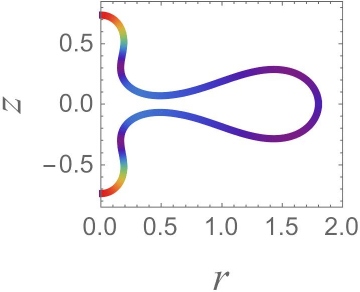}}}$ &
$\vcenter{\hbox{\includegraphics[scale=0.51]{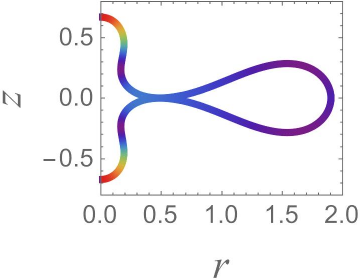}}}$ \\
{\small (a) $\medbullet$ $r^v_{01}=0.5$} & {\small b) $r_0=0.6$} & {\small (c) $r_0=0.8$} & {\small (d) $r_0=1$}
& {\small (e) $r_0=1.45$} & {\small (f) $r_0=1.8$} & {\small (g) $\blacksquare$ $r^v_{02}=1.9$}
  \end{tabular}
\end{center}
  \includegraphics[scale=0.6]{fig6g}
\caption{(a)-(e) Sequence of vesicle configurations with $v=1$ and $c_s=2^{4/3}$ of the first branch ($\medbullet$ $r^v_{01} < r_0 < r^v_{02}$ $\blacksquare$) as $r_0$ is increased. The scaled bending energy density is color coded.}
\label{fig:21}
\end{figure*}

\begin{figure*}
\begin{center}
\begin{tabular}{ccccccc}
$\vcenter{\hbox{\includegraphics[scale=0.45]{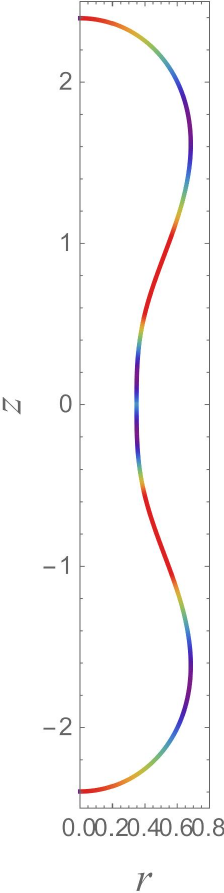}}}$ &
$\vcenter{\hbox{\includegraphics[scale=0.45]{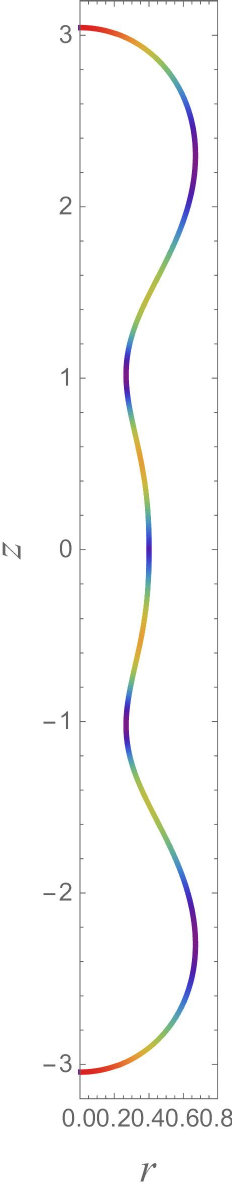}}}$ &
$\vcenter{\hbox{\includegraphics[scale=0.46]{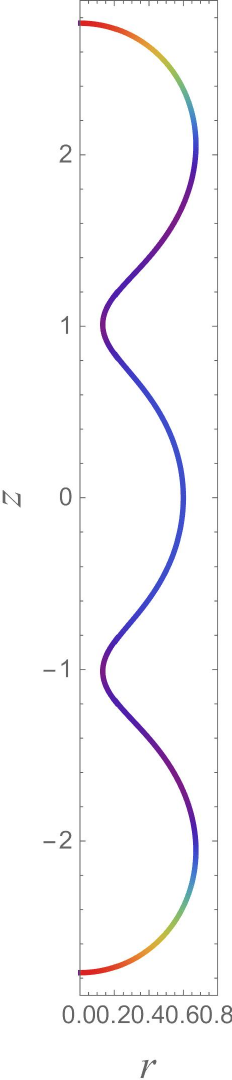}}}$ &
$\vcenter{\hbox{\includegraphics[scale=0.5]{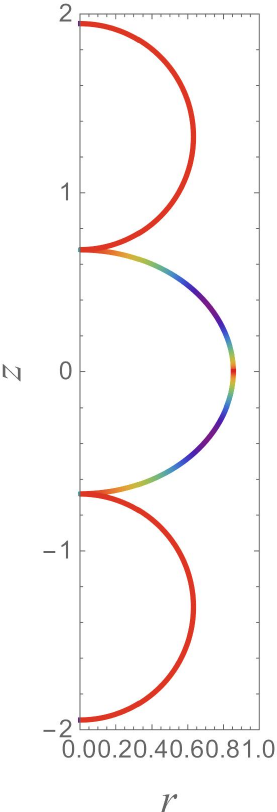}}}$ &
$\vcenter{\hbox{\includegraphics[scale=0.475]{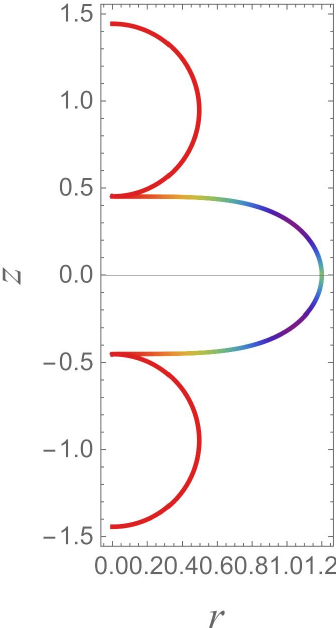}}}$ & 
$\vcenter{\hbox{\includegraphics[scale=0.5]{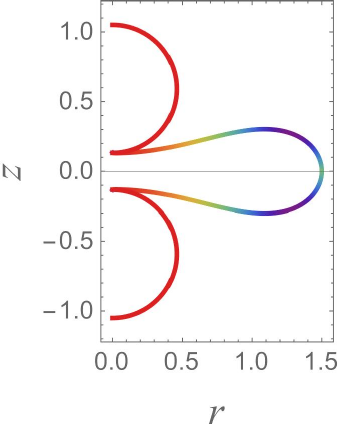}}}$ &
$\vcenter{\hbox{\includegraphics[scale=0.5]{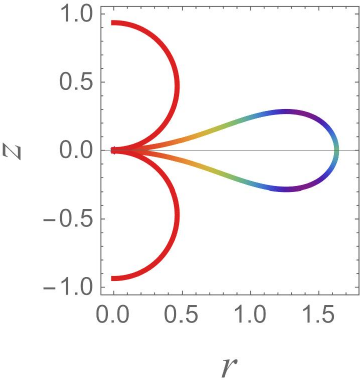}}}$ \\ \\
{\small (a) $\blacktriangle$ $r^v_{03}=0.35$} & {\small b) $r_0=0.4$} & {\small (c) $r_0=0.6$} & {\small (d) $r_0=0.857$} & {\small (e) $r_0=1.2$} & {\small (f) $r_0=1.5$} & {\small (g) $\Diamondblack$ $r^v_{04}=1.63$}
  \end{tabular}
\end{center}
  \includegraphics[scale=0.6]{fig6g}
\caption{(a)-(e) Sequence of vesicle configurations with $v=1$ and $c_s=2^{4/3}$ of the second branch ($\blacktriangle$ $r^v_{03} < r_0 < r^v_{04}$ $\Diamondblack$) as $r_0$ is increased. The scaled bending energy density is color coded.}
\label{fig:22}
\end{figure*}

\begin{figure}[htb]
 \centering
\includegraphics[width=0.475\textwidth]{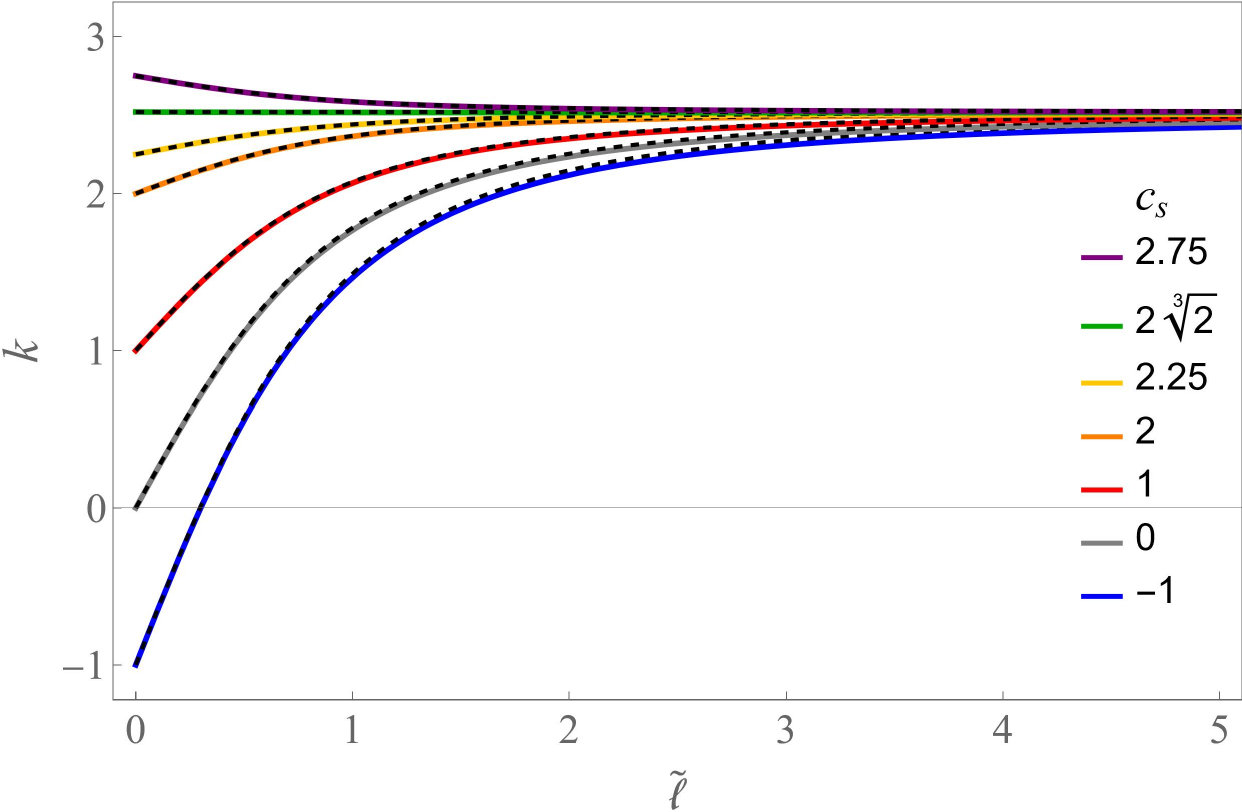}
\caption{Mean curvature along the neck as a function of the scaled arclength of configurations with $v=1$.  At the equator its value is given by $c_s$ and away from the equator its value tends to the mean curvature of the quasispherical membrane, $k_v=2^{4/3}$. The solid lines correspond to the numerical data and the dashed lines to scaled first order correction $\tilde{k}_{(1)}/r_0$.
} \label{fig:23}
\end{figure}

\begin{figure}[htb]
 \includegraphics[width=0.475\textwidth]{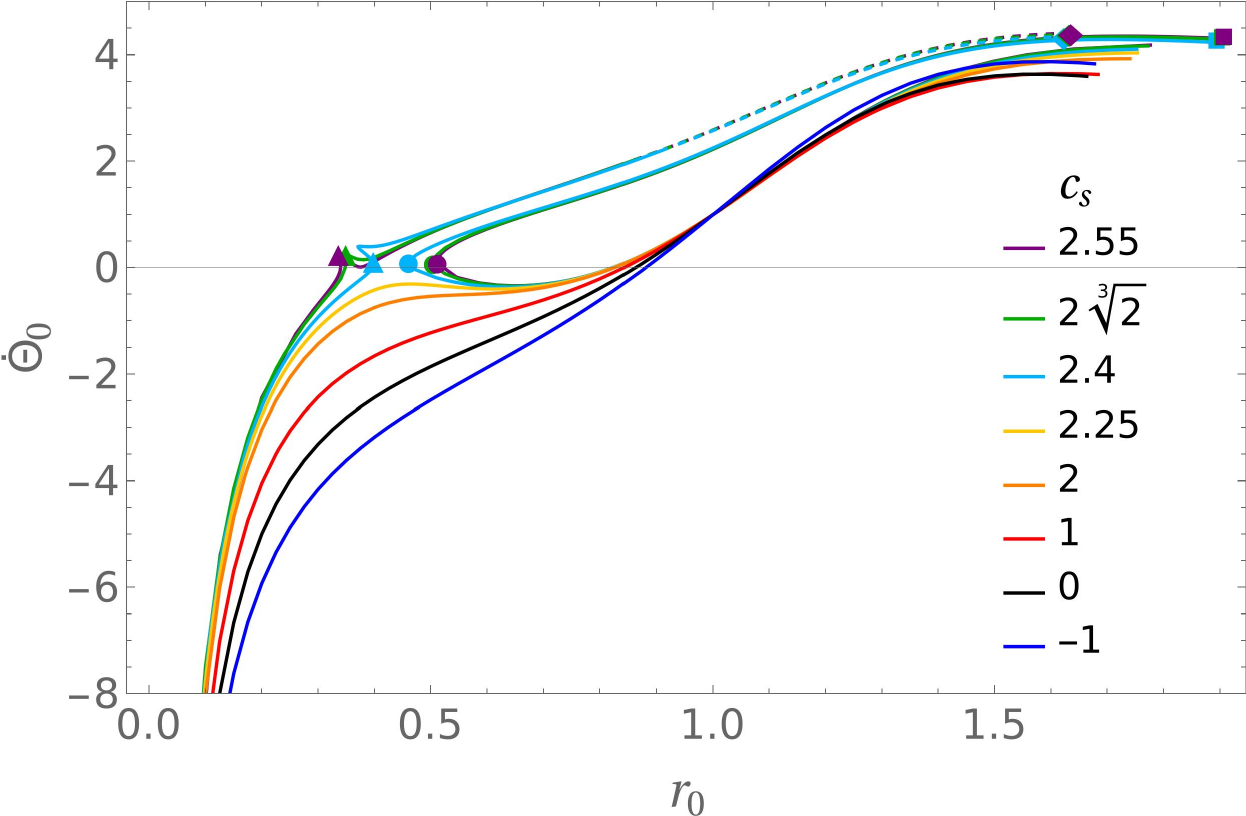}
\includegraphics[width=0.475\textwidth]{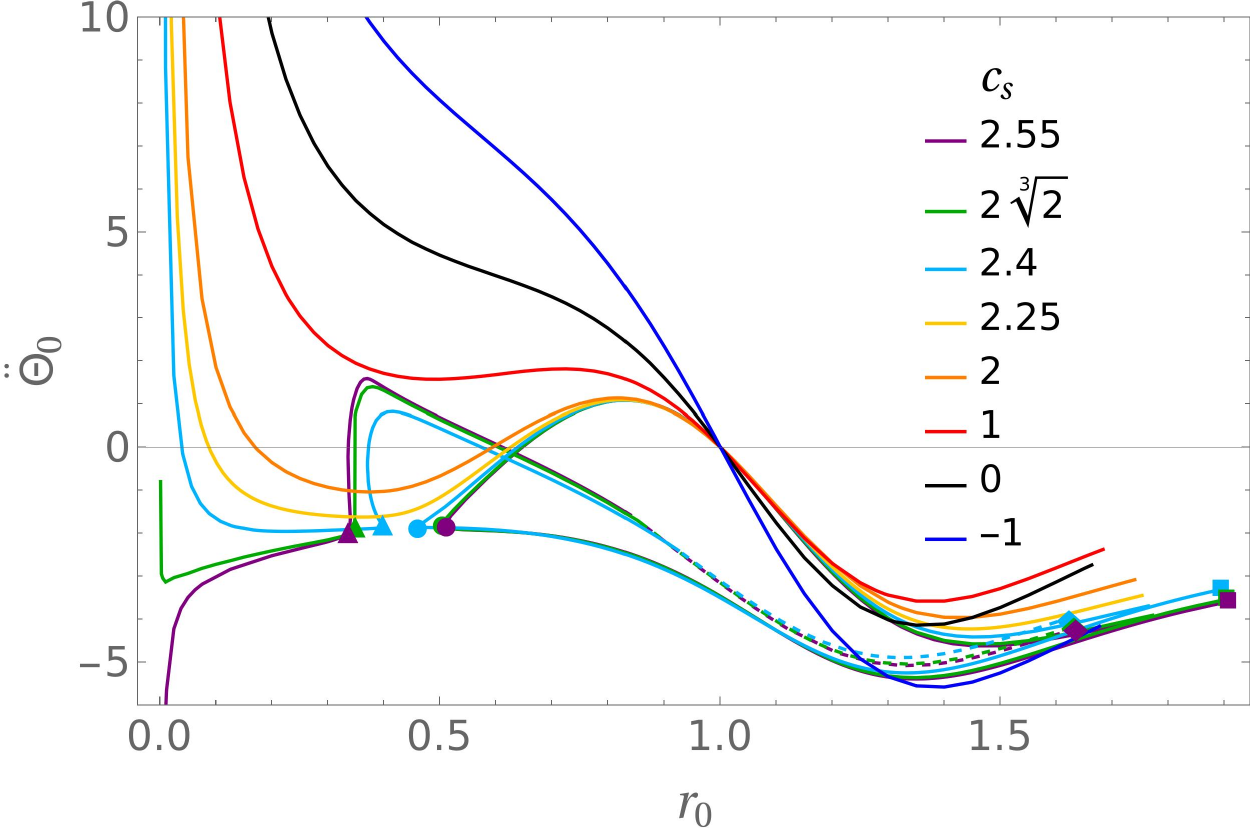}
\caption{Dependence of $\dot{\Theta}_0$ and $\ddot{\Theta}_0$ on $r_0$ for membranes with $v=1$. The first and second branches are demarcated with the symbols $(\medbullet,\blacksquare)$ and $(\blacktriangle,\Diamondblack)$, respectively. The dashed lines represent configurations with very small necks.}
\label{fig:24}
\end{figure}

\begin{figure}[htb]
\includegraphics[width=0.475\textwidth]{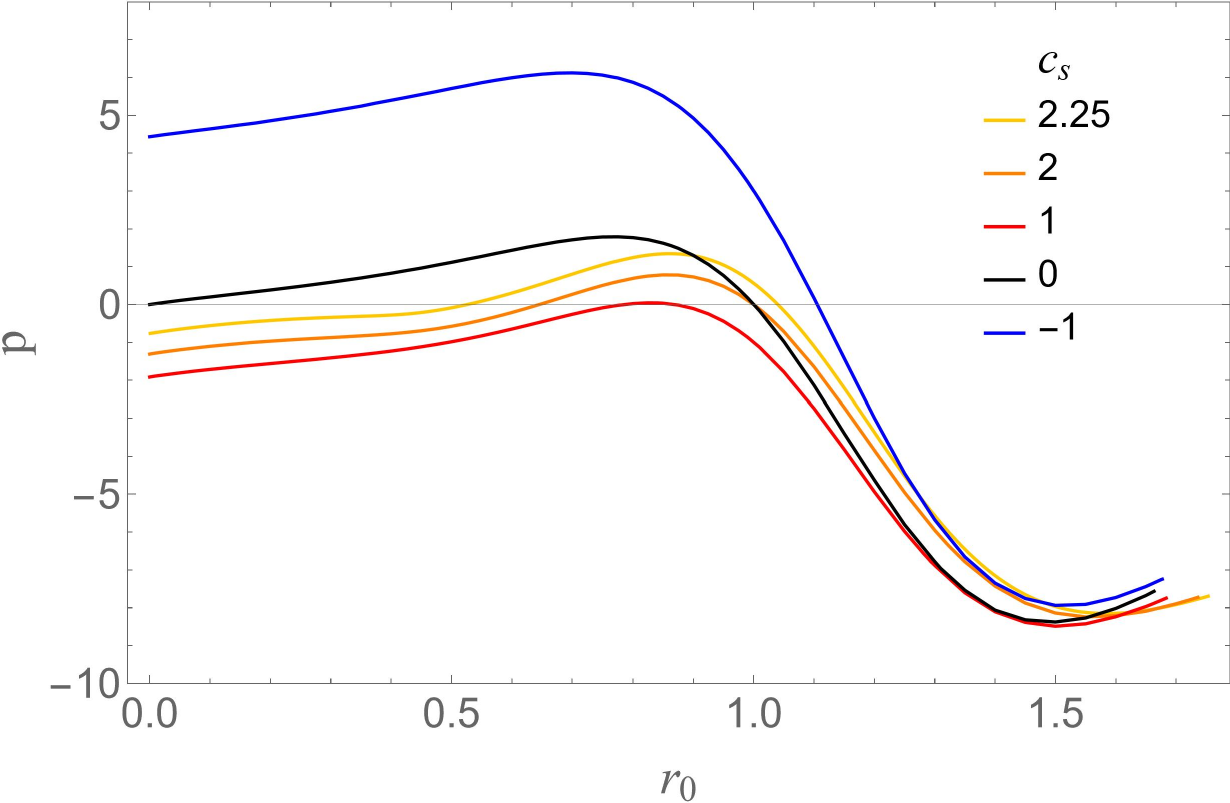}
\includegraphics[width=0.475\textwidth]{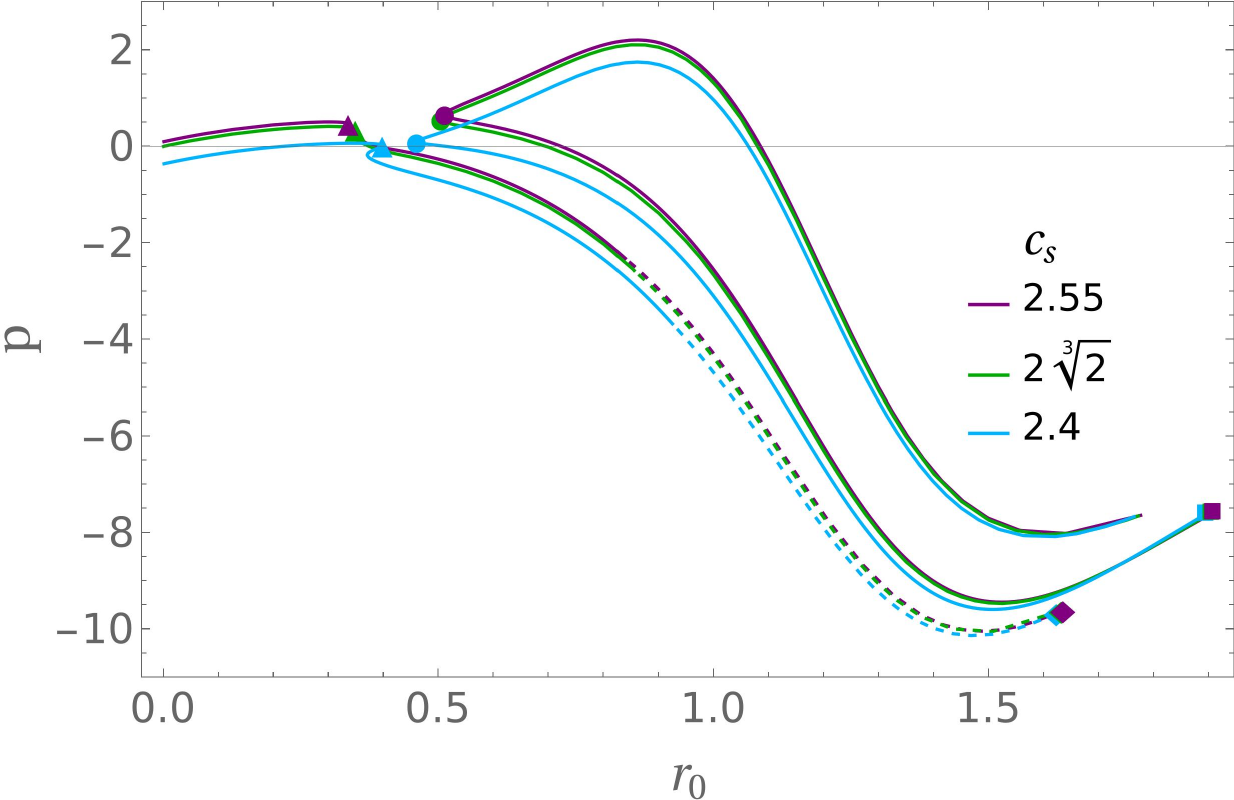}
\caption{Scaled pressure difference $\mrmp$ as a function of $r_0$ for membranes with $v=1$. The first and second branches are demarcated with the symbols $(\medbullet,\blacksquare)$ and $(\blacktriangle,\Diamondblack)$, respectively. The dashed lines represent configurations with very small necks.}
\label{fig:25}
\end{figure}

\begin{figure}[htb]
\includegraphics[width=0.475\textwidth]{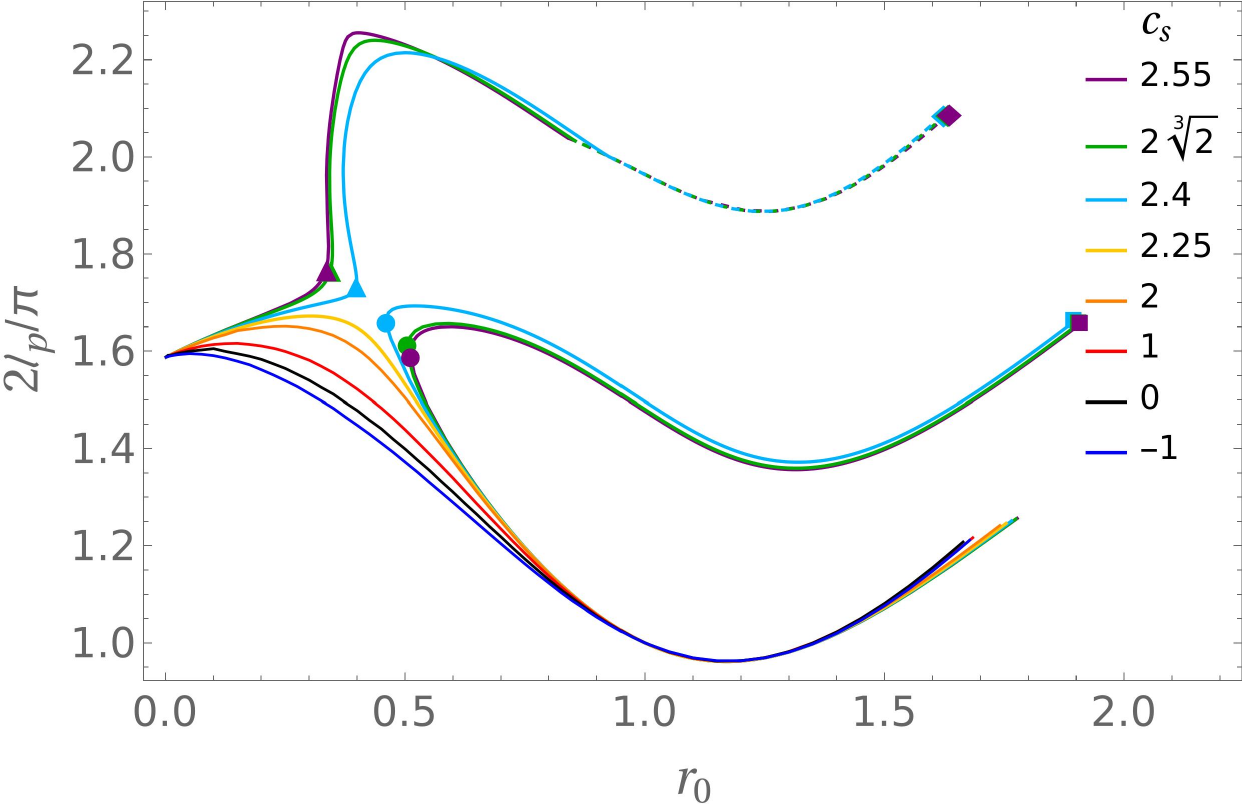}
\includegraphics[width=0.475\textwidth]{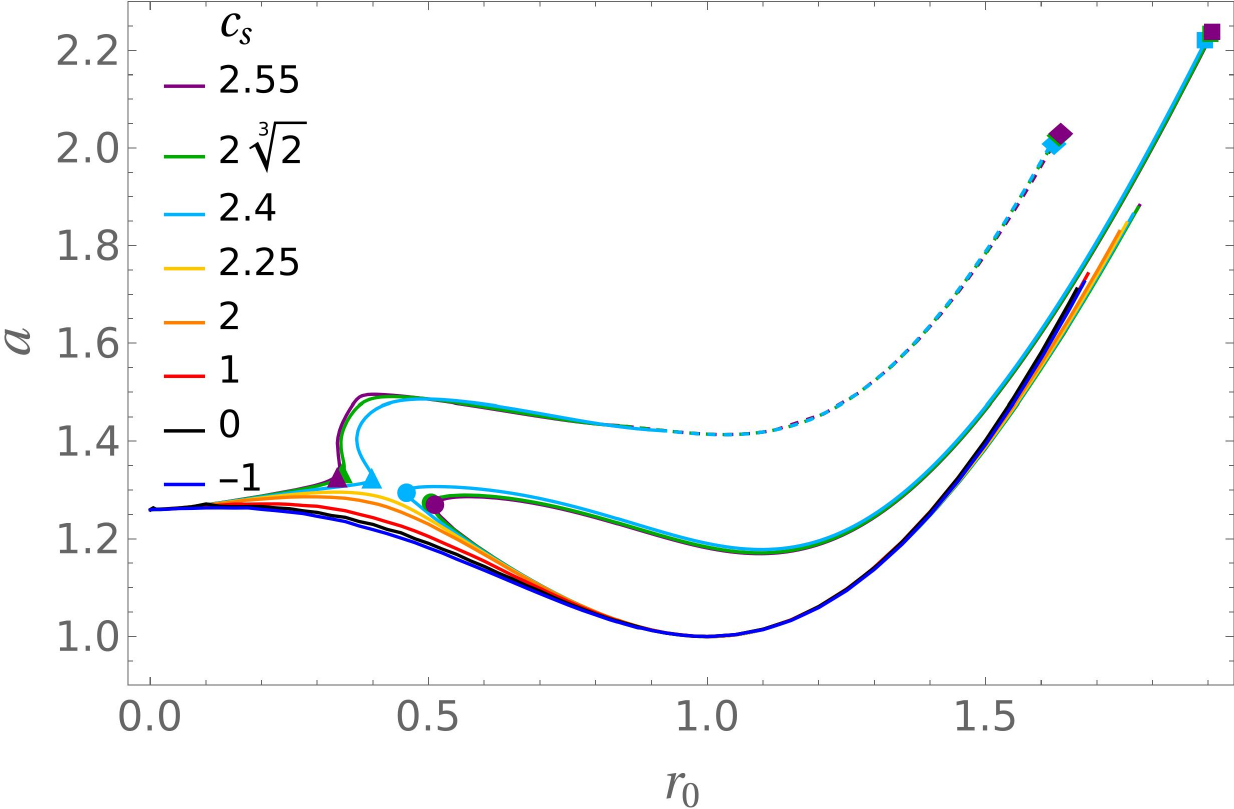}
\caption{Total length of the generating curve $2\ell_p$ and scaled area $a$ as functions of $r_0$ for membranes with $v=1$. The first and second branches are demarcated with the symbols $(\medbullet,\blacksquare)$ and $(\blacktriangle,\Diamondblack)$, respectively. The dashed lines represent configurations with very small necks.
} \label{fig:26}
\end{figure}

\begin{figure}[htb]
 \includegraphics[width=0.475\textwidth]{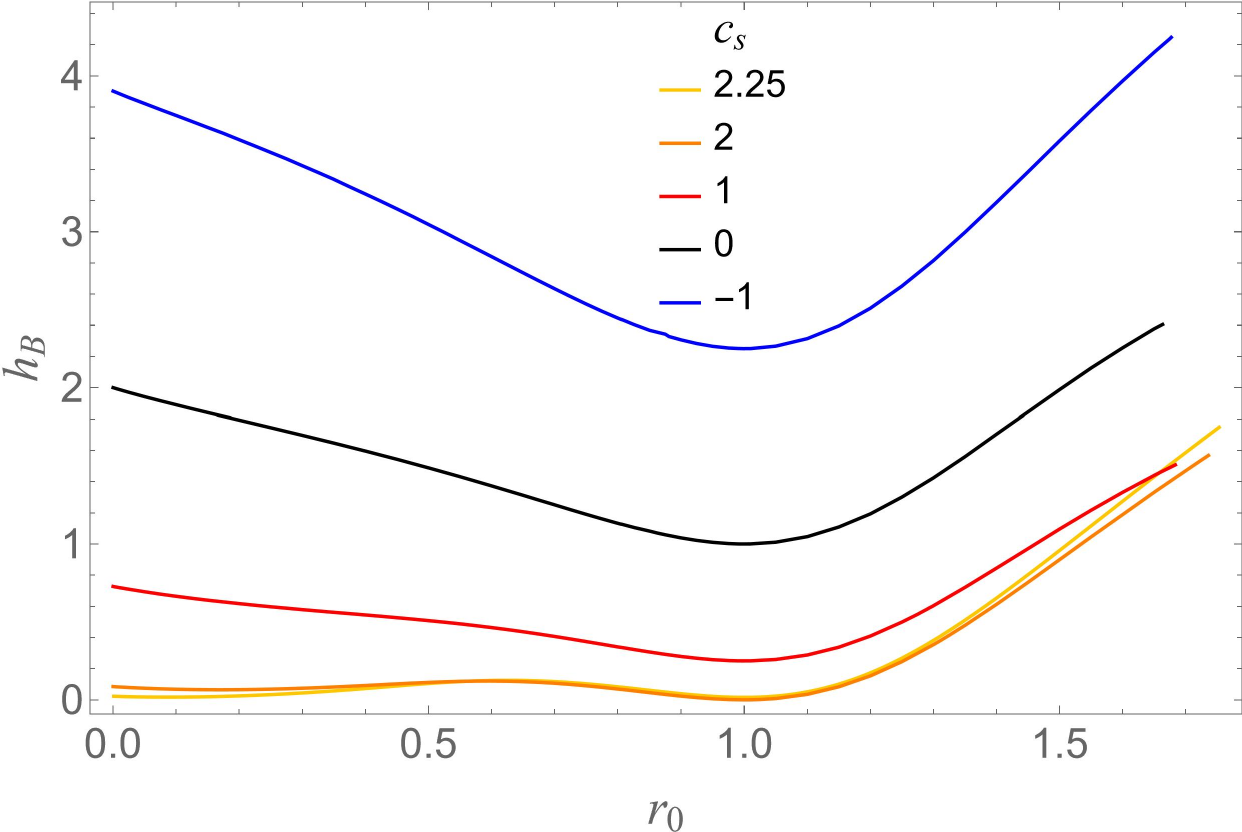}
\includegraphics[width=0.475\textwidth]{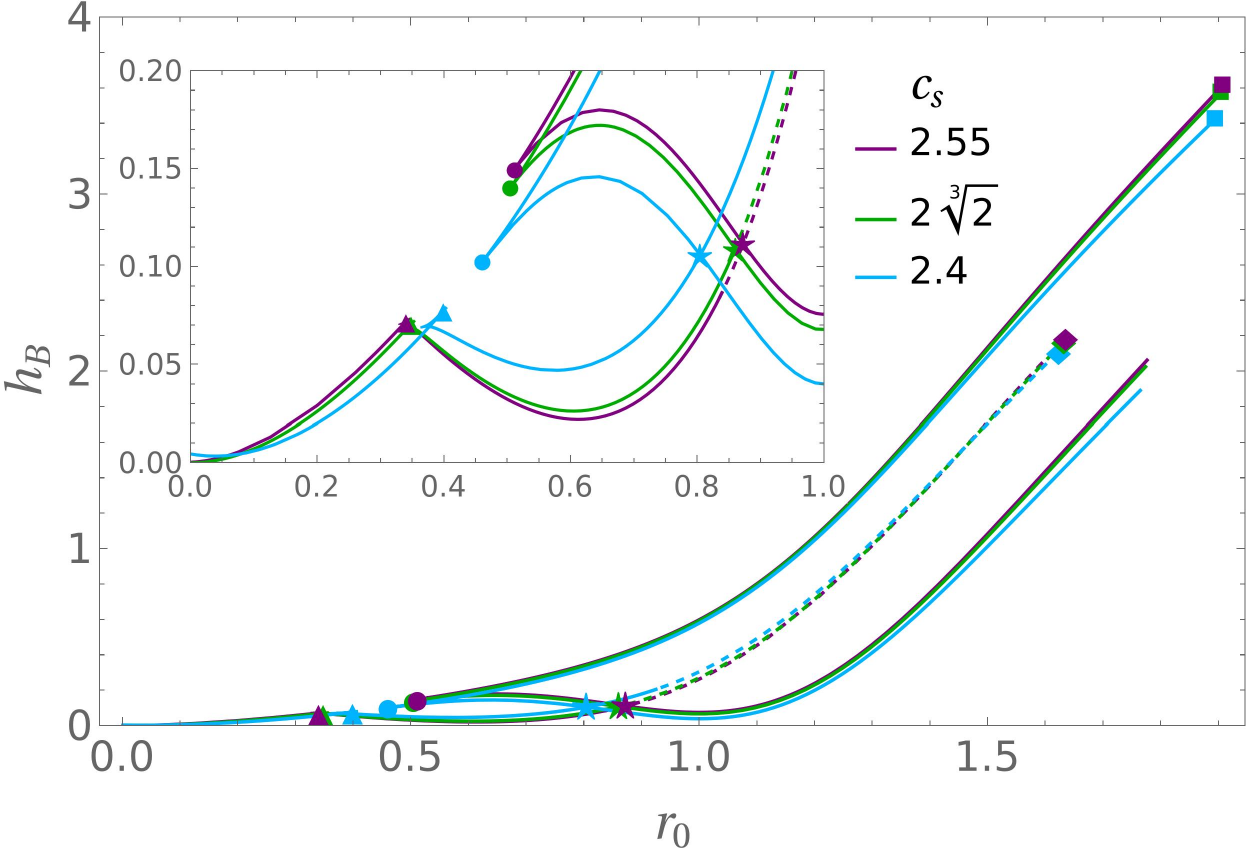}
\caption{Scaled total energy $h_B$ as a function of $r_0$ for membranes with $v=1$. The first and second branches are demarcated with the symbols $(\medbullet,\blacksquare)$ and $(\blacktriangle,\Diamondblack)$, respectively. The dashed lines represent configurations with very small necks.
} \label{fig:27}
\end{figure}

\begin{figure}[htb]
 \includegraphics[width=0.46\textwidth]{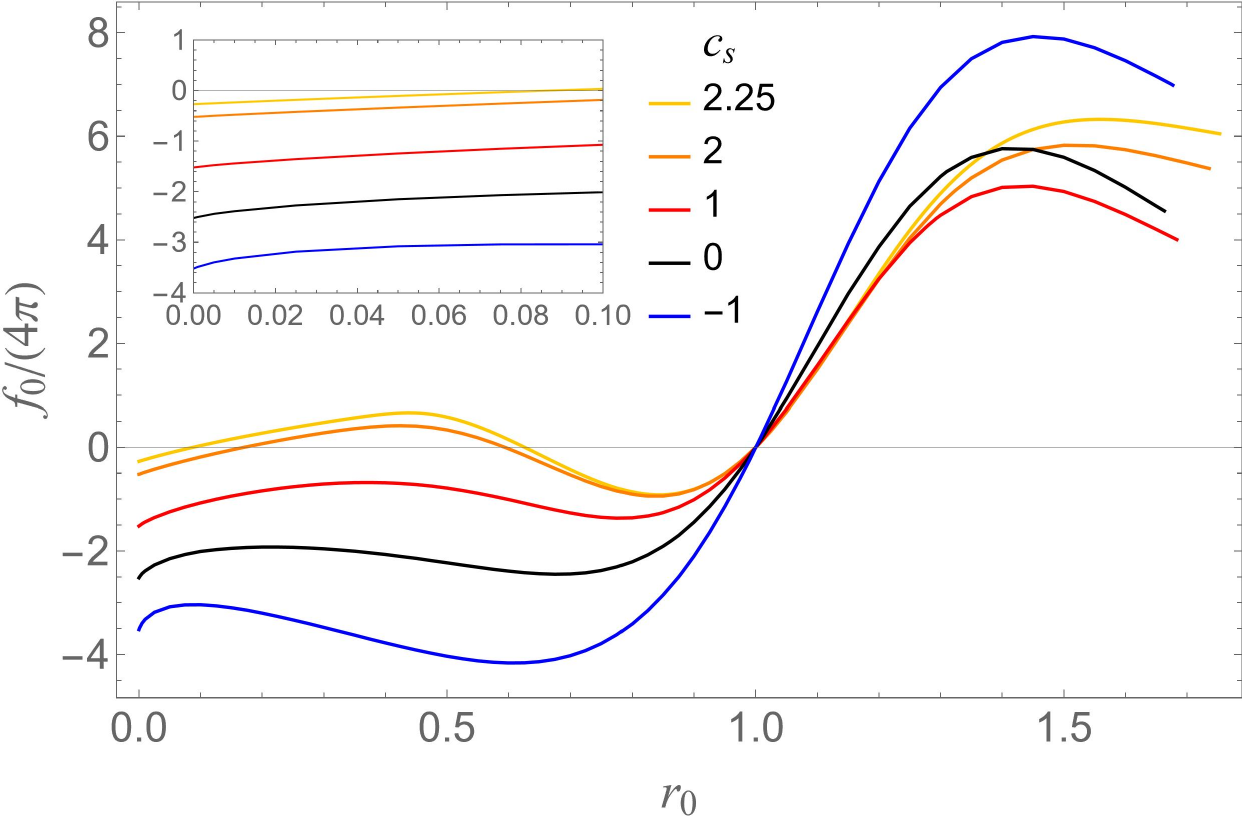}
\includegraphics[width=0.46\textwidth]{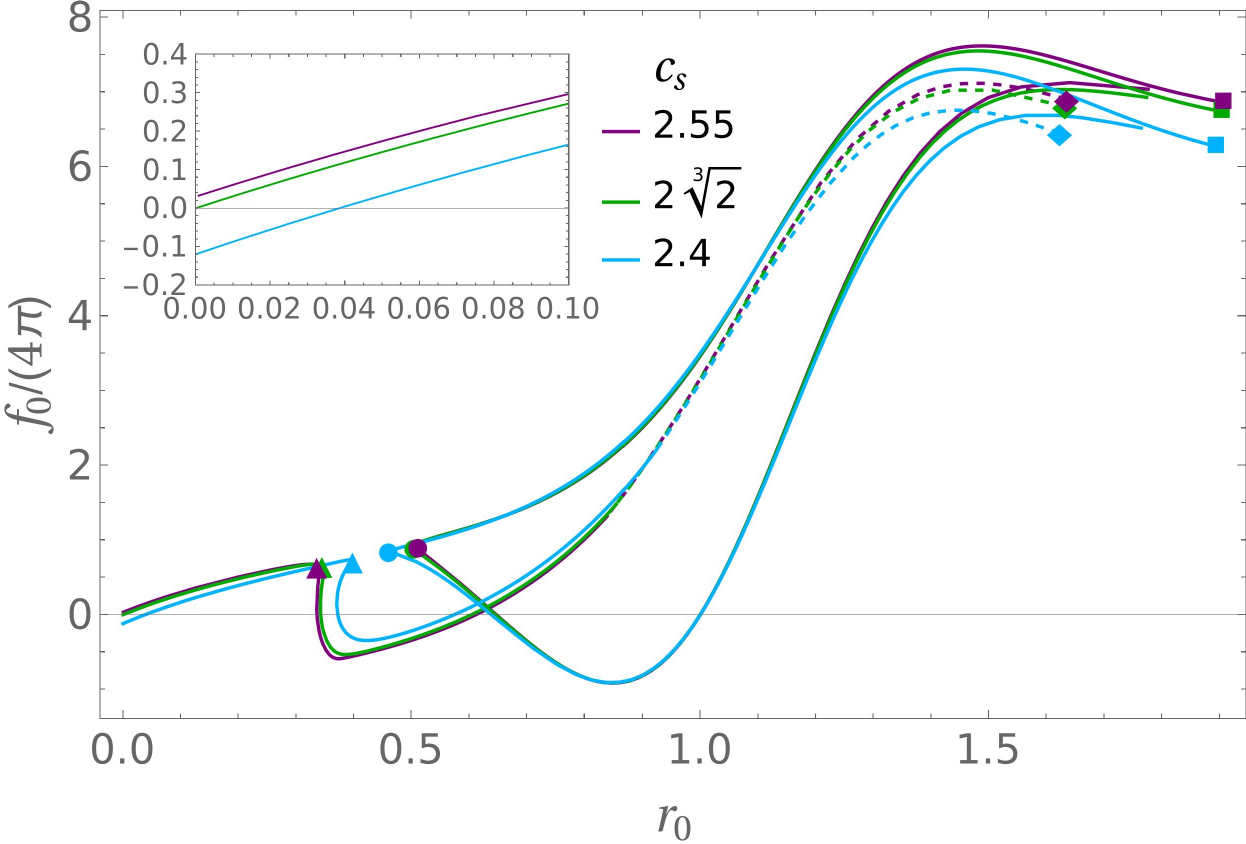}
\caption{Scaled total force $f_0$ as function of $r_0$ for membranes with $v=1$. The first and second branches are demarcated with the symbols $(\medbullet,\blacksquare)$ and $(\blacktriangle,\Diamondblack)$, respectively. The dashed lines represent configurations with very small necks.
} \label{fig:28}
\end{figure}

In this section we consider membranes whose enclosed volume is fixed, so $V=V_S$ or $v=1$. This case could represent the physical situation in which the control parameter is the temperature, whereas the difference pressure across the membrane is fixed. In this setting the membrane area varies freely on account that its thermal expansivity is large compared to that of the aqueous medium, so that the area changes faster than the volume as the temperature is varied \cite{Julicher1993, Julicher1994}. Thus, $\mu=0$ and Eq.\ (\ref{EL:kappa}) reduces to
\begin{equation}
\cos \Theta \, \dot{\kappa} + \sin \Theta  \kappa \left( \frac{\kappa}{2} -\frac{\sin \Theta}{r}+ c_s \right) + \frac{\mrmp}{2} r = 0 \,.
\end{equation}
Evaluating at a parallel with $\Theta_0$ and $r_0$ to determine $\mrmp$, we get
\begin{equation}
 \frac{\mrmp}{2}= -\frac{\cos \Theta_0}{r_0}  \dot{\kappa}_0 - \frac{\sin \Theta_0}{r_0} \kappa_0 \left( \frac{\kappa_0}{2} -\frac{\sin \Theta_0}{r_0} + c_s \right).
\end{equation}
Similar to the case of vesicles with fixed area, for values close to $c_s=0$ and up to $c_s=2.25$ we obtained a single sequence of configurations. Constricted and stretched vesicles with fixed volume are plotted in Figs.\ \ref{fig:19} and \ref{fig:20}, respectively. For values of the spontaneous curvature in the interval $2.25<c_s<2.55$ two bifurcations of the solutions arise. The branches originating from them do not meet, so the sequence of configurations becomes discontinuous, unlike the case of fixed area where the branches were connected. The configurations corresponding to both branches are plotted in Figs.\ \ref{fig:21} and \ref{fig:22}.
\vskip0pc \noindent
For values of the spontaneous curvature close to $c_s=0$ a constricted vesicle exhibits a behavior similar to the case of fixed area. As $r_0$ is reduced the vesicle initially adopts a prolate shape (Fig.\ \ref{fig:19}(a)); after the equatorial region becomes cylindrical, there is a transition to a dumbbell shape (Figs.\ \ref{fig:19}(b)-(d)), and in the limit of maximum constriction, $r_0\rightarrow 0$, the membrane consists of two quasispherical vesicles connected by an infinitesimal neck, (Figs. \ref{fig:19}(e)). However, in this case the radius of the quasispherical vesicles is $r_v= 2^{-1/3}$, so their total length, area and mean curvature are $\ell^v_{p} = 2^{-1/3} \pi$, $a^v = 2^{-2/3}$ and $k^v = 2^{4/3}$. Thus, in this case the asymptotic value of the first order correction of the mean curvature difference is
\begin{equation}
 \kappa_{(1)A} = 2^{4/3} - c_s\,.
\end{equation}
The mean curvature of the neck, $k^v_{mc}:=\tilde{k}_{(1)}/r_0$, is
\begin{equation} \label{eq:k1vfix}
 k^v_{mc}= 2^{4/3} \frac{\tell}{\tilde{r}_{(0)}} +c_s \left(1-\frac{\tell}{\tilde{r}_{(0)}} \right)  \,,
\end{equation}
plotted in Fig.\ \ref{fig:23} along with the numerical data, showing a good agreement, like the case of vesicles with $a=1$.
\begin{table}
 \begin{tabular}{||c|c|c|c|c|c|c|c|c||}
 \hline \hline
 $c_s$ & -1 & 0 & 1 & 2 & 2.25 & 2.4 & $2^{4/3}$ & 2.55 \\
 \hline \hline
$r^v_{0 \, sc}$ & 1.68 & 1.66 & 1.68 & 1.74 & 1.75 & 1.76 & 1.77 & 1.78\\
 \hline
 $\dot{\Theta}^v_{0 sc}$ & 3.83 & 3.62 & 3.63 & 3.93 & 4.04 & 4.11 & 4.17 & 4.19\\
 \hline
 $\ddot{\Theta}^v_{0 sc}$ &-4.17 & -2.74 & -2.38 & -3.08 & -3.45 & -3.69 & -3.91 & -3.96 \\
 \hline
$\mrmp^v_{mc}$ & 4.44 & 0 & -1.91 & -1.3 & -0.76 & -0.36 & 0 & 0.1 \\
  \hline
$\mrmp^v_{sc}$ & -7.25 & -7.56 & -7.75 & -7.7 & -7.69 & -7.67 & -7.66 & -7.64 \\
\hline
$\frac{2}{\pi} \ell^v_{p\,sc}$ & 1.21 & 1.20 & 1.22 & 1.23 & 1.24 & 1.25 & 1.25 & 1.26\\
\hline
$a^v_{sc}$ & 1.73 & 1.71 & 1.74 & 1.83 & 1.84 & 1.86 & 1.87 & 1.88 \\
\hline
$h^v_{B mc}$ & 3.9 & 2 & 0.73 & 0.08 & 0.02 & 0.004 & 0 & 0.0003 \\
\hline
$h^v_{B sc}$ & 4.25 & 2.4 & 1.51 & 1.58  & 1.74 & 1.89 & 2.03 & 2.06 \\
\hline
$\frac{1}{4\pi}f^v_{0 mc}$ & -3.52 & -2.52 & -1.52 & -0.52 & -0.27 & -0.12 & 0 & 0.03\\
\hline
$\frac{1}{4\pi}f^v_{0 sc}$ & 6.98 & 4.56 & 4 & 5.36 & 6.05 & 6.52 & 6.93 & 7.04 \\
\hline \hline
 \end{tabular}
 \caption{Numerical values of quantities of vesicles with $v=1$ and different values of $c_s$ in the limits of maximum constriction and self-contact.} \label{TableIII}
\end{table}
The initial behavior of a stretched vesicle is similar to that of a vesicle with fixed area, it adopts an oblate shape (Figs. \ref{fig:20}(a)-(c)), and as $r_0$ is increased the vesicle becomes flat at the poles, (Fig. \ref{fig:20}(d)). However, unlike the case of fixed area, in order to satisfy the volume constraint it does not flatten more and more as the radius is increased. Instead, the neighborhood of the pole lowers (where $\dot{z} \leq 0$) and as $r_0$ is increased the poles approach each other, (Figs. \ref{fig:20}(e)-(f)), until they get in contact at $r^v_{0sc}$ (given in Table \ref{TableIII}), while the equatorial region remains round (similar to the outer region of a torus), such that in this limit the generating curve has a droplike shape and the vesicle resembles a discocyte (Fig.\ \ref{fig:20}(g)).
\\
Although there are solutions for $r_0$ greater than the equatorial radius of the discocyte, we do not consider them, because the poles cross each other and the generating curve self-intersects, which would be not only unphysical, for it does not comply with the self-avoidance of the membrane, but it would also change the topology of the corresponding axisymmetric membrane. It is conceivable that for slightly larger values of $r_0$ the membrane adopts the shape of a stretched discocyte, such that the flat central region (where the membrane is in self-contact) becomes larger as $r_0$ is increased, while the profile curve of the outer region maintains its droplike shape. This kind of self-contact configurations could be studied by considering a linear repulsive force \cite{Stoop2011}, however it is likely that the self-exerted force breaks the up-down or the axial symmetries \cite{Boue2006, Kahraman2012, Bouzar2015}, so such analysis lies beyond the scope of this paper.
\begin{table}
 \begin{tabular}{||c|c|c|c||}
 \hline \hline
$c_s$ & $2.4$ & $2^{4/3}$ & $2.55$ \\
 \hline \hline
$\medbullet$ $r^v_{01}$ & 0.46 & 0.5 & 0.512 \\
 \hline
$\blacksquare$ $r^v_{02}$ & 1.89 & 1.9 & 1.91 \\
\hline
$\blacktriangle$ $r^v_{03}$ & 0.4 & 0.35 & 0.34 \\
 \hline
$\Diamondblack$  $r^v_{04}$ & 1.62& 1.63 & 1.64 \\
\hline
$-$ $r^v_{05}$ & 0.93 & 0.86 & 0.84 \\
\hline
$\bigstar$ $r^v_{06}$ & 0.8 & 0.86 & 0.87 \\
\hline \hline
 \end{tabular}
\caption{Numerical values of $r_0$ for vesicles with $v=1$ corresponding to the bifurcations and their branches, which are indicated with the corresponding symbols in Figs. \ref{fig:21}, \ref{fig:22}, and \ref{fig:24}-\ref{fig:28}.} \label{TableIV}
\end{table}
\vskip0pc \noindent
As the spontaneous curvature is increased two bifurcations appear in the sequence of constricted configurations. Relevant values of the equatorial radius of configurations pertaining to these bifurcations and their branches, denoted by $r^v_{0i}$, $i=1,2,\dots,6$ and shown with markers in the plots of the vesicle quantities below, are presented in Table \ref{TableIV}. The first bifurcation occurs at $r^v_{01}$, where a branch originates and extends in the interval $r^v_{01}< r_0 < r^v_{02}$, whose configurations are shown in Fig.\ \ref{fig:21}. At $r^v_{01}$ the vesicle has a dumbbell shape whose central region is almost flat (Fig.\ \ref{fig:21}(a)), but as $r_0$ is increased it develops a central protrusion (Fig.\ \ref{fig:21}(b)) which swells while remaining connected to the rounded extremes by almost cylindrical regions (Fig.\ \ref{fig:21}(c)-(d)). By further increasing $r_0$ the central protrusion flattens, so it adopts a disclike shape (Fig.\ \ref{fig:21}(e)), then its central region lowers (Fig.\ \ref{fig:21}(f)), until it gets in self-contact, such that the final configuration at $r^v_{02}$ resembles a discocyte connected to a dumbbell (Fig.\ \ref{fig:21}(g)). The second bifurcation occurs at $r^v_{03}$, where another branch begins and continues along the interval $r^v_{03} < r_0 < r^v_{04}$, whose configurations are illustrated in Fig.\ \ref{fig:22}. At $r^v_{03}$ the vesicle also has a dumbbell (Fig.\ \ref{fig:22}(a)) and as $r_0$ is increased the central region bulges (Fig.\ \ref{fig:22}(b)), similar to the sequence of configurations of the first branch. However, in this case the vesicle develops two necks, above and below the central region, whose radii decrease as $r_0$ increases, becoming vanishingly small at $r^v_{05}$. As a result, the vesicle adopts a unduloidlike shape (Fig.\ \ref{fig:22}(c)), after which its central region keeps swelling (Fig.\ \ref{fig:22}(d)), while joined to quasispherical vesicles above and below by infinitesimal necks, following a transition from a disclike shape to a discocyte at $r^v_{04}$ (Figs.\ \ref{fig:22}(e)-(g)), where the small necks get in self contact. The quasispherical vesicles satisfy Eq.\ (\ref{eq:spheresclrl}) with $\mu=0$, so their radii is given by one of its solutions \cite{Miao1991}, namely
\begin{equation}
r_\mathfrak{s} = \frac{c_s^2-\sqrt{c_s^4-8 \mrmp c_s}}{2 \mrmp} \,.
\end{equation}
The values of the parameters for vesicles with $v=1$ are shown in Figs.\ \ref{fig:24}-\ref{fig:28}, where the values corresponding to configurations delimiting the first and second branches (shown in Figs.\ \ref{fig:21} and \ref{fig:22}) are indicated with the symbols $(\medbullet,\blacksquare)$ and ($\blacktriangle,\Diamondblack$), respectively (cf. Table \ref{TableIV}), and the values corresponding to configurations with vanishingly small necks are shown with a dashed line. Also, some values of the parameters for configurations in the limits of maximum constriction and self contact mentioned below are given in Table \ref{TableIII}.
\\
The plots of the initial values of the first and second derivatives of $\Theta$ are shown in Fig.\ \ref{fig:24}. For values close to $c_s=0$, as the membrane is constricted, $\dot{\Theta}_0$ decreases monotonically, diverging to $-\infty$ in the limit of maximum constriction. For stretched membranes $\dot{\Theta}_0$ increases, reaching a value $\dot{\Theta}_{0 \,sc}$ for the discocytes. For $c_s=-1,0$ as $r_0$ is decreased $\ddot{\Theta}_0$ increases monotonically, but as $c_s$ is increased it oscillates between a maximum and a minimum after which it diverges to $+\infty$ ($-\infty$ for $c_s>2^{4/3}$) in the limit of maximum constriction $r_0 \rightarrow 0$. As the vesicle is stretched $\ddot{\Theta}_0$ decreases, reaches a minimum and then increases up to the value $\ddot{\Theta}_{0 \,sc}$ for the discocytes. For $c_s>2.25$, as $r_0$ is increased $\dot{\Theta}_0$ increases along both branches, ending at their corresponding self touching configurations. Along the first branch as $r_0$ is increased $\ddot{\Theta}_0$ first decreases towards a minimum and then it increases, whereas along the second branch it first increases rapidly, then it decreases to a minimum, after which it increases.
\vskip0pc \noindent
The pressure difference $\mrmp$ is plotted in Fig.\ \ref{fig:25}. For the initial spherical configuration $\mrmp = c_s(c_s-2)$. For values close to $c_s=0$, as the vesicle is constricted $\mrmp$ increases reaching a maximum, after which it decreases reaching a value $\mrmp^v_{mc}$ in the limit of maximum constriction. This limit value of $\mrmp$ can be determined analytically from the scaling relation, which in this case reads
\begin{equation}
 r_0^2 \ddot{\Theta}_0 = -c_s^2 a + 2 c_s m + \mrmp \,.
\end{equation}
In the limit $r_0 \rightarrow 0$, we have $r_0^2 \ddot{\Theta}_0 \rightarrow 0$, $a \rightarrow 2a_v=2^{1/3}$ and $m \rightarrow k_v a_v = 2^{2/3}$; thus the pressure difference is given by
\begin{equation}
 \mrmp^v_{mc} = 2^{1/3} c_s (c_s - 2^{4/3} ) \,,
\end{equation}
which vanishes for $c_s=0,2^{4/3}$. This result is in good agreement with the numerical data presented in Table \ref{TableIII}. For stretched vesicles $\mrmp$ decreases to a minimum and then it increases reaching the value $\mrmp^v_{sc}$ for the discocytes. For $c_s>2.25$ as $r_0$ is increased $\mrmp$ has a similar behavior along both branches, it first decreases to a minimum and then increases.
\vskip0pc \noindent
The scaled total length of the generating curve $2 \ell_p$ and the total area $a$,  plotted in Fig.\ \ref{fig:26}, exhibit a similar behavior. For constricted vesicles with values close to $c_s=0$, as $r_0$ is decreased $2 \ell_p$ and $a$ increase, reaching a maximum and then they decrease towards the limit of maximum constriction, where they reach the values $2^{2/3} \pi$ and $2a_v=2^{1/3}$ respectively. For stretching, as $r_0$ is increased, $2 \ell_p$  decreases, reaches a minimum, after which it increases to the value $2 \ell^v_{p\, sc}$ for the discocytes, whereas $a$ increases to the value $a^v_{sc}$ for the same configurations. For $c_s>2.25$, as $r_0$ is increased the total length and the total area exhibit a similar oscillatory behavior along both branches, they first increase up to a maximum (more rapidly for the second branch), then decrease towards a minimum after which they increase.
\vskip0pc \noindent
The scaled total energy $h_B$ and the total force $f_0$ are plotted in Figs.\ \ref{fig:27} and \ref{fig:28}. For values close to $c_s=0$, the initial spherical states, whose energy is given in Eq.\ (\ref{hBsphere}), correspond to a local minimum of $h_B$. Thus, for either constricted or stretched membranes $h_B$ increases, reaching the values $h^v_{B mc}$ and $h^v_{B sc}$ at the limit of maximum constriction and for the discocytes, respectively. The exact value in the limit of maximum constriction corresponds to the total bending energy of two spherical vesicles of radius $r_v$, given by
\begin{equation}
 h^v_{Bmc}= \frac{(2^{4/3}-c_s)^2}{2^{5/3}}\,.
\end{equation}
As $c_s$ is increased $h_B$ develops a maximum for constricted membranes and for $c_s>2.25$ it changes nonsmoothly at both bifurcations. As $r_0$ is increased $h_B$ increases monotonically along the first branch, whereas it first decreases to a minimum and then increases along the second branch. The configurations of the second branch and of the original sequence of constricted vesicles have the same bending energy at $r^v_{06}$ (indicated with the symbol $\bigstar$ in Fig. \ref{fig:27}), and for $r_0 < r^v_{06}$ the configurations of the second branch have the lowest energy. Furthermore, for $c_s=2^{4/3}, 2.55$ in the interval $r_0 \in [r^v_{05}, r^v_{06}]$ these configurations of minimum total energy have vanishingly small necks.
\\
The force vanishes for the initial spherical configuration. For values close to $c_s=0$ the force on constricted membranes exhibits an oscillatory behavior, it decreases to a minimum, then it increases to a maximum after which it finally achieves the value $f^v_{0 \, mc}$ at the limit of maximum constriction. The exact value of the total force in the limit of maximum constriction is given by
\begin{equation}
\frac{f^v_{0mc}}{4\pi} = -(2^{4/3} - c_s) \,.
\end{equation}
Both, $h^v_{B\,mc}$ and $f^v_{0\,mc}$ vanish for $c_s=2^{4/3}$ and agree with the numerical data presented in Table \ref{TableIII}. For stretched membranes, the total force increases toward a maximum and then it decreases reaching the value $f^v_{0 \,sc}$ for the discocytes. For $c_s > 2.25$, as $r_0$ is increased $f_0$ increases to a maximum and then it decreases along the first branch, whereas it first decreases to a minimum, then increases to a maximum, after which it decreases again along the second branch.

\section{Concluding remarks and discussion} \label{sectdiscussion}

We examined the reaction of homogeneous axisymmetric membranes to the force exerted by a rigid ring located at its equator. We considered membranes with different values of spontaneous curvature and with their area or volume fixed. For either constraint, starting with values of the spontaneous curvature about zero we obtained a single sequence of configurations, but as the spontaneous curvature increases close to the mean curvature of the quasispherical vesicles in the limit of maximum constriction, value for which the total force vanishes, we found bifurcations in the solutions of the EL equation, which originate new branches of the sequence of configurations. We demonstrated that locally the effect of the external force is to introduce a discontinuity in the derivatives of the meridional curvature of the membrane across the ring, whereas globally it introduces a source in the scaling relation between the total area, volume and mean curvature of the vesicle.
\\
Constricted vesicles have a similar behavior for both constraints, differing only by their size. As the equatorial radius is reduced, they exhibit a transition from prolate to dumbbell shapes, which in turn terminates in two quasispherical vesicles connected by a small catenoid-like neck. Although we employed a perturbative analysis to examine the geometry of such small necks, as well as the external force required to hold them, remarkably the first order corrections suffice to determine them accurately in comparison with the numerical results. Unlike previous treatments where a small neck is assumed to have constant mean curvature, we found that it is described with a good degree of precision by a deformed catenoid whose mean curvature varies. Together, the external force and the area or volume constraints, compel the mean curvature of the neck to interpolate between the spontaneous curvature at the equator (kissing condition) and the mean curvature of the quasispherical vesicles. A similar situation occurs in the formation of a neck in the spherical confinement of a vesicle, which seems to be described by a catenoidal shape, but by analyzing its mean curvature it is found to be nonvanishing \cite{Kahraman2012}.
\\
Stretched vesicles with fixed area behave differently to vesicles with fixed volume as the equatorial radius is increased. In the former case, the vesicle flattens progressively until it becomes disclike, limit that we examined using an asymptotic analysis. These stretched configurations require an increasing total force, which diverges for the disc. For the latter case, the vesicle also starts to flatten, but after a critical stretching the poles begin to approach until they get in contact, so the vesicle achieves a discocyte shape. Along such sequence the force is bounded, it increases reaching a maximum and then decreases to a finite value for the discocyte shape. A similar configuration with a discocyte shape can be obtained from the inversion of the catenoid with respect to a sphere centered at the origin. However in such case the force is not equatorial, but  applied at the poles, where it gives rise to logarithmic singularities in the curvature \cite{CastroGuven2007PRE, CastroGuven2007JPA, GuvenVazquez2013}.
\\
The increase of the spontaneous curvature elicits the appearance of bifurcations in the solutions corresponding to constricted membranes. The branches stemming from them are connected for vesicles with fixed area, but disconnected for vesicles with fixed volume. In general, the vesicles corresponding to these branches tend to develop lobes in the regions near the poles, while the central region bulges. For large enough values of the spontaneous curvature, as the equatorial radius is increased the vesicle develops very small necks connecting the lobes with the central region, akin to the remote constriction of a membrane \cite{Bozic2014}. We found for both constraints that below a critical value of the equatorial radius the lowest energy configurations belong to one of the branches, some of which have very small necks.
\\
There are several possible extensions and refinements of our framework. Here we determined the configurations of the membranes employing the spontaneous curvature model, but it would be interesting to generalize our results using the bilayer couple model \cite{Svetina1989} or the area difference elasticity model \cite{Bozic1992, Svetina2014}. Also, we considered configurations satisfying the area or the volume constraint separately, so one of the two Lagrange multipliers vanishes, but by considering both of them nonvanishing one could have access to a broader class of configurations.
\\
We focused on relatively small values of the spontaneous curvature, for which the vesicle follows a single sequence of configurations and up to values where bifurcations occur for the first time. However, for higher values more bifurcations and branches arise, some of them showing a complex behavior, which complicates the identification of the conformations with the lowest energy. In this regard, though laborious, it would be useful to classify such lowest energy configurations using a phase diagram. Furthermore, to assess their stability one could analyze the spectrum of the second variation of the energy. It might be interesting to analyze the stability of the configurations corresponding to the branches of the bifurcations, so as to confirm that for some values of the equatorial radius they exhibit instabilities towards configurations with lower energy, whereas they become the stable configurations for the values where they have the lowest energy.
\\
Here we assumed that the deformed membrane remains axisymmetric as well as symmetric with respect to the equatorial plane. However, one cannot discount the possibility of new asymmetrical states, possibly stable, which do not possess equatorial mirror symmetry, such as the shapes observed in the exobudding and endobudding processes of a vesicle \cite{Bozic2014, Lipowsky2024, LipowskyBook}, or like the configurations observed in the transition from symmetric to asymmetric dumbbells due to a periodic change of spontaneous curvature \cite{Christ2021}. For instance it is possible that constricted membranes with prolate shapes exhibit an instability by slipping through the ring and adopting an asymmetric shape. However, once the membrane adopts a dumbbell shape and a neck forms, it should become stable. To relax the axial symmetry one may look at the configurations of a spherical membrane deformed by a noncircular loop, say an elliptic one, which would constrict and stretch the membrane along orthogonal directions, producing a force dipole on the equator.
\\
We considered a ring of infinite bending rigidity, so it remains circular regardless of the force exerted by the membrane, which for instance diverges in the unphysical limit of the disclike configuration. More realistically, the ring will have a finite bending rigidity, and under confinement it will not remain circular, but adjust to its confinement \cite{Vetter2014}. Even if the ring rigidity is large but finite, since the total force ought to remain finite, one would expect that axial symmetry breaking occurs before the disclike shape is attained. Likewise, it is conceivable that if the equator of the discocyte shape is stretched, the planar central region enlarges and the force increases up to a point that, in order to minimize its bending energy, the membrane adopts a nonaxisymmetric configuration.
\\
The opposing limit corresponding to an elastic loop of fixed length confined within a rigid sphere was considered in Ref.\ \cite{GuvenVazquez2012}. The ground state of the loop was found to assume a saddle shape. In general both the loop and the membrane will deform in response to the constraint imposed by the confinement, which involves playing two competing tendencies against each other. Technically, the determination of the shape will involve the nontrivial task of minimizing the sum of the membrane and loop bending energies. Solving the coupled EL equations of the membrane-loop complex is not likely to be tractable analytically. The ground state of a confined loop would again be expected to assume a saddle shape in a deformed geometry that resembles an ellipsoid with its long axis aligned along the direction that the transmitted force is maximum. However, less obvious possibilities should not be discounted. In contrast, in a membrane with a constricted equator, the circular loop will minimize the total bending energy so that the behavior should not change. One would also still expect that a finite force will be required to constrict the membrane completely.
\\
Instead of considering the deformation of the membrane due to a ring, one could also study its deformation by an open polymer, which also plays a role in shaping the morphology of membranes. In particular, the protein reticulin is suspected to be implicated in determining the morphology of the endoplasmic reticulum. To examine the equilibrium shapes of a polymer-membrane complex, one could consider a semiflexible rigid polymer segment of fixed length, be it straight, a hairpin, or the arc of a circle, bounded to a fluid membrane.

\section*{Acknowledgements}

We have benefited from conversations with Sa\v{s}a Svetina. B. B. acknowledges partial support by the Slovenian Research Agency through the research program P1-0055. P. V.-M. acknowledges support by Secretaria de Ciencia, Humanidades, Tecnología e Innovación (SECIHTI) under programs Investigadores por México (Grant 439-2018) and Sistema Nacional de Investigadores (131142) and is grateful for the hospitality of the Josef Stefan Institute and the Institute of Biophysics of the University of Ljubljana.



\begin{appendix}

\section{Axisymmetric surfaces}\label{App:axisym}

An axially symmetric membrane can be parametrized by the arc length $l$ along its meridian and the azimuthal angle $\varphi$. In terms of the cylindrical basis $\{ \hat{\bm{\rho}} , \hat{\bm{\varphi}},\hat{\bfz} \}$, given by
\begin{subequations} \label{cylvectbasis}
\begin{eqnarray}
\hat{\bm \rho} &=& \cos \varphi \, \hat{\bf x}+ \sin \varphi \, \hat{\bf y}\,, \\
\hat{\bm \varphi} &=& -\sin \varphi \, \hat{\bf x} +\cos \varphi \, \hat{\bf y}\,,
\end{eqnarray}
\end{subequations}
the embedding functions and the two tangent vectors adapted to this parametrization are spanned as
\begin{subequations} \label{Xlphiaxisym}
\begin{eqnarray}
\mathbf{X}(l,\varphi) &=& R(l) \, \hat{\bm \rho} +  Z(l) \, \hat{\bf z}\,, \\
\mathbf{e}_l &=& R' \, \hat{\bm \rho} + Z' \, \hat{\bf z} \,, \label{eq:vel} \\
\mathbf{e}_\varphi &=& R \hat{\bm \varphi}\,, \label{eq:vephi}
\end{eqnarray}
\end{subequations}
where $R(l)$ and $Z(l)$ are the radial and height coordinates; the prime indicates derivation with respect to $l$. Since, $R(l)$ and $Z(l)$ are parametrized by arclength, their derivatives are related by $R'(l)^2 + Z'(l)^2 = 1$. This condition permits us to define
\begin{equation} \label{eq:RZprime}
R' = \cos \Theta\,, \quad Z'= \sin \Theta\,,
\end{equation}
where $\Theta$ is the angle that the tangent along the meridian ${\bf e}_l$ makes with the radial direction $\hat{\bm \rho}$.
\\
In this parametrization the metric tensor is diagonal
\begin{equation}
g_{l l}= 1 \,, \quad g_{\varphi l}= g_{l\varphi}= 0 \,, \quad g_{\varphi \varphi}= R^2\,.
\end{equation}
The line element on the surface is given by $d s^2 = dl^2 +  R(l)^2 d\varphi^2$, whereas the area element is $dA=R d\varphi dl$, so the area of a region with $\varphi \in [0,2\pi]$ and $l \in [l_i,l_f]$ is
\begin{equation} \label{areasymsurf}
A= 2 \pi \displaystyle \int \limits_{l_i}^{l_f} dl R \,.
\end{equation}
The outward unit normal vector, defined by $\mathbf{n} = g^{-1/2}\, \mathbf{e}_\varphi \times
\mathbf{e}_l$, reads
\begin{equation} \label{eq:naxsym}
\mathbf{n} = \sin\Theta \, \hat{\bm \rho}-\cos\Theta \, \hat{\bf z}\,.
\end{equation}
The enclosed volume is determined by the integrating of the support function over the surface, $V=1/3 \int dA {\bf X} \cdot {\bf n}$. For an axisymmetric surface
\begin{equation}
 V = \frac{2 \, \pi}{3} \displaystyle\int\limits^{l_f}_{l_i} dl R \left( \sin
\Theta R -\cos \Theta Z \right)\,.
\end{equation}
In the case of a closed surface symmetric with respect to the plane $Z=0$ and satisfying $Z(0)=0$ and $R(l_f)=0$, this equation can be simplified by integrating by parts,
\begin{equation} \label{volsymsurf}
 V = 2 \, \pi \displaystyle\int\limits^{l_f}_{0} dl \, \sin \Theta R^2\,.
\end{equation}
The extrinsic curvature tensor is diagonal
\begin{equation} \label{KabSAxs}
K_{ll} = \Theta' \,, \quad  K_{\varphi l}= K_{l\varphi}= 0 \quad K_{\varphi \varphi}=R \sin\Theta\,.
\end{equation}
The shape operator $K^a_{\phantom{a}b}= g^{ac} K_{cb}$ is also diagonal, so its eigenvectors  lie along meridians and parallels and the principal curvatures are the corresponding eigenvalues. These curvatures also correspond to the normal curvatures along meridians and parallels, obtained from the projections of the curvature tensor onto the outward conormal and the tangent vectors to a parallel, which are given by
\begin{equation} \label{vLvTparallel}
 \bfL = -\bfe_l \,, \quad \bfT = \hat{\bm{\varphi}}= \frac{1}{R} \bfe_\varphi \,,
\end{equation}
so the components are $\rmL^l=-1$ and  $\rmT^\varphi=1/R$. Thus, the  projections of $K_{ab}$ onto $\bfL$ and $\bfT$ are
\begin{subequations}
\begin{eqnarray}
C_\perp &=& \rmL^a \rmL^b K_{ab}=\Theta' \,, \\
C_\parallel &=& \rmT^a \rmT^b K_{ab} = \frac{\sin \Theta}{R} \,.
\end{eqnarray}
\end{subequations}
The geodesic torsion vanishes along a parallel $\tau_g=\rmT^a \rmL^b K_{ab}=0$, because it is a principal curve. Thus the tangential component $f_{\parallel \perp}$ of the stress tensor projected along the parallel outward, given in Eq.\ (\ref{Eq:fprppar}) vanishes. Thus, the projected stress tensor is orthogonal to the parallel, $\bff_\perp = f_{\perp \perp} \bfL +f_\perp \bfn$. Using the expressions of the vectors $\bfL$ and $\bfT$, Eqs.\ (\ref{vLvTparallel}), along with the expressions of the adapted basis (\ref{eq:vel}), (\ref{eq:vephi}) and (\ref{eq:naxsym}), we get that the projections onto the unit vectors $\hat{\bm{\rho}}$ and $\hat{\bfz}$ are
\begin{subequations}
\begin{eqnarray}
{\bf f}_\perp \cdot \hat{\bm \rho} &=& \sin \Theta f_\perp- \cos \Theta f_ {\perp \perp} \,, \label{fperprproj}\\
{\bf f}_\perp \cdot \hat{\bfz} &=& -\cos \Theta f_\perp - \sin \Theta f_{\perp \perp} \,. \label{fperpzproj}
\end{eqnarray}
\end{subequations}

\section{Hamiltonian formulation}\label{App:HamDer}

The bending energy (scaled by $1/\rmk_B$), given in Eq.\ (\ref{eq:Hbending}), of an axially symmetric geometry with mirror symmetry with respect to the equatorial plane is\footnote{We do not consider the Gaussian energy density, because it is a total derivative, $\int dA K_G = -2 \pi \int dl (\cos \Theta)'$, so it does not contribute to the total energy of the bulk.}
\begin{equation}
\frac{L_{B}}{2 \pi}= 2 \int_{0}^{l_p} d l \scrL_{B} \,,
\end{equation}
where the energy density is given by $\scrL_{B}= R K_D^2/2$, with $K_D$ the mean curvature difference, defined by
\begin{eqnarray}
K_D &=& \Theta' + \frac{\sin \Theta}{R} - C_s \,.
\end{eqnarray}
In order to implement in the variational principle the constraints of fixed area and volume, given by Eqs.\ (\ref{areasymsurf}) and (\ref{volsymsurf}), we introduce two global Lagrange multipliers $\sigma$ and $\rmP$. Another two local Lagrange multipliers $\lambda_R$ and $\lambda_Z$ are also introduced to enforce the relations between derivatives of $R$ and $Z$ and the angle $\Theta$, given by Eq.\ (\ref{eq:RZprime}), which together imply that $l$ is arclength. As shown below, these two Lagrange multipliers are related to the stresses on the membrane.  Thus, omitting the constant terms $-\sigma A_0 +P V_0$, the Lagrangian is
\begin{equation} \label{def:Laxisim}
 \frac{L}{2 \pi} =  2 \int_0^{l_p} d l \scrL \,,
\end{equation}
where the constrained Lagrangian density is defined by
\begin{eqnarray}
\scrL &=& \scrL_{B} + \bar{\sigma} \, R  -\frac{\bar{\rmP}}{2} R^2 \sin \Theta \nonumber \\
&+& R \bar{\lambda}_R (R'-\cos \Theta) + R \bar{\lambda}_Z (Z'-\sin \Theta) \,.
\end{eqnarray}
The conjugate momenta to $\Theta$ and to the coordinates $R$ and $Z$ are
\begin{subequations} \label{eq:PThetaRZ}
\begin{eqnarray}
P_\Theta &:=& \frac{\partial\scrL }{\partial \Theta'} = R K_D \,, \\
P_R  &:=& \frac{\partial\scrL }{\partial R'} =R \lambda_R\,, \\
P_Z &:=& \frac{\partial\scrL }{\partial Z'} = R \lambda_Z \,.
\end{eqnarray}
\end{subequations}
The corresponding Hamiltonian density is
\begin{eqnarray} \label{def:AxisymHam}
\scrH &=& P_\Theta \left(\frac{P_\Theta}{2 R}
-\frac{\sin \Theta}{R} + C_s \right) - \bar{\sigma} R + \frac{\bar{\rmP}}{2} R^2 \sin \Theta \nonumber \\
&+&  P_R \cos \Theta +P_Z \sin \Theta \,.
\end{eqnarray}
Since $\scrH$ does not depend explicitly on $l$, it is conserved, $\scrH'=0$. Moreover, in order to make $\scrL$ stationary with respect to total length variations, the Hamiltonian must vanish
\begin{equation} \label{eq:Heq0}
\scrH=0 \,,
\end{equation}
which avoids the introduction of an unphysical constraint fixing the interval of $l$, (the arc length at the pole, $l_p$, is not fixed) \cite{Julicher1994, CastroGuven2007PRE, CastroGuven2007JPA}.
\\
The Hamilton equations for $\Theta, R$ and $Z$ are identified as
\begin{subequations} \label{hameqscoords}
\begin{eqnarray}
 \Theta' &=& \frac{P_\Theta-\sin \Theta}{R} + C_s \,, \label{Thetap}\\
 \quad R' &=& \cos \Theta \,, \label{Rp}\\
 \quad Z' &=& \sin \Theta\,, \label{Zp}
\end{eqnarray}
 \end{subequations}
whereas those for the corresponding momenta are
\begin{subequations}
\begin{eqnarray}
 P_\Theta' &=& \left(\frac{P_\Theta}{R} -\frac{\bar{\rmP}}{2} R^2 -P_Z \right) \cos \Theta + P_R \sin \Theta \,, \quad \label{PThetap}\\
 P_R'&=& \frac{P_\Theta}{R^2} \left(\frac{P_\Theta}{2} - \sin \Theta \right) -\bar{\rmP} R \sin \Theta 
 + \bar{\sigma}\,, \label{PRp}\\
 P_Z' &=& 0 \,. \label{PZp}
\end{eqnarray}
\end{subequations}
The momentum $P_Z$ is constant because $Z$ is a cyclic coordinate.
\\
Multipliying Eqs.\ (\ref{PThetap}) and (\ref{def:AxisymHam}) by $\cos \Theta$ and $\sin \Theta$ respectively and taking their difference in order to eliminate $P_R$, we get
\begin{align} \label{PzPthRTh}
&R \cos \Theta \left( \frac{P_\Theta}{R}\right)' + \sin \Theta \left( P_\Theta \left(\frac{P_\Theta}{2 R} -\frac{\sin \Theta}{R} + C_s \right) - \bar{\sigma} \,R \right) \nonumber \\
&+ \frac{\bar{\rmP}}{2} R^2 + P_Z =0\,.
\end{align}
Dividing by $R$, using Eq.\ (\ref{eq:PThetaRZ}) to replace $P_\Theta$ and $P_Z$ in favor of $K_D$ and $\lambda_Z$, we obtain a first order differential equation for the mean curvature difference
\begin{align} \label{EqKD}
&\cos \Theta K_D' + \sin \Theta \left( K_D \left(\frac{K_D}{2} -\frac{\sin \Theta}{R} + C_s \right) - \bar{\sigma} \right) \nonumber \\
&+ \frac{\bar{\rmP}}{2} R + \bar{\lambda}_Z =0\,.
\end{align}
Thus, in order to determine the equilibrium configurations, we can solve the equivalent system of first order differential equations, given by Eqs.\ (\ref{hameqscoords}) and (\ref{EqKD}) for $R$, $Z$, $\Theta$, and $K_D$, where the equations for the last two variables can be recast as
\begin{subequations} \label{AppEq:EDThetaKD}
 \begin{eqnarray}
\Theta ' &=& K_D - \frac{\sin \Theta }{R} + C_s \,,   \\
K'_D &=& - \tan \Theta \left( K_D \left(\frac{K_D}{2} -\frac{\sin \Theta}{R} + C_s \right) - \bar{\sigma} \right) \nonumber \\
&&-\sec \Theta \left( \frac{\bar{\rmP}}{2} R + \bar{\lambda}_Z \right) \,.
\end{eqnarray}
\end{subequations}
Substituting Eqs.\ (\ref{fproys}) for the projections of the stress tensor in Eq.\ (\ref{EqKD}), the first integral of Eq.\ (\ref{Eq:ELfree}) is expressed as a balance of stresses
\begin{equation} \label{AxysimEL}
 \cos \Theta f_\perp + \sin \Theta f_{\perp \perp} +  \frac{\rmP}{2}\,R + \lambda_Z=0\,.
\end{equation}
Substituting Eq.\ (\ref{fperpzproj}), in Eq.\ (\ref{AxysimEL}), we determine the Lagrange multiplier $\lambda_Z$
\begin{equation} \label{PZeq}
\lambda_Z = \bff_\perp \cdot \hat{\bfz} - \frac{\rmP}{2} R \,.
\end{equation}
The momentum $P_Z$ can be determined from the boundary terms in the variation of $L$ in Eq.\ (\ref{def:Laxisim}), which are given by $\int ds \delta Q'$, where
\begin{equation} \label{HamderdeltaQaxisim}
\delta Q = P_\Theta \delta \Theta + P_R \delta R + P_Z \delta Z\,.
\end{equation}
For a surface with spherical topology, the angle $\Theta$ and the radial coordinate $R$ are fixed at the poles, thus $\delta \Theta$ and $\delta R$ vanish, whereas $\delta Z$ is arbitrary at the poles so $P_Z$ must vanish \cite{Julicher1994}, so the first integral of the EL equation, given by Eq.\ (\ref{eq:shapeaxi}), is reproduced.
\\
An important example of a surface with a net vertical force is provided by the Clifford Torus, whose radial, height and functions are $R = R_T + r_c \cos \theta$ and $Z=r_c \sin \theta$,  where $\theta=l/r_c$ ($\Theta = \theta+\pi/2$) and the two constant radii are related by $R_T = \sqrt{2} \, r_c$, which satisfies Eq.\ (\ref{PZeq}) for the following values of the parameters
\begin{subequations}
\begin{eqnarray}
\bar{\sigma} &=& C_s \left(\frac{2}{r_c} - \frac{C_s}{2} \right) \,,
\\
\bar{P} &=& \frac{2 C_s}{r_c^2} \,,
\\
R \bar{\lambda}_Z&=&-\frac{1}{r_c} -2 C_s \,.
\end{eqnarray}
\end{subequations}
We see that although $\sigma$ and $P$ vanish on a toroidal vesicle with null spontaneous curvature, it requires a nonvanishing vertical force.
\\
The other Lagrange multiplier $\lambda_R$ can be determined from Eq.\ (\ref{PThetap}), which divided by $R$ and using Eqs.\ (\ref{eq:PThetaRZ}) for the momenta, can be recast as
\begin{equation}
\lambda_R  \sin \Theta =  K_D' + \cos \Theta \left( \frac{\bar{\rmP}}{2} R + \lambda_Z\right)\,.
\end{equation}
Recall that $\bar{f}_\perp=K'$, using Eq.\ (\ref{AxysimEL}) to substitute $\lambda_Z$ and using Eq.\ (\ref{fperprproj}) for the radial projection of ${\bf f}_\perp$, we  identify $\lambda_R$ as the radial force density on the membrane
\begin{equation} \label{def:lambdaRprojfrho}
\lambda_R  =  \sin \Theta f_\perp -\cos \Theta f_{\perp \perp} = \bff_\perp \cdot \hat{\bm \rho}\,.
\end{equation}
By taking into account that at the poles $R(l_p)=0$, the constraint of fixed radius at equator, $R(0)=R_0$, can be expressed as $ \int^{l_p}_0 dl R(l)'= -R_0$, which is implemented with a constant Lagrange multiplier $\lambda_n$, so $\scrL \rightarrow \scrL + (R_0/2) \bar{\lambda}_n R'$. Since the added term is a total derivative, it does not modify either the Hamilton or the EL equations for $R$. However it modifies the momentum conjugate to $R$ in the boundary term in Eq.\ (\ref{HamderdeltaQaxisim}) corresponding to $R$, $P_R \rightarrow P_R + (R_0/2) \bar{\lambda}_n$, so at the equator where $\delta \Theta=0$ and $\delta Z=0$, we have $R_0 \bar{\lambda}_n = -2 P_R(0)$, or by using Eq.\ (\ref{def:lambdaRprojfrho}) as well as that $\Theta_0=\pi/2$, we obtain
\begin{equation}
\bar{\lambda}_n = - 2 \bar{\lambda}_R(0) = -2 \bar{f}_\perp (0)= -2 \Theta''_0\,,
\end{equation}
which reproduces the force density, given in Eq.\ (\ref{def:lambdan}), and whose integration along the ring provides the total constraining force, $\bar{F}=-4\pi R_0 \Theta''_0$.

\end{appendix}

\bibliography{bibliography}

\begin{thebibliography}{50}%
\makeatletter
\providecommand \@ifxundefined [1]{%
 \@ifx{#1\undefined}
}%
\providecommand \@ifnum [1]{%
 \ifnum #1\expandafter \@firstoftwo
 \else \expandafter \@secondoftwo
 \fi
}%
\providecommand \@ifx [1]{%
 \ifx #1\expandafter \@firstoftwo
 \else \expandafter \@secondoftwo
 \fi
}%
\providecommand \natexlab [1]{#1}%
\providecommand \enquote  [1]{``#1''}%
\providecommand \bibnamefont  [1]{#1}%
\providecommand \bibfnamefont [1]{#1}%
\providecommand \citenamefont [1]{#1}%
\providecommand \href@noop [0]{\@secondoftwo}%
\providecommand \href [0]{\begingroup \@sanitize@url \@href}%
\providecommand \@href[1]{\@@startlink{#1}\@@href}%
\providecommand \@@href[1]{\endgroup#1\@@endlink}%
\providecommand \@sanitize@url [0]{\catcode `\\12\catcode `\$12\catcode
  `\&12\catcode `\#12\catcode `\^12\catcode `\_12\catcode `\%12\relax}%
\providecommand \@@startlink[1]{}%
\providecommand \@@endlink[0]{}%
\providecommand \url  [0]{\begingroup\@sanitize@url \@url }%
\providecommand \@url [1]{\endgroup\@href {#1}{\urlprefix }}%
\providecommand \urlprefix  [0]{URL }%
\providecommand \Eprint [0]{\href }%
\providecommand \doibase [0]{https://doi.org/}%
\providecommand \selectlanguage [0]{\@gobble}%
\providecommand \bibinfo  [0]{\@secondoftwo}%
\providecommand \bibfield  [0]{\@secondoftwo}%
\providecommand \translation [1]{[#1]}%
\providecommand \BibitemOpen [0]{}%
\providecommand \bibitemStop [0]{}%
\providecommand \bibitemNoStop [0]{.\EOS\space}%
\providecommand \EOS [0]{\spacefactor3000\relax}%
\providecommand \BibitemShut  [1]{\csname bibitem#1\endcsname}%
\let\auto@bib@innerbib\@empty
\bibitem [{\citenamefont {Thalmann}\ and\ \citenamefont
  {Marques}(2019)}]{ThalmannBook}%
  \BibitemOpen
  \bibfield  {author} {\bibinfo {author} {\bibfnamefont {F.}~\bibnamefont
  {Thalmann}}\ and\ \bibinfo {author} {\bibfnamefont {C.~M.}\ \bibnamefont
  {Marques}},\ }\bibinfo {title} {Theory of polymer-membrane interactions},\
  in\ \href {https://doi.org/10.1201/9781315152516} {\emph {\bibinfo
  {booktitle} {The Giant Vesicle Book}}},\ \bibinfo {editor} {edited by\
  \bibinfo {editor} {\bibfnamefont {R.}~\bibnamefont {Dimova}}\ and\ \bibinfo
  {editor} {\bibfnamefont {C.}~\bibnamefont {Marques}}}\ (\bibinfo  {publisher}
  {CRC Press},\ \bibinfo {address} {Boca Raton},\ \bibinfo {year} {2019})\ pp.\
  \bibinfo {pages} {229--261}\BibitemShut {NoStop}%
\bibitem [{\citenamefont {Heinrich}\ \emph {et~al.}(1999)\citenamefont
  {Heinrich}, \citenamefont {Bozic},\ and\ \citenamefont
  {Svetina}}]{Heinrich1999}%
  \BibitemOpen
  \bibfield  {author} {\bibinfo {author} {\bibfnamefont {V.}~\bibnamefont
  {Heinrich}}, \bibinfo {author} {\bibfnamefont {B.}~\bibnamefont {Bozic}},\
  and\ \bibinfo {author} {\bibfnamefont {S.~a. Z.~B.}\ \bibnamefont
  {Svetina}},\ }\href {https://doi.org/10.1016/S0006-3495(99)77362-5}
  {\bibfield  {journal} {\bibinfo  {journal} {Biophysical Journal}\ }\textbf
  {\bibinfo {volume} {4}},\ \bibinfo {pages} {2056} (\bibinfo {year}
  {1999})}\BibitemShut {NoStop}%
\bibitem [{\citenamefont {Powers}\ \emph {et~al.}(2002)\citenamefont {Powers},
  \citenamefont {Huber},\ and\ \citenamefont {Goldstein}}]{Powers2002}%
  \BibitemOpen
  \bibfield  {author} {\bibinfo {author} {\bibfnamefont {T.}~\bibnamefont
  {Powers}}, \bibinfo {author} {\bibfnamefont {G.}~\bibnamefont {Huber}},\ and\
  \bibinfo {author} {\bibfnamefont {R.}~\bibnamefont {Goldstein}},\ }\href
  {https://doi.org/10.1103/PhysRevE.65.041901} {\bibfield  {journal} {\bibinfo
  {journal} {Phys. Rev. E}\ }\textbf {\bibinfo {volume} {65}},\ \bibinfo
  {pages} {041901} (\bibinfo {year} {2002})}\BibitemShut {NoStop}%
\bibitem [{\citenamefont {Der{\'e}nyi}\ \emph {et~al.}(2002)\citenamefont
  {Der{\'e}nyi}, \citenamefont {J{\"u}licher},\ and\ \citenamefont
  {Prost}}]{Derenyi2002}%
  \BibitemOpen
  \bibfield  {author} {\bibinfo {author} {\bibfnamefont {I.}~\bibnamefont
  {Der{\'e}nyi}}, \bibinfo {author} {\bibfnamefont {F.}~\bibnamefont
  {J{\"u}licher}},\ and\ \bibinfo {author} {\bibfnamefont {J.}~\bibnamefont
  {Prost}},\ }\href {https://doi.org/10.1103/PhysRevLett.88.238101} {\bibfield
  {journal} {\bibinfo  {journal} {Phys. Rev. Lett.}\ }\textbf {\bibinfo
  {volume} {88}},\ \bibinfo {pages} {238101} (\bibinfo {year}
  {2002})}\BibitemShut {NoStop}%
\bibitem [{\citenamefont {Kozlov}(2001)}]{Kozlov2001}%
  \BibitemOpen
  \bibfield  {author} {\bibinfo {author} {\bibfnamefont {M.~M.}\ \bibnamefont
  {Kozlov}},\ }\href {https://doi.org/10.1034/j.1600-0854.2001.020107.x}
  {\bibfield  {journal} {\bibinfo  {journal} {Traffic}\ }\textbf {\bibinfo
  {volume} {2}},\ \bibinfo {pages} {51} (\bibinfo {year} {2001})}\BibitemShut
  {NoStop}%
\bibitem [{\citenamefont {Morlot}\ and\ \citenamefont
  {Roux}(2013)}]{Morlot2013}%
  \BibitemOpen
  \bibfield  {author} {\bibinfo {author} {\bibfnamefont {S.}~\bibnamefont
  {Morlot}}\ and\ \bibinfo {author} {\bibfnamefont {A.}~\bibnamefont {Roux}},\
  }\href {https://doi.org/10.1146/annurev-biophys-050511-102247} {\bibfield
  {journal} {\bibinfo  {journal} {Annual Review of Biophysics}\ }\textbf
  {\bibinfo {volume} {42}},\ \bibinfo {pages} {629} (\bibinfo {year}
  {2013})}\BibitemShut {NoStop}%
\bibitem [{\citenamefont {Yang}\ \emph {et~al.}(2017)\citenamefont {Yang},
  \citenamefont {Du},\ and\ \citenamefont {Tu}}]{Yang2017}%
  \BibitemOpen
  \bibfield  {author} {\bibinfo {author} {\bibfnamefont {P.}~\bibnamefont
  {Yang}}, \bibinfo {author} {\bibfnamefont {Q.}~\bibnamefont {Du}},\ and\
  \bibinfo {author} {\bibfnamefont {Z.~C.}\ \bibnamefont {Tu}},\ }\href
  {https://doi.org/10.1103/PhysRevE.95.042403} {\bibfield  {journal} {\bibinfo
  {journal} {Phys. Rev. E}\ }\textbf {\bibinfo {volume} {95}},\ \bibinfo
  {pages} {042403} (\bibinfo {year} {2017})}\BibitemShut {NoStop}%
\bibitem [{\citenamefont {Harker-Kirschneck}\ \emph {et~al.}(2022)\citenamefont
  {Harker-Kirschneck}, \citenamefont {Hafner}, \citenamefont {Yao},
  \citenamefont {Vanhille-Campos}, \citenamefont {Jiang}, \citenamefont
  {Pulschen}, \citenamefont {Hurtig}, \citenamefont {Hryniuk}, \citenamefont
  {Culley}, \citenamefont {Henriques}, \citenamefont {Baum},\ and\
  \citenamefont {{\v S}ari{\'c}}}]{Harker2021}%
  \BibitemOpen
  \bibfield  {author} {\bibinfo {author} {\bibfnamefont {L.}~\bibnamefont
  {Harker-Kirschneck}}, \bibinfo {author} {\bibfnamefont {A.~E.}\ \bibnamefont
  {Hafner}}, \bibinfo {author} {\bibfnamefont {T.}~\bibnamefont {Yao}},
  \bibinfo {author} {\bibfnamefont {C.}~\bibnamefont {Vanhille-Campos}},
  \bibinfo {author} {\bibfnamefont {X.}~\bibnamefont {Jiang}}, \bibinfo
  {author} {\bibfnamefont {A.}~\bibnamefont {Pulschen}}, \bibinfo {author}
  {\bibfnamefont {F.}~\bibnamefont {Hurtig}}, \bibinfo {author} {\bibfnamefont
  {D.}~\bibnamefont {Hryniuk}}, \bibinfo {author} {\bibfnamefont
  {S.}~\bibnamefont {Culley}}, \bibinfo {author} {\bibfnamefont
  {R.}~\bibnamefont {Henriques}}, \bibinfo {author} {\bibfnamefont
  {B.}~\bibnamefont {Baum}},\ and\ \bibinfo {author} {\bibfnamefont
  {A.}~\bibnamefont {{\v S}ari{\'c}}},\ }\bibfield  {journal} {\bibinfo
  {journal} {Proceedings of the National Academy of Sciences}\ }\textbf
  {\bibinfo {volume} {119}},\ \href {https://doi.org/10.1073/pnas.2107763119}
  {10.1073/pnas.2107763119} (\bibinfo {year} {2022})\BibitemShut {NoStop}%
\bibitem [{\citenamefont {Bahrami}\ and\ \citenamefont
  {Bahrami}(2019)}]{Bahrami2019}%
  \BibitemOpen
  \bibfield  {author} {\bibinfo {author} {\bibfnamefont {A.}~\bibnamefont
  {Bahrami}}\ and\ \bibinfo {author} {\bibfnamefont {A.~H.}\ \bibnamefont
  {Bahrami}},\ }\href {https://doi.org/10.1088/1361-6528/ab1ed5} {\bibfield
  {journal} {\bibinfo  {journal} {Nanotechnology}\ }\textbf {\bibinfo {volume}
  {30}},\ \bibinfo {pages} {345101} (\bibinfo {year} {2019})}\BibitemShut
  {NoStop}%
\bibitem [{\citenamefont {Spakowitz}\ and\ \citenamefont
  {Wang}(2003)}]{Spakowitz2003}%
  \BibitemOpen
  \bibfield  {author} {\bibinfo {author} {\bibfnamefont {A.}~\bibnamefont
  {Spakowitz}}\ and\ \bibinfo {author} {\bibfnamefont {Z.-G.}\ \bibnamefont
  {Wang}},\ }\href {https://doi.org/10.1103/PhysRevLett.91.166102} {\bibfield
  {journal} {\bibinfo  {journal} {Phys. Rev. Lett.}\ }\textbf {\bibinfo
  {volume} {91}},\ \bibinfo {pages} {166102} (\bibinfo {year}
  {2003})}\BibitemShut {NoStop}%
\bibitem [{\citenamefont {Stoop}\ \emph {et~al.}(2011)\citenamefont {Stoop},
  \citenamefont {Najafi}, \citenamefont {Wittel}, \citenamefont {Habibi},\ and\
  \citenamefont {Herrmann}}]{Stoop2011}%
  \BibitemOpen
  \bibfield  {author} {\bibinfo {author} {\bibfnamefont {N.}~\bibnamefont
  {Stoop}}, \bibinfo {author} {\bibfnamefont {J.}~\bibnamefont {Najafi}},
  \bibinfo {author} {\bibfnamefont {F.~K.}\ \bibnamefont {Wittel}}, \bibinfo
  {author} {\bibfnamefont {M.}~\bibnamefont {Habibi}},\ and\ \bibinfo {author}
  {\bibfnamefont {H.~J.}\ \bibnamefont {Herrmann}},\ }\href
  {https://doi.org/10.1103/PhysRevLett.106.214102} {\bibfield  {journal}
  {\bibinfo  {journal} {Phys. Rev. Lett.}\ }\textbf {\bibinfo {volume} {106}},\
  \bibinfo {pages} {214102} (\bibinfo {year} {2011})}\BibitemShut {NoStop}%
\bibitem [{\citenamefont {Guven}\ and\ \citenamefont
  {V{\'a}zquez-Montejo}(2012)}]{GuvenVazquez2012}%
  \BibitemOpen
  \bibfield  {author} {\bibinfo {author} {\bibfnamefont {J.}~\bibnamefont
  {Guven}}\ and\ \bibinfo {author} {\bibfnamefont {P.}~\bibnamefont
  {V{\'a}zquez-Montejo}},\ }\href {https://doi.org/10.1103/PhysRevE.85.026603}
  {\bibfield  {journal} {\bibinfo  {journal} {Phys. Rev. E}\ }\textbf {\bibinfo
  {volume} {85}},\ \bibinfo {pages} {026603} (\bibinfo {year}
  {2012})}\BibitemShut {NoStop}%
\bibitem [{\citenamefont {Evans}\ and\ \citenamefont
  {Skalak}(1980)}]{EvansSkalak1980}%
  \BibitemOpen
  \bibfield  {author} {\bibinfo {author} {\bibfnamefont {E.~A.}\ \bibnamefont
  {Evans}}\ and\ \bibinfo {author} {\bibfnamefont {R.}~\bibnamefont {Skalak}},\
  }\href {https://doi.org/10.1016/0014-5793(82)80251-2} {\emph {\bibinfo
  {title} {{Mechanics and Thermodynamics of Biomembranes}}}}\ (\bibinfo
  {publisher} {CRC Press, Boca Raton, Florida},\ \bibinfo {year}
  {1980})\BibitemShut {NoStop}%
\bibitem [{\citenamefont {Jenkins}(1977)}]{Jenkins1977}%
  \BibitemOpen
  \bibfield  {author} {\bibinfo {author} {\bibfnamefont {J.}~\bibnamefont
  {Jenkins}},\ }\href {https://doi.org/10.1137/0132063} {\bibfield  {journal}
  {\bibinfo  {journal} {SIAM Journal on Applied Mathematics}\ }\textbf
  {\bibinfo {volume} {32}},\ \bibinfo {pages} {755} (\bibinfo {year}
  {1977})}\BibitemShut {NoStop}%
\bibitem [{\citenamefont {Steigmann}(1999)}]{Steigman1999}%
  \BibitemOpen
  \bibfield  {author} {\bibinfo {author} {\bibfnamefont {D.~J.}\ \bibnamefont
  {Steigmann}},\ }\href {https://doi.org/10.1007/s002050050183} {\bibfield
  {journal} {\bibinfo  {journal} {Archive for Rational Mechanics and Analysis}\
  }\textbf {\bibinfo {volume} {150}},\ \bibinfo {pages} {127} (\bibinfo {year}
  {1999})}\BibitemShut {NoStop}%
\bibitem [{\citenamefont {Capovilla}\ and\ \citenamefont
  {Guven}(2002)}]{CapoGuven2002}%
  \BibitemOpen
  \bibfield  {author} {\bibinfo {author} {\bibfnamefont {R.}~\bibnamefont
  {Capovilla}}\ and\ \bibinfo {author} {\bibfnamefont {J.}~\bibnamefont
  {Guven}},\ }\href {http://stacks.iop.org/0305-4470/35/i=30/a=302} {\bibfield
  {journal} {\bibinfo  {journal} {Journal of Physics A: Mathematical and
  General}\ }\textbf {\bibinfo {volume} {35}},\ \bibinfo {pages} {6233}
  (\bibinfo {year} {2002})}\BibitemShut {NoStop}%
\bibitem [{\citenamefont {Guven}(2004)}]{Guven2004}%
  \BibitemOpen
  \bibfield  {author} {\bibinfo {author} {\bibfnamefont {J.}~\bibnamefont
  {Guven}},\ }\href {http://stacks.iop.org/0305-4470/37/i=28/a=L02} {\bibfield
  {journal} {\bibinfo  {journal} {Journal of Physics A: Mathematical and
  General}\ }\textbf {\bibinfo {volume} {37}},\ \bibinfo {pages} {L313}
  (\bibinfo {year} {2004})}\BibitemShut {NoStop}%
\bibitem [{\citenamefont {Lomholt}\ and\ \citenamefont
  {Miao}(2006)}]{Lomholt2006}%
  \BibitemOpen
  \bibfield  {author} {\bibinfo {author} {\bibfnamefont {M.~A.}\ \bibnamefont
  {Lomholt}}\ and\ \bibinfo {author} {\bibfnamefont {L.}~\bibnamefont {Miao}},\
  }\href {http://stacks.iop.org/0305-4470/39/i=33/a=005} {\bibfield  {journal}
  {\bibinfo  {journal} {Journal of Physics A: Mathematical and General}\
  }\textbf {\bibinfo {volume} {39}},\ \bibinfo {pages} {10323} (\bibinfo {year}
  {2006})}\BibitemShut {NoStop}%
\bibitem [{\citenamefont {Deserno}(2015)}]{Deserno2015}%
  \BibitemOpen
  \bibfield  {author} {\bibinfo {author} {\bibfnamefont {M.}~\bibnamefont
  {Deserno}},\ }\href
  {https://doi.org/https://doi.org/10.1016/j.chemphyslip.2014.05.001}
  {\bibfield  {journal} {\bibinfo  {journal} {Chemistry and Physics of Lipids}\
  }\textbf {\bibinfo {volume} {185}},\ \bibinfo {pages} {11} (\bibinfo {year}
  {2015})}\BibitemShut {NoStop}%
\bibitem [{\citenamefont {Guven}\ and\ \citenamefont
  {V{\'a}zquez-Montejo}(2018)}]{Guven2018}%
  \BibitemOpen
  \bibfield  {author} {\bibinfo {author} {\bibfnamefont {J.}~\bibnamefont
  {Guven}}\ and\ \bibinfo {author} {\bibfnamefont {P.}~\bibnamefont
  {V{\'a}zquez-Montejo}},\ }\bibinfo {title} {The geometry of fluid membranes:
  Variational principles, symmetries and conservation laws},\ in\ \href
  {https://doi.org/10.1007/978-3-319-56348-0_4} {\emph {\bibinfo {booktitle}
  {The Role of Mechanics in the Study of Lipid Bilayers}}},\ \bibinfo {editor}
  {edited by\ \bibinfo {editor} {\bibfnamefont {D.~J.}\ \bibnamefont
  {Steigmann}}}\ (\bibinfo  {publisher} {Springer International Publishing},\
  \bibinfo {address} {Cham},\ \bibinfo {year} {2018})\ pp.\ \bibinfo {pages}
  {167--219}\BibitemShut {NoStop}%
\bibitem [{\citenamefont {Fourcade}\ \emph {et~al.}(1994)\citenamefont
  {Fourcade}, \citenamefont {Miao}, \citenamefont {Rao}, \citenamefont
  {Wortis},\ and\ \citenamefont {Zia}}]{Fourcade1994}%
  \BibitemOpen
  \bibfield  {author} {\bibinfo {author} {\bibfnamefont {B.}~\bibnamefont
  {Fourcade}}, \bibinfo {author} {\bibfnamefont {L.}~\bibnamefont {Miao}},
  \bibinfo {author} {\bibfnamefont {M.}~\bibnamefont {Rao}}, \bibinfo {author}
  {\bibfnamefont {M.}~\bibnamefont {Wortis}},\ and\ \bibinfo {author}
  {\bibfnamefont {R.~K.~P.}\ \bibnamefont {Zia}},\ }\href
  {https://doi.org/10.1103/PhysRevE.49.5276} {\bibfield  {journal} {\bibinfo
  {journal} {Phys. Rev. E}\ }\textbf {\bibinfo {volume} {49}},\ \bibinfo
  {pages} {5276} (\bibinfo {year} {1994})}\BibitemShut {NoStop}%
\bibitem [{\citenamefont {Agudo-Canalejo}\ and\ \citenamefont
  {Lipowsky}(2016)}]{Agudo2016}%
  \BibitemOpen
  \bibfield  {author} {\bibinfo {author} {\bibfnamefont {J.}~\bibnamefont
  {Agudo-Canalejo}}\ and\ \bibinfo {author} {\bibfnamefont {R.}~\bibnamefont
  {Lipowsky}},\ }\href {https://doi.org/10.1039/C6SM01481J} {\bibfield
  {journal} {\bibinfo  {journal} {Soft Matter}\ }\textbf {\bibinfo {volume}
  {12}},\ \bibinfo {pages} {8155–8166} (\bibinfo {year} {2016})}\BibitemShut
  {NoStop}%
\bibitem [{\citenamefont {Lipowsky}(2022)}]{Lipowsky2022}%
  \BibitemOpen
  \bibfield  {author} {\bibinfo {author} {\bibfnamefont {R.}~\bibnamefont
  {Lipowsky}},\ }\href {https://doi.org/https://doi.org/10.1002/adbi.202101020}
  {\bibfield  {journal} {\bibinfo  {journal} {Advanced Biology}\ }\textbf
  {\bibinfo {volume} {6}},\ \bibinfo {pages} {2101020} (\bibinfo {year}
  {2022})}\BibitemShut {NoStop}%
\bibitem [{\citenamefont {Lipowsky}(2024)}]{Lipowsky2024}%
  \BibitemOpen
  \bibfield  {author} {\bibinfo {author} {\bibfnamefont {R.}~\bibnamefont
  {Lipowsky}},\ }\href {https://doi.org/10.1140/epje/s10189-023-00399-z}
  {\bibfield  {journal} {\bibinfo  {journal} {The European Physical Journal E}\
  }\textbf {\bibinfo {volume} {47}},\ \bibinfo {pages} {4} (\bibinfo {year}
  {2024})}\BibitemShut {NoStop}%
\bibitem [{\citenamefont {Bo\v{z}i\v{c}}\ \emph {et~al.}(2014)\citenamefont
  {Bo\v{z}i\v{c}}, \citenamefont {Guven}, \citenamefont {V{\'a}zquez-Montejo},\
  and\ \citenamefont {Svetina}}]{Bozic2014}%
  \BibitemOpen
  \bibfield  {author} {\bibinfo {author} {\bibfnamefont {B.}~\bibnamefont
  {Bo\v{z}i\v{c}}}, \bibinfo {author} {\bibfnamefont {J.}~\bibnamefont
  {Guven}}, \bibinfo {author} {\bibfnamefont {P.}~\bibnamefont
  {V{\'a}zquez-Montejo}},\ and\ \bibinfo {author} {\bibfnamefont
  {S.}~\bibnamefont {Svetina}},\ }\href
  {https://doi.org/10.1103/PhysRevE.89.052701} {\bibfield  {journal} {\bibinfo
  {journal} {Phys. Rev. E}\ }\textbf {\bibinfo {volume} {89}},\ \bibinfo
  {pages} {052701} (\bibinfo {year} {2014})}\BibitemShut {NoStop}%
\bibitem [{\citenamefont {do~Carmo}(2016)}]{DoCarmoBook}%
  \BibitemOpen
  \bibfield  {author} {\bibinfo {author} {\bibfnamefont {M.~P.}\ \bibnamefont
  {do~Carmo}},\ }\href@noop {} {\emph {\bibinfo {title} {{Differential Geometry
  of Curves and Surfaces}}}},\ \bibinfo {edition} {2nd}\ ed.\ (\bibinfo
  {publisher} {Dover},\ \bibinfo {address} {New York},\ \bibinfo {year}
  {2016})\BibitemShut {NoStop}%
\bibitem [{\citenamefont {Canham}(1970)}]{Canham1970}%
  \BibitemOpen
  \bibfield  {author} {\bibinfo {author} {\bibfnamefont {P.}~\bibnamefont
  {Canham}},\ }\href {https://doi.org/10.1016/S0022-5193(70)80032-7} {\bibfield
   {journal} {\bibinfo  {journal} {Journal of Theoretical Biology}\ }\textbf
  {\bibinfo {volume} {26}},\ \bibinfo {pages} {61} (\bibinfo {year}
  {1970})}\BibitemShut {NoStop}%
\bibitem [{\citenamefont {Helfrich}(1973)}]{Helfrich1973}%
  \BibitemOpen
  \bibfield  {author} {\bibinfo {author} {\bibfnamefont {W.}~\bibnamefont
  {Helfrich}},\ }\href
  {http://zfn.mpdl.mpg.de/data/Reihe_C/28/ZNC-1973-28c-0693.pdf} {\bibfield
  {journal} {\bibinfo  {journal} {Z. Naturforsch. C}\ }\textbf {\bibinfo
  {volume} {28}},\ \bibinfo {pages} {693} (\bibinfo {year} {1973})}\BibitemShut
  {NoStop}%
\bibitem [{\citenamefont {Willmore}(1982)}]{Willmore}%
  \BibitemOpen
  \bibfield  {author} {\bibinfo {author} {\bibfnamefont {T.~J.}\ \bibnamefont
  {Willmore}},\ }\href@noop {} {\emph {\bibinfo {title} {{Total curvature in
  Riemannian geometry}}}},\ edited by\ \bibinfo {editor} {\bibfnamefont
  {E.}~\bibnamefont {Horwood}}\ (\bibinfo  {publisher} {Halsted Press},\
  \bibinfo {address} {New York},\ \bibinfo {year} {1982})\BibitemShut {NoStop}%
\bibitem [{\citenamefont {Seifert}(1997)}]{SeifertAP}%
  \BibitemOpen
  \bibfield  {author} {\bibinfo {author} {\bibfnamefont {U.}~\bibnamefont
  {Seifert}},\ }\href {https://doi.org/10.1080/00018739700101488} {\bibfield
  {journal} {\bibinfo  {journal} {Advances in Physics}\ }\textbf {\bibinfo
  {volume} {46}},\ \bibinfo {pages} {13} (\bibinfo {year} {1997})}\BibitemShut
  {NoStop}%
\bibitem [{\citenamefont {Lipowsky}(2019)}]{LipowskyBook}%
  \BibitemOpen
  \bibfield  {author} {\bibinfo {author} {\bibfnamefont {R.}~\bibnamefont
  {Lipowsky}},\ }\bibinfo {title} {Understanding giant vesicles: A theoretical
  perspective},\ in\ \href {https://doi.org/10.1201/9781315152516} {\emph
  {\bibinfo {booktitle} {The Giant Vesicle Book}}},\ \bibinfo {editor} {edited
  by\ \bibinfo {editor} {\bibfnamefont {R.}~\bibnamefont {Dimova}}\ and\
  \bibinfo {editor} {\bibfnamefont {C.}~\bibnamefont {Marques}}}\ (\bibinfo
  {publisher} {CRC Press},\ \bibinfo {address} {Boca Raton},\ \bibinfo {year}
  {2019})\ pp.\ \bibinfo {pages} {73--168}\BibitemShut {NoStop}%
\bibitem [{\citenamefont {Guven}\ \emph {et~al.}(2014)\citenamefont {Guven},
  \citenamefont {Valencia},\ and\ \citenamefont
  {V{\'a}zquez-Montejo}}]{GuvValVaz2014}%
  \BibitemOpen
  \bibfield  {author} {\bibinfo {author} {\bibfnamefont {J.}~\bibnamefont
  {Guven}}, \bibinfo {author} {\bibfnamefont {D.~M.}\ \bibnamefont
  {Valencia}},\ and\ \bibinfo {author} {\bibfnamefont {P.}~\bibnamefont
  {V{\'a}zquez-Montejo}},\ }\href {https://doi.org/10.1088/1751-8113}
  {\bibfield  {journal} {\bibinfo  {journal} {Journal of Physics A:
  Mathematical and Theoretical}\ }\textbf {\bibinfo {volume} {47}},\ \bibinfo
  {pages} {355201} (\bibinfo {year} {2014})}\BibitemShut {NoStop}%
\bibitem [{\citenamefont {Svetina}\ and\ \citenamefont
  {\v{Z}ek\v{s}}(1989)}]{Svetina1989}%
  \BibitemOpen
  \bibfield  {author} {\bibinfo {author} {\bibfnamefont {S.}~\bibnamefont
  {Svetina}}\ and\ \bibinfo {author} {\bibfnamefont {B.}~\bibnamefont
  {\v{Z}ek\v{s}}},\ }\href {https://doi.org/10.1007/BF00257107} {\bibfield
  {journal} {\bibinfo  {journal} {European Biophysics Journal}\ }\textbf
  {\bibinfo {volume} {17}},\ \bibinfo {pages} {101} (\bibinfo {year}
  {1989})}\BibitemShut {NoStop}%
\bibitem [{\citenamefont {J\"ulicher}\ and\ \citenamefont
  {Lipowsky}(1996)}]{Julicher1996}%
  \BibitemOpen
  \bibfield  {author} {\bibinfo {author} {\bibfnamefont {F.}~\bibnamefont
  {J\"ulicher}}\ and\ \bibinfo {author} {\bibfnamefont {R.}~\bibnamefont
  {Lipowsky}},\ }\href {https://doi.org/10.1103/PhysRevE.53.2670} {\bibfield
  {journal} {\bibinfo  {journal} {Phys. Rev. E}\ }\textbf {\bibinfo {volume}
  {53}},\ \bibinfo {pages} {2670} (\bibinfo {year} {1996})}\BibitemShut
  {NoStop}%
\bibitem [{\citenamefont {Zhong-can}\ and\ \citenamefont
  {Helfrich}(1989)}]{Zhong-Can1989}%
  \BibitemOpen
  \bibfield  {author} {\bibinfo {author} {\bibfnamefont {O.-Y.}\ \bibnamefont
  {Zhong-can}}\ and\ \bibinfo {author} {\bibfnamefont {W.}~\bibnamefont
  {Helfrich}},\ }\href {https://doi.org/10.1103/PhysRevA.39.5280} {\bibfield
  {journal} {\bibinfo  {journal} {Phys. Rev. A}\ }\textbf {\bibinfo {volume}
  {39}},\ \bibinfo {pages} {5280} (\bibinfo {year} {1989})}\BibitemShut
  {NoStop}%
\bibitem [{\citenamefont {Farago}\ and\ \citenamefont
  {Santangelo}(2005)}]{Farago2005}%
  \BibitemOpen
  \bibfield  {author} {\bibinfo {author} {\bibfnamefont {O.}~\bibnamefont
  {Farago}}\ and\ \bibinfo {author} {\bibfnamefont {C.~D.}\ \bibnamefont
  {Santangelo}},\ }\href {https://doi.org/10.1063/1.1835952} {\bibfield
  {journal} {\bibinfo  {journal} {The Journal of Chemical Physics}\ }\textbf
  {\bibinfo {volume} {122}},\ \bibinfo {eid} {044901} (\bibinfo {year}
  {2005})}\BibitemShut {NoStop}%
\bibitem [{\citenamefont {Illya}\ and\ \citenamefont
  {Deserno}(2008)}]{Illya2008}%
  \BibitemOpen
  \bibfield  {author} {\bibinfo {author} {\bibfnamefont {G.}~\bibnamefont
  {Illya}}\ and\ \bibinfo {author} {\bibfnamefont {M.}~\bibnamefont
  {Deserno}},\ }\href {https://doi.org/10.1529/biophysj.108.131300} {\bibfield
  {journal} {\bibinfo  {journal} {Biophysical Journal}\ }\textbf {\bibinfo
  {volume} {95(9)}},\ \bibinfo {pages} {4163} (\bibinfo {year}
  {2008})}\BibitemShut {NoStop}%
\bibitem [{\citenamefont {J\"ulicher}\ and\ \citenamefont
  {Lipowsky}(1993)}]{Julicher1993}%
  \BibitemOpen
  \bibfield  {author} {\bibinfo {author} {\bibfnamefont {F.}~\bibnamefont
  {J\"ulicher}}\ and\ \bibinfo {author} {\bibfnamefont {R.}~\bibnamefont
  {Lipowsky}},\ }\href {https://doi.org/10.1103/PhysRevLett.70.2964} {\bibfield
   {journal} {\bibinfo  {journal} {Phys. Rev. Lett.}\ }\textbf {\bibinfo
  {volume} {70}},\ \bibinfo {pages} {2964} (\bibinfo {year}
  {1993})}\BibitemShut {NoStop}%
\bibitem [{\citenamefont {J{\"u}licher}\ and\ \citenamefont
  {Seifert}(1994)}]{Julicher1994}%
  \BibitemOpen
  \bibfield  {author} {\bibinfo {author} {\bibfnamefont {F.}~\bibnamefont
  {J{\"u}licher}}\ and\ \bibinfo {author} {\bibfnamefont {U.}~\bibnamefont
  {Seifert}},\ }\href {https://doi.org/10.1103/PhysRevE.49.4728} {\bibfield
  {journal} {\bibinfo  {journal} {Phys. Rev. E}\ }\textbf {\bibinfo {volume}
  {49}},\ \bibinfo {pages} {4728} (\bibinfo {year} {1994})}\BibitemShut
  {NoStop}%
\bibitem [{\citenamefont {Bou\'e}\ \emph {et~al.}(2006)\citenamefont {Bou\'e},
  \citenamefont {Adda-Bedia}, \citenamefont {Boudaoud}, \citenamefont
  {Cassani}, \citenamefont {Couder}, \citenamefont {Eddi},\ and\ \citenamefont
  {Trejo}}]{Boue2006}%
  \BibitemOpen
  \bibfield  {author} {\bibinfo {author} {\bibfnamefont {L.}~\bibnamefont
  {Bou\'e}}, \bibinfo {author} {\bibfnamefont {M.}~\bibnamefont {Adda-Bedia}},
  \bibinfo {author} {\bibfnamefont {A.}~\bibnamefont {Boudaoud}}, \bibinfo
  {author} {\bibfnamefont {D.}~\bibnamefont {Cassani}}, \bibinfo {author}
  {\bibfnamefont {Y.}~\bibnamefont {Couder}}, \bibinfo {author} {\bibfnamefont
  {A.}~\bibnamefont {Eddi}},\ and\ \bibinfo {author} {\bibfnamefont
  {M.}~\bibnamefont {Trejo}},\ }\href
  {https://doi.org/10.1103/PhysRevLett.97.166104} {\bibfield  {journal}
  {\bibinfo  {journal} {Phys. Rev. Lett.}\ }\textbf {\bibinfo {volume} {97}},\
  \bibinfo {pages} {166104} (\bibinfo {year} {2006})}\BibitemShut {NoStop}%
\bibitem [{\citenamefont {Kahraman}\ \emph {et~al.}(2012)\citenamefont
  {Kahraman}, \citenamefont {Stoop},\ and\ \citenamefont
  {Müller}}]{Kahraman2012}%
  \BibitemOpen
  \bibfield  {author} {\bibinfo {author} {\bibfnamefont {O.}~\bibnamefont
  {Kahraman}}, \bibinfo {author} {\bibfnamefont {N.}~\bibnamefont {Stoop}},\
  and\ \bibinfo {author} {\bibfnamefont {M.~M.}\ \bibnamefont {Müller}},\
  }\href {https://doi.org/10.1209/0295-5075/97/68008} {\bibfield  {journal}
  {\bibinfo  {journal} {Europhysics Letters}\ }\textbf {\bibinfo {volume}
  {97}},\ \bibinfo {pages} {68008} (\bibinfo {year} {2012})}\BibitemShut
  {NoStop}%
\bibitem [{\citenamefont {Bouzar}\ \emph {et~al.}(2015)\citenamefont {Bouzar},
  \citenamefont {Menas},\ and\ \citenamefont {M\"uller}}]{Bouzar2015}%
  \BibitemOpen
  \bibfield  {author} {\bibinfo {author} {\bibfnamefont {L.}~\bibnamefont
  {Bouzar}}, \bibinfo {author} {\bibfnamefont {F.}~\bibnamefont {Menas}},\ and\
  \bibinfo {author} {\bibfnamefont {M.~M.}\ \bibnamefont {M\"uller}},\ }\href
  {https://doi.org/10.1103/PhysRevE.92.032721} {\bibfield  {journal} {\bibinfo
  {journal} {Phys. Rev. E}\ }\textbf {\bibinfo {volume} {92}},\ \bibinfo
  {pages} {032721} (\bibinfo {year} {2015})}\BibitemShut {NoStop}%
\bibitem [{\citenamefont {Miao}\ \emph {et~al.}(1991)\citenamefont {Miao},
  \citenamefont {Fourcade}, \citenamefont {Rao}, \citenamefont {Wortis},\ and\
  \citenamefont {Zia}}]{Miao1991}%
  \BibitemOpen
  \bibfield  {author} {\bibinfo {author} {\bibfnamefont {L.}~\bibnamefont
  {Miao}}, \bibinfo {author} {\bibfnamefont {B.}~\bibnamefont {Fourcade}},
  \bibinfo {author} {\bibfnamefont {M.}~\bibnamefont {Rao}}, \bibinfo {author}
  {\bibfnamefont {M.}~\bibnamefont {Wortis}},\ and\ \bibinfo {author}
  {\bibfnamefont {R.~K.~P.}\ \bibnamefont {Zia}},\ }\href
  {https://doi.org/10.1103/PhysRevA.43.6843} {\bibfield  {journal} {\bibinfo
  {journal} {Phys. Rev. A}\ }\textbf {\bibinfo {volume} {43}},\ \bibinfo
  {pages} {6843} (\bibinfo {year} {1991})}\BibitemShut {NoStop}%
\bibitem [{\citenamefont {Castro-Villarreal}\ and\ \citenamefont
  {Guven}(2007{\natexlab{a}})}]{CastroGuven2007PRE}%
  \BibitemOpen
  \bibfield  {author} {\bibinfo {author} {\bibfnamefont {P.}~\bibnamefont
  {Castro-Villarreal}}\ and\ \bibinfo {author} {\bibfnamefont {J.}~\bibnamefont
  {Guven}},\ }\href {https://doi.org/10.1103/PhysRevE.76.011922} {\bibfield
  {journal} {\bibinfo  {journal} {Phys. Rev. E}\ }\textbf {\bibinfo {volume}
  {76}},\ \bibinfo {pages} {011922} (\bibinfo {year}
  {2007}{\natexlab{a}})}\BibitemShut {NoStop}%
\bibitem [{\citenamefont {Castro-Villarreal}\ and\ \citenamefont
  {Guven}(2007{\natexlab{b}})}]{CastroGuven2007JPA}%
  \BibitemOpen
  \bibfield  {author} {\bibinfo {author} {\bibfnamefont {P.}~\bibnamefont
  {Castro-Villarreal}}\ and\ \bibinfo {author} {\bibfnamefont {J.}~\bibnamefont
  {Guven}},\ }\href {http://stacks.iop.org/1751-8121/40/i=16/a=002} {\bibfield
  {journal} {\bibinfo  {journal} {Journal of Physics A: Mathematical and
  Theoretical}\ }\textbf {\bibinfo {volume} {40}},\ \bibinfo {pages} {4273}
  (\bibinfo {year} {2007}{\natexlab{b}})}\BibitemShut {NoStop}%
\bibitem [{\citenamefont {Guven}\ and\ \citenamefont
  {V{\'a}zquez-Montejo}(2013)}]{GuvenVazquez2013}%
  \BibitemOpen
  \bibfield  {author} {\bibinfo {author} {\bibfnamefont {J.}~\bibnamefont
  {Guven}}\ and\ \bibinfo {author} {\bibfnamefont {P.}~\bibnamefont
  {V{\'a}zquez-Montejo}},\ }\href {https://doi.org/10.1103/PhysRevE.87.042710}
  {\bibfield  {journal} {\bibinfo  {journal} {Phys. Rev. E}\ }\textbf {\bibinfo
  {volume} {87}},\ \bibinfo {pages} {042710} (\bibinfo {year}
  {2013})}\BibitemShut {NoStop}%
\bibitem [{\citenamefont {Bozic}\ \emph {et~al.}(1992)\citenamefont {Bozic},
  \citenamefont {Svetina}, \citenamefont {Zeks},\ and\ \citenamefont
  {Waugh}}]{Bozic1992}%
  \BibitemOpen
  \bibfield  {author} {\bibinfo {author} {\bibfnamefont {B.}~\bibnamefont
  {Bozic}}, \bibinfo {author} {\bibfnamefont {S.}~\bibnamefont {Svetina}},
  \bibinfo {author} {\bibfnamefont {B.}~\bibnamefont {Zeks}},\ and\ \bibinfo
  {author} {\bibfnamefont {R.}~\bibnamefont {Waugh}},\ }\href
  {https://doi.org/10.1016/S0006-3495(92)81903-3} {\bibfield  {journal}
  {\bibinfo  {journal} {Biophys J.}\ }\textbf {\bibinfo {volume} {61}},\
  \bibinfo {pages} {963} (\bibinfo {year} {1992})}\BibitemShut {NoStop}%
\bibitem [{\citenamefont {Svetina}\ and\ \citenamefont
  {\v{Z}ek\v{s}}(2014)}]{Svetina2014}%
  \BibitemOpen
  \bibfield  {author} {\bibinfo {author} {\bibfnamefont {S.}~\bibnamefont
  {Svetina}}\ and\ \bibinfo {author} {\bibfnamefont {B.}~\bibnamefont
  {\v{Z}ek\v{s}}},\ }\href
  {https://doi.org/https://doi.org/10.1016/j.cis.2014.01.010} {\bibfield
  {journal} {\bibinfo  {journal} {Advances in Colloid and Interface Science}\
  }\textbf {\bibinfo {volume} {208}},\ \bibinfo {pages} {189} (\bibinfo {year}
  {2014})}\BibitemShut {NoStop}%
\bibitem [{\citenamefont {Christ}\ \emph {et~al.}(2021)\citenamefont {Christ},
  \citenamefont {Litschel}, \citenamefont {Schwille},\ and\ \citenamefont
  {Lipowsky}}]{Christ2021}%
  \BibitemOpen
  \bibfield  {author} {\bibinfo {author} {\bibfnamefont {S.}~\bibnamefont
  {Christ}}, \bibinfo {author} {\bibfnamefont {T.}~\bibnamefont {Litschel}},
  \bibinfo {author} {\bibfnamefont {P.}~\bibnamefont {Schwille}},\ and\
  \bibinfo {author} {\bibfnamefont {R.}~\bibnamefont {Lipowsky}},\ }\href
  {https://doi.org/10.1039/D0SM00790K} {\bibfield  {journal} {\bibinfo
  {journal} {Soft Matter}\ }\textbf {\bibinfo {volume} {17}},\ \bibinfo {pages}
  {319} (\bibinfo {year} {2021})}\BibitemShut {NoStop}%
\bibitem [{\citenamefont {Vetter}\ \emph {et~al.}(2014)\citenamefont {Vetter},
  \citenamefont {Wittel},\ and\ \citenamefont {Herrmann}}]{Vetter2014}%
  \BibitemOpen
  \bibfield  {author} {\bibinfo {author} {\bibfnamefont {R.}~\bibnamefont
  {Vetter}}, \bibinfo {author} {\bibfnamefont {F.~K.}\ \bibnamefont {Wittel}},\
  and\ \bibinfo {author} {\bibfnamefont {H.~J.}\ \bibnamefont {Herrmann}},\
  }\href {http://dx.doi.org/10.1038/ncomms5437} {\bibfield  {journal} {\bibinfo
   {journal} {Nature Communications}\ }\textbf {\bibinfo {volume} {5}},\
  \bibinfo {pages} {5437} (\bibinfo {year} {2014})}\BibitemShut {NoStop}%
\end{thebibliography}%

\end{document}